\documentstyle[11pt,fullpage]{article}
\def\rf#1{(\ref{eq:#1})}
\def\lab#1{\label{eq:#1}}
\def\nonu{\nonumber}
\def\br{\begin{eqnarray}}
\def\er{\end{eqnarray}}
\def\be{\begin{equation}}
\def\ee{\end{equation}}

\def\foot#1{\footnotemark\footnotetext{#1}}
\def\lb{\lbrack}
\def\rb{\rbrack}
\def\llangle{\left\langle}
\def\rrangle{\right\rangle}

\def\llb{\left\lbrack}
\def\rrb{\right\rbrack}

\def\lcurl{\left\{}
\def\rcurl{\right\}}
\def\({\left(}
\def\){\right)}
\def\v{\vert}                     
\def\bv{\bigm\vert}               
\def\Bgv{\;\Bigg\vert}            
\def\bgv{\bigg\vert}              
\def\lskip{\vskip\baselineskip\vskip-\parskip\noindent}
\def\mskp{\par\vskip 0.3cm \par\noindent}
\def\sskp{\par\vskip 0.15cm \par\noindent}
\def\bc{\begin{center}}
\def\ec{\end{center}}

\newcommand\partder[2]{{{\partial {#1}}\over{\partial {#2}}}}

\newcommand\sbr[2]{\left\lbrack\,{#1}\, ,\,{#2}\,\right\rbrack} 
\newcommand\Sbr[2]{\Bigl\lbrack\,{#1}\, ,\,{#2}\,\Bigr\rbrack} 
 
\def\a{\alpha}
\def\b{\beta}

\def\d{\delta}

\def\eps{\epsilon}
\def\vareps{\varepsilon}

\def\h{{1\over 2}}

\def\l{\lambda}

\def\m{\mu}

\def\o{\over}
\def\om{\omega}
\def\O{\Omega}

\def\vp{\varphi}
\def\P{\Phi}
\def\pa{\partial}

\def\bpa{{\bar \partial}}
\def\pr{\prime}

\def\s{\sigma}
\def\S{\Sigma}
\def\t{\tau}
\def\th{\theta}

\def\Th{\Theta}

\def\wti{\widetilde}

\newcommand\sumi[1]{\sum_{#1}^{\infty}}   
\newcommand\twomat[4]{\left(\begin{array}{cc}  
{#1} & {#2} \\ {#3} & {#4} \end{array} \right)}

\newcommand\nxnrmat[9]{\left(\begin{array}{cccc}  
{#1} & \cdots & {#2} & {#3}\\ \vdots & \ddots & \vdots & \vdots \\
{#4} & \cdots & {#5} & {#6}\\
{#7} & \cdots & {#8} & {#9} \end{array} \right)}

\newcommand\ndcol[3]{\left(\begin{array}{cccc}  
{#1} \\ \vdots \\ {#2} \\ {#3} \end{array} \right)}
\newcommand\nmcol[5]{\left(\begin{array}{cccccccc}  
{#1} \\ \vdots \\ {#2} \\ {#3} \\ {#4}\\ \vdots \\
{#5} \end{array} \right)}

\def\cA{{\cal A}}
\def\cB{{\cal B}}
\def\cC{{\cal C}}

\def\cE{{\cal E}}

\def\cG{{\cal G}}
\def\cH{{\cal H}}

\def\cK{{\cal K}}
\def\cL{{\cal L}}
\def\cM{{\cal M}}

\def\cO{{\cal O}}

\def\cT{{\cal T}}

\def\cW{{\cal W}}

\def\lie{{\cal G}}
\def\dlie{{\cal G}^{\ast}}
\def\elie{{\widetilde \lie}}
\def\edlie{{\elie}^{\ast}}

\font \msb=msbm10 scaled \magstep1
\newcommand{\IR}{\mbox{\msb R} }

\newcommand{\IZ}{\mbox{\msb Z} }

\newfam\euffam
\font\sixeuf=eufm6
\font\eighteuf=eufm8
\font\twelveeuf=eufm10 scaled\magstep1
 \textfont\euffam=\twelveeuf
 \scriptfont\euffam=\eighteuf
   \scriptscriptfont\euffam=\sixeuf
\def\euf{\fam\euffam\twelveeuf}
\def\sg{{\euf s}}

\def\one{\hbox{{1}\kern-.25em\hbox{l}}}
\def\0#1{\relax\ifmmode\mathaccent"7017{#1}%
        \else\accent23#1\relax\fi}

\def\mark{\noindent{\bf Remark.}\quad}

\def\proof{\par{\it Proof}. \ignorespaces} \def\endproof{{$\Box$}\par}

\def\Ouc{{\cal O}_{(U_0 ,c)}}           
\def\Gs{G_{stat}}                            
\def\As{{\lie}_{stat}}                             
\def\Suc#1{\Sigma \Bigl( #1 ; (U_0 ,c) \Bigr)}        
\def\suc#1{{\hat \sigma} (#1 ; (U_0 ,c))}
\def\sh{\hat s}                              
\def\ssh#1{{\hat \sigma}^{#1}}

\def\yt{Y_t (g)}

\def\Tu#1{{\widetilde \Theta}^{#1}}
\def\Td#1{{\widetilde \Theta}_{#1}}
\def\Z{\widetilde Z}

\def\DN{{\left\lbrack D {\widetilde \Theta} \right\rbrack}^2_N}

\def\Tor{{\wti {\rm SDiff}}\, (T^2 )}
\def\Lh#1{{\hat {\cal L}}({#1})}

\def\DO{DOP (S^1 )}                           
\def\DA{{\cal DOP} (S^1 )}                    
\def\dDA{{\cal DOP}^{\ast} (S^1 )}                  

\def\sta{\, ,\,}

\newcommand\DB{{Darboux-B\"{a}cklund}~}

\def\Res{{\rm Res}}

\newcommand\X{{\widehat {\cal X}} \(\l, \mu\)}  
\def\bt{{\bar t}}
\def\pai{\partial^{-1}\!}

\newcommand\st[2]{\stackrel{(#1 )}{#2}}

\newcommand\stta[2]{\stackrel{(#1 )}{t_{#2}}}

\newcommand{\ct}[1]{\cite{#1}}
\newcommand{\bi}[1]{\bibitem{#1}}
\newcommand\PRL[3]{{\sl Phys. Rev. Lett.} {\bf#1} (#2) #3}
\newcommand\NPB[3]{{\sl Nucl. Phys.} {\bf B#1} (#2) #3}

\newcommand\CMP[3]{{\sl Commun. Math. Phys.} {\bf #1} (#2) #3}

\newcommand\PLA[3]{{\sl Phys. Lett.} {\bf #1A} (#2) #3}
\newcommand\PLB[3]{{\sl Phys. Lett.} {\bf #1B} (#2) #3}

\newcommand\JMP[3]{{\sl J. Math. Phys.} {\bf #1} (#2) #3}

\newcommand\AoP[3]{{\sl Ann. of Phys.} {\bf #1} (#2) #3}

\newcommand\PR[3]{{\sl Phys. Reports} {\bf #1} (#2) #3}

\newcommand\LMP[3]{{\sl Letters in Math. Phys.} {\bf #1} (#2) #3}
\newcommand\IJMPA[3]{{\sl Int. J. Mod. Phys.} {\bf A#1} (#2) #3}
\newcommand\TMP[3]{{\sl Theor. Math. Phys.} {\bf #1} (#2) #3}
\newcommand\JPA[3]{{\sl J. Physics} {\bf A#1} (#2) #3}
\newcommand\JSM[3]{{\sl J. Soviet Math.} {\bf #1} (#2) #3}
\newcommand\MPLA[3]{{\sl Mod. Phys. Lett.} {\bf A#1} (#2) #3}

\newcommand\PHSA[3]{{\sl Physica} {\bf A#1} (#2) #3}

\newcommand\JGP[3]{{\sl J. Geom. Phys.} {\bf #1} (#2) #3}

\begin{document}
\noindent\null\hfill {hep-th/0009194}
 
\vskip .3in
 
\begin{center}
{\large {\bf Gauging of Geometric Actions and 
Integrable Hierarchies of KP Type}}
\end{center}
 
\vskip .3in
 
\begin{center}
E. Nissimov${}^{1}$ and S. Pacheva${}^{1}$
\footnotetext[1]{E-mail: nissimov@inrne.bas.bg , svetlana@inrne.bas.bg} \\
{\small Institute for Nuclear Research and Nuclear Energy, 
Bulgarian Academy of Sciences} \\
{\small Boul. Tsarigradsko Chausee 72, BG-1784 ~Sofia, Bulgaria}
\end{center}
\lskip
\begin{abstract}
This work consist of two interrelated parts. First, we derive massive
gauge-invariant generalizations of geometric actions on coadjoint orbits of
arbitrary (infinite-dimensional) groups $G$ with central extensions, with
gauge group $H$ being certain (infinite-dimensional) subgroup of $G$.
We show that there exist generalized ``zero-curvature'' representation
of the pertinent equations of motion on the coadjoint orbit. Second,
in the special case of $G$ being Kac-Moody group the equations of
motion of the underlying gauged WZNW geometric action are identified as
additional-symmetry flows of generalized Drinfeld-Sokolov integrable
hierarchies based on the loop algebra ${\widehat \cG}$. For
${\widehat \cG} = {\widehat {SL}}(M+R)$ the latter hiearchies are equivalent
to a class of constrained (reduced) KP hierarchies called ${\sl cKP}_{R,M}$,
which contain as special cases a series of well-known integrable systems
(mKdV, AKNS, Fordy-Kulish, Yajima-Oikawa {\sl etc.}). We describe in some
detail the loop algebras of additional (non-isospectral) symmetries of
${\sl cKP}_{R,M}$ hierarchies. Apart from gauged WZNW models, certain
higher-dimensional nonlinear systems such as Davey-Stewartson and $N$-wave
resonant systems are also identified as additional symmetry flows of
${\sl cKP}_{R,M}$ hierarchies. Along the way we exhibit explicitly the 
interrelation between the Sato pseudo-differential operator formulation and 
the algebraic (generalized) Drinfeld-Sokolov formulation of ${\sl cKP}_{R,M}$
hierarchies. Also we present the explicit derivation of the general \DB solutions
of ${\sl cKP}_{R,M}$ preserving their additional (non-isospectral) symmetries,
which for $R=1$ contain among themselves solutions to the gauged 
$SL(M+1)/U(1)\times SL(M)\,$ WZNW field equations.
\end{abstract}
\lskip 
{\large {\bf 1. Introduction}}
\mskp
Wess-Zumino-Novikov-Witten (WZNW) models \ct{WZNW} originally appeared in
the literature as effective actions incorporating the anomalies in gauge
theories with quantized chiral fermionic matter fields. Subsequently, with the
advent of string theory WZNW models and their gauged versions
\ct{gauged-WZNW} (embracing a large class of $D=2$ conformal field theories
which provide Lagrangian description of Goddard-Kent-Olive coset construction
\ct{GKO}) have been recognized as fundamental building blocks of the latter
describing various ground states of string dynamics.

Standard WZNW models and their gauged versions are themselves special cases
(when the dynamical fields take values in Kac-Moody groups) of a much
broader class of geometric dynamical (Hamiltonian) models on coadjoint
orbits of arbitrary (infinite-dimensional) groups with central extensions.
These coadjoint-orbit models can also be viewed as anomalous quantum effective 
actions of matter fields with gauged (inifinite-dimensional) Noether
symmetry groups. Moreover, all basic properties of standard (gauged) WZNW models,
in particular, the fundamental Polyakov-Wiegmann group composition law \ct{PW},
have their natural extensions \ct{ANPZ-90,kovhid} in the general setting of 
geometric coadjoint-orbit models. Non-trivial examples of such more general 
geometric models include the effective action of induced $D=2$ gravity 
\ct{D2grav,Po90,AFS-AS} and its (extended) supersymmetric generalizations 
\ct{ny-prep,ANPZ-90}; the effective WZNW-type action of the toroidal membrane 
\ct{torus,shorty}; the effective WZNW-type action of induced 
${\bf W_{1+\infty}}$-gravity \ct{dop-tmf}, WZNW models based on two-loop 
Kac-Moody algebras \ct{AFGZ-91}, etc. (see Section 3 below).

In the framework of the general formalism for geometric actions on coadjoint 
orbits of (infinite-dimensional) groups with central extensions proposed in 
ref.\ct{ANPZ-90} each physical model is fully characterized by few
fundamental ingredients: (i) pairing $\langle \cdot \v \cdot \rangle$ between
Lie algebra $\lie$ and its dual $\dlie$; (ii) (non-)trivial Lie algebra
two-cocycle $\omega ( \cdot ,\cdot )$ yielding the central extension and
$\dlie$-valued group one-cocycle associated with the latter; (iii)
the fundamental $\lie$-valued Maurer-Cartan form on $G$.
These ingredients carry the whole information about the symmetry structure of 
the models in question and enter into the general recipe for constructing 
the pertinent coadjoint-orbit geometric action. Classical $r$-matrices and 
Yang-Baxter equations appear naturally in this geometric setting 
\ct{rtex,ny-prep}.

Standard (gauged) WZNW models are integrable in a sense that their
pertinent equations of motion admit ``zero-curvature'' Lax representation
\ct{Leznov-Saveliev}. 
In certain cases gauged WZNW field equations can be shown to belong to
a whole integrable hierarchy of infinitely many soliton-like nonlinear
evolution equations ({\sl i.e.}, integrability in a strong sense). 
Namely, under certain conditions the equations of motion
of gauged WZNW models can be identified as additional (non-isospectral)
symmetry flows of KP-type integrable hierarchies (or generalized
Drinfeld-Sokolov hierarchies from algebraic point of view; see Section 7
below). These properties justify the natural question whether one can
extend the notion of integrability or ``zero-curvature'' Lax representation
of ordinary gauged WZNW models to the general case of gauged geometric actions
on coadjoint orbits of arbitrary (infinite-dimensional) groups with central 
extensions. The affirmative answer to this question is provided in Section 5 
below.

Going back to the identification of ordinary gauged WZNW field equations as
additional symmetry flows of KP-type integrable hierarchies, one observes that
now it is possible to employ the standard \DB techniques from Sato
pseudo-differential operator approach to KP-type hierarchies for a systematic
generation of soliton-like solutions of gauged WZNW equations of motion.

To this end let us recall that Kadomtsev-Petviashvili (KP) hierarchy of 
integrable soliton evolution equations, together with its reductions and 
multi-component (matrix) generalizations, describe a variety of physically 
important nonlinear phenomena (for a review, see {\sl e.g.} 
\ct{KP-gen,Dickey-book}). Constrained (reduced) KP models are intimately 
connected with the matrix models in non-perturbative string theory of 
elementary particles at ultra-high energies (\ct{matrix-models} and references 
therein). They provide an unified description of a number of basic soliton 
equations such as (modified) Korteveg-de-Vries, nonlinear Schr\"{o}dinger 
(AKNS hierarchy in general), Yajima-Oikawa, coupled Boussinesq-type equations 
etc.. Dispersionless limits of KP models were recently found \ct{Wiegmann-cKP} 
to play a fundamental role in the description of interface dynamics 
(the so called Laplacian growth problem). Furthermore, 
multi-component (matrix) KP hierarchies are known to contain such physically
interesting systems as 2-dimensional Toda lattice, Davey-Stewartson,
$N$-wave resonant system etc.. It has been shown recently in ref.\ct{Van-de-Leur}
that multi-component KP tau-functions provide solutions to the basic
Witten-Dijkgraaf-Verlinde-Verlinde equations in topological field theory.

Multi-component (matrix) KP hierarchies can be identified as ordinary (scalar)
one-component KP hierarchies supplemented with a special set of commuting
additional symmetry flows, namely, the Cartan subalgebra of the underlying
loop algebra of additional symmetries. This construction was initially proposed 
in refs.\ct{multi-comp-KP} and it is further elaborated on in subsection 6.4 
below. In particular, Davey-Stewartson \ct{Dav-Stew} and $N$-wave resonant 
systems are shown to arise as symmetry flows of ordinary ${\sl cKP}_{R,M}$ 
hierarchies (see subsection 6.5 below).
The above identification leads to new systematic methods of
constructing soliton-like solutions of multi-component KP hierachies by
employing the well-established techniques of \DB transformations in ordinary
one-component KP hierarchies. These issues are discussed in detail in
Section 8. We present there the explicit construction of \DB orbits of
solutions for the tau-function of ${\sl cKP}_{R,M}$ hierarchies which
simultaneously preserve the additional (non-isospectral) symmetries of the
latter. The pertinent tau-functions are given in terms of generalized
Wronskian-like determinants which contain among themselves solutions of the
$SL(M+1)/U(1)\times SL(M)\,$ gauged WZNW field equations. A subclass of the 
generalized Wronskian-like determinant solutions appears in the form of
{\em multiple Wronskians} which contain as special cases the well-known
(multi-)dromion solutions \ct{Boiti-etal,H-H} of Davey-Stewartson equations. 
\mskp
The plan of exposition in the present paper is as follows. In Section 2 we
briefly recapitulate the main ingredients of the general formalism \ct{ANPZ-90} 
for construction of geometric actions on coadjoint orbits of arbitrary
(infinite-dimensional) groups with central extensions. Section 3 contains
various physically interesting nontrivial examples of such geometric 
coadjoint-orbit actions generalizing the ordinary WZNW actions. In Section 4
we describe the general procedure for gauging of arbitrary coadjoint-orbit 
geometric actions generalizing the usual procedure for gauging of WZNW models.
Section 5 provides the ``zero-curvature'' representation of the equations of
motion of gauged geometric actions on arbitrary group coadjoint orbits.
Section 6 is devoted, after a brief review of Sato pseudo-differential
operator formuation of KP-type integrable hierarchies, to the construction of 
the full loop-algebra of additional (non-isospectral) symmetries for 
constrained (reduced) KP hierarchies. In particular, multi-component (matrix) 
KP hierarchies are obtained out of ordinary one-component (scalar) KP
hierarchies endowed with appropriate infinite sets of commuting additional 
symmetry flows. In Section 7 it is first shown that gauged WZNW equations of
motion can be viewed as additional non-isospectral) symmetry flows of 
generalized Drinfeld-Sokolov hierarchies containing among themselves the class
of ${\sl cKP}_{R,M}$ constrained KP hierarchies. Further, in the case of 
${\sl cKP}_{R,M}$ hierarchies we explicitly demonstrate the
interrelation between Sato pseudo-differential operator and algebraic
(generalized) Drinfeld-Sokolov formulations. Section 8 contains a detailed
systematic construction of \DB solutions for the tau-functions of all
${\sl cKP}_{R,M}$ hierarchies preserving their additional symmetries. As a
byproduct one obtains in this way solutions for the equations of motion of 
gauged $SL(M+1)/U(1)\times SL(M)\,$ WZNW models.

The results of Section 6 have been previously reported in a short form in 
ref.\ct{virflow}.
\lskip
{\large {\bf 2. General Formalism for Geometric Actions on Coadjoint Orbits}}
\lskip
\underline{\em 2.1 Basic Ingredients}
\mskp
Consider arbitrary (infinite-dimensional) group $G$ with a Lie 
algebra $\lie$ and its dual space $\dlie$. The adjoint and coadjoint actions of
$G$ and $\lie$ on $\lie$ and $\dlie$ are given by:
\be
Ad(g) (X ) = gXg^{-1} \quad ,\quad ad(X_1) X_2 = \Sbr{X_1}{X_2}
\lab{adjoint-act}
\ee
\be
\llangle Ad^{*}(g) U \v X \rrangle = \llangle U \v Ad(g^{-1}) X \rrangle 
\quad ,\quad
\llangle ad^{*}(X_1)U \v X_2 \rrangle = - \llangle U \v ad(X_1) X_2 \rrangle
\lab{coadjoint-act}
\ee
Here $g \in G$ and $X,\, X_{1,2} \in \lie \; ,\; U \in \dlie$ are arbitrary 
elements, whereas $\langle \cdot \v \cdot \rangle\,$ indicates the natural 
bilinear form ``pairing" $\lie$ and $\dlie$ .

Our primary interest is in infinite-dimensional Lie algebras with a
central extension $\elie = \lie \oplus \IR$ of $\lie \;$ and, 
correspondingly, an extension $\edlie = \dlie \oplus \IR$
of the dual space $\dlie$. The central extension is given
by a linear operator $ \sh : \lie \longrightarrow \dlie$ satisfying:
\be
\sh (\lb X_1 ,X_2 \rb ) = ad^{\ast}(X_1) \sh (X_2) - ad^{\ast}(X_2) \sh (X_1)  
\lab{jacob}
\ee
which defines a nontrivial two-cocycle on the Lie algebra $\lie$ :
\be
\om (X_1,X_2) \equiv - \l \llangle \sh (X_1 )\bv X_2 \rrangle
\qquad {\rm for}\;\; \forall \, X_{1,2} \in \lie
\lab{acocycle}
\ee
where $\l$ is a numerical normalization constant. The Jacobi identity \rf{jacob}
can be integrated ($X_2 \longrightarrow g= \exp X_2 $) to get a unique nontrivial
$\dlie$-valued group one-cocycle $S(g)$ in terms of the Lie-algebra cocycle 
operator $\sh$ (provided $H^1 (G) = \emptyset \, ,\, {\rm dim} H^2 (G) = 1 \,$; 
see \ct{Ki82}) :
\be
ad^{\ast} (X ) S(g)  = Ad^{*}(g)\, \sh \( Ad(g^{-1}) X \)
- \sh (X)  \qquad {\rm for} \;\;\forall \, X \in \lie
\lab{kir}
\ee
satisfying the relations :
\be
\sh (X) = {d \o d t} S( e^{tX}) \Bgv_{t=0} \;\;\;
,\;\;\; S(g_1 g_2 ) = S(g_1 ) + Ad^{*}(g_1 )S(g_2 )  \lab{cocycle} 
\ee
One can easily generalize the adjoint and coadjoint actions of $G$ and 
$\lie$ to the case with a central extension (acting on elements 
$(X,n),(X_{1,2},n_{1,2}) \in \elie\,$ and $(U,c) \in \edlie $; see e.g. 
\ct{ANPZ-90}) :
\br
{\wti A}d (g) \,(X,n) &=& \left( Ad (g) X \;, \; n + \lambda
\llangle S (g^{-1}) \, \v \,X \rrangle \right)
\lab{eadj}  \\
{\wti ad} (X_1 ,n_1 ) \; (X_2 ,n_2 ) 
&\equiv&  \llb (X_1 ,n_1 ) \, , \, (X_2 ,n_2 ) \rrb =
\Bigl( ad(X_1)\; X_2 \; ,\; -
\l \llangle \sh (X_1) \v X_2 \rrangle  \Bigr)
\lab{ealgadj} 
\er
\be
{\wti Ad}^{\ast} (g) \, (U , c) = \( Ad^{\ast}(g) U + c \l \, S (g)
 \;, \; c \) \;\;\; ,\;\;\;
{\wti ad}^{\ast} (X ,n) (U,c) = \left( ad^{\ast} (X ) \,U + \,c \,\l
\, \sh (X)\; , \; 0 \;\right)  
\lab{ecoadj}
\ee
Also, the bilinear form $\langle \cdot \v \cdot \rangle$ on 
$\dlie \otimes \lie$ can be extended to a bilinear form on 
$\edlie \otimes \elie$ as :
\be
\llangle (U,c) \bv (\xi ,n) \rrangle =
\llangle U \bv \xi \rrangle + c\, n     \lab{ebilin}
\ee
From physical point of view the interpretation of the $\lie$-cocycle $\sh $ is 
that of ``anomaly'' of the Lie algebra (i.e., existence of a c-number term in 
the commutator \rf{ealgadj}), whereas the group cocycle $S(g)$ is the integrated
``anomaly', {\sl i.e.}, the ``anomaly'' for finite group transformations 
(see Eqs.\rf{eadj} and \rf{cocycle}).

Another basic geometric object is the fundamental $\lie$-valued Maurer-Cartan
one-form $Y(g)$ on $G$ with values in $\cG$ satisfying:
\be
d \, Y(g) = \h \llb \, Y(g) \, , \, Y(g) \, \rrb\
\lab{MC-form}
\ee
It is related to the group one-cocycle $S(g)$ through the equation :
\be
d \, S (g ) = ad^{\ast} (Y(g))\, S(g) + \sh (Y(g))
\lab{comvary}
\ee
and possesses group one-cocycle property similar to that
of $S(g)$ \rf{cocycle} :
\be
Y(g_1 g_2 ) = Y(g_1 ) + Ad(g_1 ) Y(g_2 ) \lab{ycocycle}
\ee
Using \rf{kir} one can rewrite relation \rf{comvary} in another useful form:
\be
d \, S (g ) = - Ad^\ast (g) \sh \bigl( Y(g^{-1})\bigr)
\lab{comvary-1}
\ee

The group- and algebra-cocycles $S(g)$ and $\sh (X)$ can be generalized to 
include trivial (co-boundary) parts ($\,(U_0 ,c)$ being an arbitrary point in the
extended dual space $\edlie \,$) :
\br
\S (g) & \equiv& \Suc{g}  = c\l S(g) + Ad^{\ast}(g)U_0 - U_0
\lab{Sigmab}  \\
\ssh{} (X)& \equiv& \suc{\xi} = ad^{\ast} (X) U_0 + c\l \,\sh (X )
= {d \o d t} \S (e^{ t \xi}) \Bgv_{t=0} 
\lab{sigmab} 
\er
The generalized cocycles \rf{Sigmab} and \rf{sigmab} satisfy the same relations 
as \rf{cocycle}, \rf{comvary} and \rf{kir}.
\lskip
\underline{\em 2.2 Coadjoint Orbits}
\mskp
The coadjoint orbit of $G$, passing through the point $(U_0 ,c)$ of the dual 
space $\edlie$, is defined as (cf. \rf{ecoadj}) :
\be
\Ouc \equiv \biggl\{ \Bigl( U (g),c \Bigr) \in \edlie \; ;\; 
U(g) =  U_0 + \S (g) = Ad^{*}(g) U_0 + c\l S(g) \biggr\} \lab{orbit}
\ee
The orbit \rf{orbit} is a right coset $\Ouc \simeq G / {\Gs}$ where $\Gs\;$ 
is the stationary subgroup of the point $(U_0 ,c)$ w.r.t. the coadjoint action 
\rf{ecoadj} :
\be
\Gs = \biggl\{ k \in G \; ;\; \S(k) \equiv 
c\l S(k) + Ad^{\ast}(k)U_0 - U_0 = 0 \biggr\} \lab{gstat}
\ee
The Lie algebra corresponding to $\Gs$ is :
\be
\As \equiv \biggl\{ X_0 \in \lie \; ;\; {\hat \s} (X_0 )
\equiv ad^{*} (X_0 ) U_0 + c\l \,\sh (X_0 ) = 0 \biggr\}
\lab{algstat}
\ee
Now, using the basic geometric objects from sect. 2.1,
we can express the Kirillov-Knstant symplectic form $\O_{KK}\,$ \ct{KK}
on $\Ouc$ for any infinite-dimensional (centrally extended) group $G$
in a simple compact form \ct{ANPZ-90} . Namely,
introducing the centrally extended objects :
\be
{\wti \S}(g) \equiv (\S (g),c) \in \edlie \;\;\; ,\;\;\;
{\wti Y}(g) \equiv \Bigl( Y(g), m_Y (g) \Bigr) \in \elie
\lab{centr}  
\ee
\be
d{\wti \S}(g) = {\wti ad}^{\ast} \Bigl({\wti Y}(g)\Bigr)
{\wti \S}(g) \;\;\; ,\;\;\; d{\wti Y}(g) = \h \llb
{\wti Y}(g)\, ,\, {\wti Y}(g) \rrb  
\lab{tilde}
\ee
we obtain (using \rf{ebilin} and \rf{comvary}) :
\be
\O_{KK} = -d \biggl( \llangle {\wti \S}(g) \bv {\wti Y}(g)
\rrangle \biggr) = -\h \llangle d {\wti \S}(g) \bv {\wti
Y}(g) \rrangle                  
\lab{KK}
\ee
\mskp
\underline{\em 2.3 Geometric Actions and Symmetries}
\mskp
The geometric action on a coadjoint orbit $\Ouc$ of
arbitrary infinite-dimensional (centrally extended) group
$G$ can now be written down compactly as \ct{ANPZ-90,kovhid} :
\be
W [g] = \int d^{-1} \O_{KK} = 
- \int \llangle {\wti \S}(g) \bv {\wti Y}(g)\rrangle                  
\lab{action}
\ee
or, in more detail, introducing the explicit expressions
\rf{centr}, \rf{tilde}, \rf{Sigmab} and \rf{sigmab} :
\be
W [g] = \int \llangle U_0 \bv Y(g^{-1}) \rrangle -
c\l \int \biggl\lb \Bigl\langle S(g) \bv Y(g) \Bigr\rangle - \h
d^{-1} \Bigl( \Bigl\langle \sh (Y(g)) \bv Y(g) \Bigr\rangle\Bigr)
\biggr\rb  
\lab{gaction}
\ee
The integral in \rf{action}, \rf{gaction} is over
one-dimensional curve on the phase space $\Ouc$ with a
``time-evolution" parameter $t$ . Along the curve the
exterior derivative becomes $\, d = dt \pa_t$ and the
projection of the Maurer-Cartan one-form $Y(g)$ is : $Y(g)=dt \yt$.

The fundamental Poisson brackets resulting from the geometric action
\rf{gaction} read:
\be
\lcurl\llangle {\wti \S} (g) \bv {\wti X}_1 \rrangle\, ,\,
\llangle {\wti \S} (g) \bv {\wti X}_2 \rrangle \rcurl_{PB} =
-\llangle {\wti \S} (g)\bv \lb {\wti X}_1 \, , \, {\wti X}_2 \rb \rrangle
\quad ,\quad  {\wti X}_{1,2} \equiv \( X_{1,2}\, ,\, n_{1,2}\)
\lab{PB}
\ee
or in the case of orbits $\cO_{(U_0 = 0, c)}$ (cf. \rf{orbit}) :
\be
\lcurl\llangle S(g) \bv X_1 \rrangle\, ,\,
\llangle S(g) \bv X_2 \rrangle \rcurl_{PB} =
-\llangle S(g)\bv \lb X_1 \, ,\, X_2 \rb \rrangle +
c\l \, \llangle \sh (X_1) \bv X_2 \rrangle
\lab{PB-1}
\ee
In particular, \rf{PB} shows that ${\wti\S (g)}$ is an equivariant moment map.

Using the group cocycle properties of $S(g)$ and $Y(g)$ (Eqs.\rf{cocycle} and 
\rf{ycocycle}) we derive the following fundamental group composition law 
\ct{kovhid} (with $\S (g)$ as in \rf{Sigmab}) :
\be
W [g_1 g_2] = W [g_1] + W [g_2] + \int \llangle \S(g_2)\bv Y(g_1^{-1})\rrangle
\lab{pw}
\ee
Eq.\rf{pw} is a generalization of the famous Polyakov-Wiegmann composition
law \ct{PW} in WZNW models to geometric actions on coadjoint orbits of arbitrary
(infinite-dimensional) groups with central extensions.

Eq.\rf{pw} contains the whole information about the symmetries of the geometric
action \rf{gaction}. Under arbitrary left and right infinitesimal group 
translations:
\be
g \to (\one + \vareps_L) g \quad ,\quad g \to g (\one +\vareps_R) \quad ,\quad
\vareps_{L,R} \in \cG
\lab{inf-group-transl}
\ee
we obtain using \rf{pw} :
\be
\d_L W [g] = - \int \llangle \S (g) \bv d\vareps_L \rrangle
\lab{left}
\ee
{\sl i.e.}, $\S (g)\,$ \rf{Sigmab} is a Noether conserved current 
~$\pa_t \S (g) = 0$, and:
%
\be
\d_R W [g] = \int \llangle \ssh{}(\vareps_R ) \bv Y(g^{-1}) \rrangle
= - \int \llangle \ssh{} (Y(g^{-1})) \bv \vareps_R \rrangle
\lab{right}
\ee
Recalling \rf{algstat} we find ``gauge" invariance of $W [g]$ under right group
translations from the stationary subgroup $\Gs\,$ \rf{gstat} of the orbit 
$\Ouc\,$ \rf{orbit} : $\d_RW [g] = 0\,$ for $\forall \vareps_R \in \As\,$ 
\rf{algstat}. This reveals the geometric meaning of ``hidden" local symmetries 
\ct{D2grav} in models with arbitrary infinite-dimensional Noether symmetry groups.
%
\lskip 
{\large {\bf 3. Examples of Geometric Actions on Coadjoint Orbits}}
\lskip
\underline{\em 3.1 Kac-Moody Groups}
\mskp
The Kac-Moody group elements $g \simeq g(x)$ are smooth
mappings $S^1 \longrightarrow G_0\,$, where $G_0$ is a
finite-dimensional Lie group with generators $\{ T^A \}$. 
The explicit form of \rf{eadj}-\rf{ecoadj} reads in this case :
\br
Ad(g)X = g(x) X (x) g^{-1}(x) \;\;\; &,&\;\;\;
ad(X_1 ) X_2 = \lb X_1 (x)\, ,\, X_2 (x) \rb
\;\;\; , \;\;\; X_{1,2} (x) = X_{1,2}^A (x) T_A \nonu
\\
Ad^{\ast}(g) U = g(x) U(x) g^{-1}(x) \;\;\; &,& \;\;\;
ad^{\ast}(X )U = \lb X (x) \, ,\, U(x) \rb \;\;\; ,
\;\;\; U(x) =U_A (x) T^A  \nonu
\er
\be
\sh (X) = \pa_x X (x) \;\;\; ,\;\;\; S(g) = \pa_x g(x) \, g^{-1}(x) 
\;\;\; ,\;\;\; Y(g) = dg(x)\, g^{-1}(x)
\lab{KM}
\ee
Plugging \rf{KM} into \rf{gaction} one obtains the well-known WZNW action
\ct{WZNW} for $G_0$-valued chiral fields coupled to an external ``potential" 
$U_0 (x)$, whereas Eq.\rf{pw} reduces to the Polyakov-Wiegmann group composition
law for WZNW actions \ct{PW}.
\lskip
\underline{\em 3.2 Virasoro Group}
\mskp
The Virasoro group elements $g \simeq F(x)$ are smooth 
diffeomorphisms of the circle
$S^1 \;$. Group multiplication is given by composition of
diffeomorphisms in inverse order : $g_1 \cdot \;g_2 = F_2 \circ F_1 
\; (x) = F_2\Bigl( F_1 (x)\Bigr)\;$. Eqs.\rf{eadj}-\rf{ecoadj} have now the
following explicit form :
\br
Ad(F) X = {\Bigl( \pa_x F \Bigr)}^{-1} X \( F(x)\) \;\;\; &,&
Ad^{\ast} (F) U = {\Bigl( \pa_x F \Bigr)}^2 U \( F(x)\)
\nonu \\
ad (X_1) X_2 \equiv \lb X_1 ,X_2 \rb = X_1 \pa_x X_2 - (\pa_x X_1 ) X_2
\;\;\; &,&\;\;\; ad^{\ast}(X) U = X \pa_x U + 2 (\pa_x X) U 
\nonu
\er
\be
\sh (\xi ) = \pa_x^3 \xi \;\;\; , \;\;\;
 S(F) = {{\pa_x^3 F}\o {\pa_x F}} - {3\o 2}
{\({{\pa_x^2 F}\o {\pa_x F}}\)}^2 \;\;\; ,\;\;\;
Y(F) = {{dF}\o {\pa_x F}} 
\lab{coadvir}
\ee
Here $S(F)$ is the well-known Schwarzian. Plugging \rf{coadvir} into the 
general expressions \rf{gaction} and \rf{pw} one reproduces the well-known
Polyakov $D=2$ gravity action (coupled to an external stress-tensor $U_0 (x)$) :
\be
W [F] = \int dt dx \llb - U_0 (F(t ,x)) \,\pa_x F\,
\pa_t F + {c\o {48\pi}} {{\pa_t F}\o {\pa_x F}}
\left( {{\pa_x^3 F}\o {\pa_x F}} - 2 {{(\pa_x^2 F)^2}
\o {(\pa_x F)^2}} \right) \rrb   \lab{polya}
\ee
and its group composition law \ct{D2grav,Po90}.
\lskip
\underline{\em 3.3 $(N,0)\,$ Super-Virasoro Group ($N \leq 4$)}
\mskp
Here we shall use the manifestly
$(N,0)$ supersymmetric formalism. The points of the $(N,0)$
superspace are labeled as $\;(t,z),\;\; z\equiv(x,\th^i ),\; i=1,.., N$ .
The group elements are given by superconformal diffeomorphisms :
\be
z\equiv(x,\th^j)\;\longrightarrow\;
\Z\equiv \Bigl( F(x,\th^j ),\Tu{i}(x,\th^j )\Bigr)      \lab{sconf}
\ee
obeying the superconformal constraints:
\be
D^j F -i\Tu{k}D^j\Td{k} = 0 \;\;\; ,\;\;\;
D^j \Tu{l} D^k \Td{l} - \d^{jk}\DN = 0 \;\;\; ,\;\;\;
\DN \equiv {1\o N}D^m\Tu{n}D_m\Td{n}
\lab{sconstr}
\ee
with the following superspace notations: $D^i = \partder{}{\th_i} + i \th^i \pa_x
\;\; ,\;\; D^N \equiv {1\o {N!}} \eps_{i_1
\cdot\cdot\cdot i_N} D^{i_1} \cdot\cdot\cdot D^{i_N} $.
The $(N,0)$ supersymmetric analogues of \rf{coadvir} read:
\be
Ad(\Z ) X = {\Bigl( \DN \Bigr)}^{-1} X (\Z (z)) \;\;\; ,\;\;\;
Ad^{\ast}(\Z )U = {\Bigl( \DN \Bigr)}^{2-{N\o 2}} U (\Z (z)) \nonu  
\ee
\br
ad(X_1) X_2 \equiv \llb X_1 ,X-2 \rrb = X_1 \pa_x X_2 - (\pa_x X_1) X_2
- {i\o 2} D_k X_1 D^k X_2 
\nonu \\
ad^{\ast}(X)U = X \pa_x U + (2-{N\o 2}) (\pa_x X )U - {i\o 2}D_k \xi D^k U   
\nonu
\er
\be
\sh_N (X)=i^{N(N-2)} D^N \pa_x^{3-N} X \quad ,\quad
Y_N (\Z ) = \Bigl( dF + i \Tu{j} d\Td{j} \Bigr) {\left(\DN \right)}^{-1}            
\lab{coadsvir}
\ee
The associated $\dlie$-valued group one-cocycles $S_N(\Z)$ coincide with the 
well-known \ct{Scho88} $(N,0)$ super-Schwarzians. Inserting the latter and 
\rf{coadsvir} into \rf{gaction} one obtains the $(N,0)$ supersymmetric 
generalization of the Polyakov $D=2$ gravity action for any $N \leq 4$ 
\ct{ny-prep,ANPZ-90} :
\be
W_N \lb \Z \rb = \int dt\;(dz) \biggl\lb \pa_t \Bigl( \ln
\DN \Bigr) D^N \pa_x^{1-N} \Bigl( \DN \Bigr) -
U_0 (\Z ) {\left(\DN \right)}^{2- {N\o 2}}       
Y_N (\Z ) \biggr\rb    \lab{Naction}
\ee
\lskip
\underline{\em 3.4 Group of Area-Preserving Diffeomorphisms on
Torus  with Central Extension ${\wti {\rm {\bf SDiff}}} {\bf (T^2 )}$ }
\mskp
The elements of $\Tor\,$ are described by smooth diffeomorphisms 
$\; T^2 \ni \vec{x} \equiv
(x^1 ,x^2 ) \longrightarrow F^i (\vec{x}) \in T^2 \; \, (i=1,2)$, 
such that $\; \det\! \parallel {{\pa F^i}\o {\pa x^j}} 
\parallel = 1$ . The Lie algebra of $\Tor\,$ reads :
$\, \llb \Lh{\vec{x}}, \Lh{\vec{y}} \rrb =
- \eps^{ij} \pa_i \Lh{\vec{x}} \pa_j \d^{(2)}(\vec{x}-\vec{y}) - 
a^i \pa_i \d^{(2)}(\vec{x}-\vec{y})\,$, where $\; \vec{a} \equiv (a^1 ,a^2 )\;$ 
are the ``central charges" \ct{memb} . The general Eqs.\rf{eadj}-\rf{ecoadj}
now specialize to \ct{torus} :
\br
Ad(\vec{F}) X = X (\vec{F}(\vec{x})) \;\;\; &,& \;\;\;
ad(X_1)X_2 \equiv \lb X_1 ,X_2 \rb (\vec{x}) =
\eps^{ij} \pa_i X_1 (\vec{x}) \pa_j X_2 (\vec{x})  
\nonu \\
Ad^{\ast}(\vec{F}) U = U(\vec{F}(\vec{x})) \;\;\; &,&
\;\;\; ad^{\ast}(X) U = \eps^{ij} \pa_i X (\vec{x})\pa_j U (\vec{x}) 
\nonu
\er
\be
\sh (X)= a^i \pa_i X (\vec{x}) \;\;\; ,\;\;\;
S(\vec{F}) = a^i \eps_{ij} \Bigl(F^j (\vec{x}) - x^j \Bigr)
\;\;\; ,\;\;\; Y(\vec{F}) = \h \eps_{ij} F^i dF^j + d\rho
(\vec{F})   \lab{torcoad}
\ee
where $\pa_i \rho (\vec{F}) = - \h \left( \eps_{kl} F^k
\pa_i F^l + \eps_{ij} x^j \right)\,$ .

Plugging \rf{torcoad} into \rf{gaction} we get the
$\Tor$ co-orbit geometric action \ct{torus} :
\be
W_{\Tor} \lb \vec{F}\rb = - {1\o 3} \int dt\,dx^2 \left(
a^k \eps_{kl} F^l \right) \,\eps_{ij} F^i \pa_t F^j
\lab{toraction}
\ee
In \ct{shorty} it was shown that \rf{toraction} is the Wess-Zumino anomalous 
effective action for the toroidal membrane in the light-cone gauge.
\lskip
\lskip
\underline{\em 3.5 ${\bf W_{1+\infty}}$-Gravity Effective Action }
\mskp
The ${\bf W_{1+\infty}}$-algebra is isomorphic to the algebra of all
differential operators on the circle ~
$\DA =\lcurl X \equiv X (x,\pa) = \sum_{i\geq 0} X_i \pa^i\rcurl$.
Accordingly, the dual space is the space of all purely pseudo-differential
operators $\dDA = \lcurl U \equiv U(x,\pa) = \sum_{j\geq 1} u_j \pa^{-j}\rcurl$, 
where the bilinear pairing is defined by:
\be
\llangle U \bv X \rrangle = \int dx \Res (U\, X) \quad ;\quad
\Res \cA \equiv a_{-1} \;\;\; {\rm for~any}\;\; \cA = \sum_k a_k \pa^k
\lab{bilin-DOP}
\ee
The corresponding Lie group $\DO$ is defined as formal exponentiation of the
Lie algebra $\DA$ , where the group elements $g(x,\pa) = \exp X(x,\pa)$ are 
understood again in the sense of pseudo-differential operator calculus.
The relevant objects from the coajoint orbit formalism are given in this
case as follows:
\br
Ad(g) X = g(x,\pa) X(x,\pa) g^{-1}(x,\pa) \quad ,\quad
ad(X_1 ) X_2 = \lb X_1 (x,\pa)\, ,\, X_2 (x,\pa) \rb
\nonu \\
Ad^{\ast}(g) U = \( g(x,\pa) U(x,\pa) g^{-1}(x,\pa)\)_{-} \quad ,\quad
ad^{\ast}(X) U = \Sbr{X (x,\pa)}{U(x,\pa)}_{-} 
\nonu
\er
\be
\sh (X ) = - \Sbr{\ln \pa}{X (x,\pa)}_{-} \;\; ,\;\;
S(g) = -\(\Sbr{\ln \pa}{X (x,\pa)}\, g^{-1}(x,\pa)\)_{-}  \;\; ,\;\;
Y(g) = dg(x,\pa)\, g^{-1}(x,\pa)
\lab{DOP}
\ee
Everywhere in \rf{DOP} products are understood in the sense of
pseudo-differential operator calculus, and the subscript $(-)$ indicates
taking the purely pseudo-differential part.

Now, the geometric action on a coadjoint orbit of $\DO \simeq {\bf W_{1+\infty}}$,
which is the WZNW effective action of induced ${\bf W_{1+\infty}}$-gravity, 
is given as \ct{dop-tmf} (for brevity we suppress the indication of arguments in
$g=g (x,\pa)$ etc.) :
\br
W [g] \equiv W_{\DO} \lb g\rb = - \int dt dx \Res \( U_0 \, g^{-1} \, \pa_t g\)  +   
\phantom{aaaaaaaaaaaaaaaa} \nonu   \\
{c\o {4\pi}}\int \int dx \Res \Biggl( \lb \ln \pa \sta g  \rb \,
g^{-1} \, \pa_t g \, g^{-1}
-\h d^{-1} \Bigl\{ \left\lb \ln \pa \sta dg \, g^{-1}\right\rb 
\wedge \( dg \, g^{-1} \) \Bigr\}\,\Biggr) 
\lab{dop-action}
\er
whereas the group composition law reads \ct{dop-tmf} :
\be
W \lb g\, h\rb = W \lb g\rb + W \lb h\rb + 
\int dt dx \biggl\{\Bigl( \, h\, U_0 \, h^{-1}
- {c\o {4\pi}}\lb \ln \pa \sta h \rb \, h^{-1} \,\Bigr) \, g^{-1} \,\pa_t g \; 
\biggr\}    
\lab{pw-dop}
\ee
It has been shown in the first ref.\ct{dop-tmf} that the stationary subgroup
\rf{gstat} of the pertinent $\DO \simeq {\bf W_{1+\infty}}$ coadjoint orbit is
$SL(\infty)$, which thereby appears as ``hidden'' symmetry of the
${\bf W_{1+\infty}}$ geometric action \rf{dop-action}, and that the
energy-monentum componet $T_{++}$ possesses $SL(\infty)$ Sugawara form.
\lskip
{\large {\bf 4. Gauging of Geometric Actions}}
\mskp
Let us now return to the general case of arbitrary (infinite-dimensional)
groups $G$ with central extensions. Henceforth, for simplicity we will
consider geometric actions \rf{gaction} on coadjoint orbits \rf{orbit} with 
$U_0 =0$ \foot{Note that the first term in \rf{gaction} containing $U_0$ can 
be interpreted as coupling to an external background field.} and also we will 
set the normalization constant in \rf{gaction} $-c\l = 1$ :
\be
W [g] = \int \biggl\lb \Bigl\langle S(g) \bv Y(g) \Bigr\rangle - \h
d^{-1} \Bigl( \Bigl\langle \sh (Y(g)) \bv Y(g) \Bigr\rangle\Bigr)
\biggr\rb  
\lab{gaction-1}
\ee
whereupon the generalized group composition law \rf{pw} simplifies to:
\be
W [g_1 g_2] = W [g_1] + W [g_2] - \int \llangle S(g_2) \bv Y(g_1^{-1})\rrangle 
\lab{pw-1}
\ee

Now, suppose that there exist two fixed elements $E_{-} \in \cG$ and 
$\cE_{+} \in \cG^{\ast}$ such that they define splitting (as vector spaces) of the
Lie algebra $\cG$ and its dual $\cG^\ast$ with the following properties:
\be
\cG = \cH \oplus \cM  \quad ,\quad  \cG^\ast = \cH^\ast \oplus \cM^\ast
\quad ,\quad
\cM^\ast \equiv \lcurl U \bgv \langle U \bv X_H \rangle = 0 \;\; {\rm for}\;
\forall X_H \in \cH \rcurl
\lab{G-alg-dual-split}
\ee
where:
\be
\cH = {\rm Ker} \Bigl(\m \(\cE_{+}\)\Bigr)  
\quad ,\quad  E_{-} \in \cH  \quad ,\quad   
\sh (E_{-}) = 0 \;\; (\; i.e.\; E_{-} \in \cG_{stat} ~,~ {\rm cf. ~\rf{algstat}}\;)
\lab{Ker-Im}
\ee
and $\m (U) (\cdot)$ is the mapping:
\be
\m (U) \; :\; \cG \to \cG^\ast \quad ,\quad  \m (U) X =  ad^\ast (X) U  \quad
{\rm for} \;\; X \in \cG \;\; ,\;\; U \in \cG^\ast
\lab{mu-map}
\ee
Properties \rf{Ker-Im} imply that $\cH$ is a subalgebra of $\cG$ and that:
\be
\m \(\cE_{+}\) \cG \equiv ad^\ast \(\cG \) \cE_{+}  \subset \cM^\ast
\quad ,\quad  ad^\ast \(\cH\) \cM^\ast \subset \cM^\ast
\lab{comm-rel}
\ee
In particular, the first two relations \rf{Ker-Im} show that the fixed elements
$E_{-}$ and $\cE_{+}$ mutually ``commute'':
\be
ad^\ast (E_{-}) \cE_{+} = 0
\lab{E-commut}
\ee
The first two properties \rf{Ker-Im} arise as sufficient conditions for 
consistency of the equations of motion of the gauged geometric actions given 
below. The third property \rf{Ker-Im} arises as sufficient condition for
validity of the generalized ``zero-curvature'' representation on the group
coadjoint orbit of the equations of motion of the gauged geometric actions.

In the special case of $G$ being Kac-Moody group (cf. subsection 3.1 above)
where Lie algebras and dual spaces are identified, relations
\rf{G-alg-dual-split}--\rf{Ker-Im} acquire the following meaning:
\be
\cG = \cH \oplus \cM  \quad ,\quad 
\cH = {\rm Ker}\(ad (E_{+})\)
\lab{G-alg-split}
\ee
where $\cE_{+}\equiv E_{+}$ and $E_{-}$ are two mutually commuting fixed
Kac-Moody algebra elements belonging to $\cH$. In order to make contact with
integrable models, one requires in addition to \rf{G-alg-split} that $E_{+}$
is semisimple element, {\sl i.e.} :
\be
\cG = \cH \oplus \cM \quad ,\quad \cH = {\rm Ker} \bigl( ad(E_{+})\bigr)
\quad ,\quad
\cM = {\rm Im} \bigl( ad(E_{+})\bigr) \quad \longrightarrow \quad
\Sbr{\cH}{\cM} \subset \cM
\lab{semisimple}
\ee

Going back to the general case of coadjoint orbits of arbitrary
(infinite-dimensional) groups with central extensions, 
let us consider ``gauge'' fields -- the one-form $\cA_{+} \in \cH$
and $A_{-} \in \cH^{\ast}$ which are parametrized in terms of the group elements 
$h_L,\, h_R \in H$ as:
\be
\cA_{+} = Y (h_L) \quad ,\quad A_{-} = S(h_R^{-1})
\lab{A-param}
\ee
where $Y(\cdot)$ and $S(\cdot)$ are the fundamental Maurer-Cartan form
\rf{comvary}--\rf{ycocycle} and the nontrivial group cocycle \rf{kir}
restricted on the subgroup $H$.

Now we are ready to introduce the following new geometric action which is a 
``massive'' gauge-invariant generalization of \rf{gaction-1} :
\br
W\lb g, A_{+}, A_{-} \rb \equiv 
W\lb h^{-1}_L g h^{-1}_R \rb - W\lb h^{-1}_L h^{-1}_R\rb +
\int dt \Bigl\langle Ad^\ast (h_R g^{-1} h_L) \cE_{+} \bv E_{-}\Bigr\rangle 
\phantom{aaaaaaaaaaaa}
\lab{gauged-gaction-def}  \\
= \int \Bigl\lb \Bigl\langle S(g) \bv Y(g) \Bigr\rangle - 
\h d^{-1} \Bigl( \Bigl\langle \sh (Y(g)) \bv Y(g) \Bigr\rangle\Bigr)
\Bigr\rb
- \int \Bigl\lb \Bigl\langle S(g) \bv \cA_{+} \Bigr\rangle +
\Bigl\langle A_{-} \bv Y(g^{-1}) \Bigr\rangle \Bigr\rb -
\nonu
\er
\be
- \int \Bigl\lb \Bigl\langle Ad^\ast (g) A_{-} \bv \cA_{+} \Bigr\rangle
- \Bigl\langle A_{-} \bv \cA_{+}\Bigr\rangle \Bigr\rb  -
\int dt \Bigl\langle Ad^\ast (h_R g^{-1} h_L) \cE_{+} \bv E_{-}\Bigr\rangle 
\lab{gauged-gaction}
\ee
where $\cA_{+},\, A_{-}$ are as in \rf{A-param}. Along the curve of ``time''
integration in \rf{gauged-gaction} the Maurer-Cartan $\cH$-valued one-form
$\cA_{+}=Y(h_L)$ becomes $\cA_{+} = A_{+} dt$ similar to the fundamental
$\cG$-valued Maurer-Cartan form $Y(g) = Y_t (g) dt$ as pointed out above. 
Let us stress that, although the gauged geometric action \rf{gauged-gaction}
formally resembles ordinary $G/H$ gauged WZNW models where $G,\, H$ are 
Kac-Moody groups, the action \rf{gauged-gaction} is valid in the much more 
general setting of arbitrary (infinite-dimensional) groups with central 
extensions (cf. the examples in Section 3).

The action \rf{gauged-gaction} exhibits manifest ``vector-like'' gauge 
invariance under:
\be
g \to h^{-1} g h \quad ,\quad h_L \to h^{-1} h_L \quad ,\quad h_R \to h_R h 
\quad ,\quad h \in H
\lab{gauge-transf}
\ee
In particular, we note that the last ``mass'' term on the r.h.s. of 
\rf{gauged-gaction} is gauge-invariant by itself. Also, let us emphasize
that the second explicit form \rf{gauged-gaction} is obtained from the
first defining form \rf{gauged-gaction-def} by using the general group
composition law identities \rf{pw-1}. The gauged geometric action 
\rf{gauged-gaction} is a generalization of the well-known gauged WZNW actions 
\ct{gauged-WZNW} to the case of arbitrary (infinite-dimensional)
groups with central extensions (see the examples in Section 3 above).

To derive the equations of motion for \rf{gauged-gaction} we need the
infinitesimal forms of the group cocycle properties \rf{cocycle} and 
\rf{ycocycle} under left and right infinitesimal group translations
\rf{inf-group-transl} :
\br
\d_L Y(g) = d\vareps_L - ad(Y(g)) \vareps_L \quad ,\quad
\d_L S(g) = \sh (\vareps_L) + ad^\ast (\vareps_L) S(g) 
\nonu \\
\d_R S(g^{-1}) = - \sh (\vareps_R) - ad^\ast (\vareps_R) S(g^{-1})
\phantom{aaaaaaaaaaaaaaaaaa}
\lab{cocycle-inf}
\er
Using Eqs.\rf{cocycle-inf} the variations of \rf{gauged-gaction} w.r.t.
$\vareps_L = \d g\, g^{-1} \in \cG$, $\vareps_L = \d h_L\, h_L^{-1} \in \cH$
and $\vareps_R = h_R^{-1} \d h_R \in \cH$ yield the following gauge-invariant
equations of motion:
\br
\Bigl(\pa_t - ad^\ast (A_{+})\Bigr) \Bigl( S(g)+Ad^{\ast}(g) A_{-}-A_{-}\Bigr)
+ ad^\ast \bigl( Ad(g h^{-1}_R) E_{-}\bigr) \( Ad^\ast (h_L) \cE_{+}\)
\nonu \\
+ \pa_t A_{-} - \sh (A_{+}) - ad^\ast (A_{+}) A_{-} = 0
\phantom{aaaaaaaaaaaaaaaaaaaaaaaaaaa}
\lab{g-eqmotion}
\er
\be
\Bigl(\pa_t - ad^\ast (A_{+})\Bigr) 
\Bigl( S(g)+Ad^{\ast}(g) A_{-}-A_{-}\Bigr)\bgv_{\cH^\ast} +
ad^\ast \bigl( Ad(g h^{-1}_R) E_{-}\bigr) \( Ad^\ast (h_L)\cE_{+}\)\bgv_{\cH^\ast} =0
\lab{A-plus-eqmotion}
\ee
\br
\sh \Bigl( Y_t (g^{-1}) + Ad(g^{-1})A_{+} - A_{+}\Bigr) \bgv_{\cH^{\ast}} +
ad^\ast \Bigl( Y_t (g^{-1})+Ad(g^{-1})A_{+} - A_{+}\Bigr) A_{-}\bgv_{\cH^\ast}
\nonu \\
+ ad^\ast ( Ad (h^{-1}_R) E_{-})\bigl( Ad^\ast (g^{-1} h_L)\cE_{+}\bigr) 
\bgv_{\cH^\ast} = 0
\phantom{aaaaaaaaaaaaaaaaaaaaaaaaaaa}
\lab{A-minus-eqmotion}
\er

Projecting Eq.\rf{g-eqmotion} along $\cH^\ast$ and taking into account 
\rf{A-plus-eqmotion} we get that the expression in the second line of
\rf{g-eqmotion} (which lies entirely in the subspace $\cH^\ast$) vanishes 
separately:
\be
\pa_t A_{-} - \sh (A_{+}) - ad^\ast (A_{+}) A_{-} = 0
\lab{zero-curv-gen}
\ee
Eq.\rf{zero-curv-gen} has the meaning of a generalized ``zero-curvature''
equation where we remind that $A_{+} \in \cH$ whereas $A_{-} \in \cH^\ast$
(cf. \rf{A-param}).
Indeed, in the special case of Kac-Moody groups \rf{KM} it reduces
to the ordinary zero-curvature equation on the subalgebra $\cH$ :
\be
\pa_t A_{-} - \pa_x A_{+} -\Sbr{A_{+}}{A_{-}}=0 \quad ,\quad
A_{-}=\pa_x h_R^{-1}\, h_R \;\; ,\;\; A_{+} = \pa_t h_L\, h_L^{-1}
\lab{zero-curv-KM}
\ee

Now, going back to general coadjoint-orbit actions and taking into account 
\rf{A-param} we see that Eq.\rf{zero-curv-gen} implies
$h_L = h_R^{-1} =h$ where $h$ is arbitrary element of the gauge subgroup $H$
reflecting the residual gauge freedom. Therefore, we are entitled to choose
the gauge fixing:
\be
h = \one \quad \longrightarrow \quad A_{+}=0 \;\; ,\;\; A_{-}=0
\lab{gauge-fix}
\ee
which simplifies the rest of the equations of motions to:
\be
\pa_t S(g) + ad^\ast \bigl( Ad(g) E_{-}\bigr)\cE_{+} = 0
\lab{g-eqmotion-gf}
\ee
\be
\sh \bigl( Y_t (g^{-1})\bigr) \bgv_{\cH^\ast} +
ad^\ast (E_{-})\bigl( Ad^\ast (g^{-1}) \cE_{+}\bigr) \bgv_{\cH^\ast} = 0
\lab{A-minus-eqmotion-gf}
\ee
Using relation \rf{comvary-1} we can rewrite Eq.\rf{g-eqmotion-gf} in an
equivalent form:
\be
\sh \bigl( Y_t (g^{-1})\bigr) -
ad^\ast (E_{-})\bigl( Ad^\ast (g^{-1}) \cE_{+}\bigr) = 0
\lab{g-eqmotion-gf-1}
\ee

Comparing Eq.\rf{A-minus-eqmotion-gf} and the projected along $\cH^\ast$
Eq.\rf{g-eqmotion-gf-1}, we deduce that both terms on the l.h.s. of
\rf{A-minus-eqmotion-gf} must separately vanish:
\be
\sh (Y_t (g^{-1}) \bgv_{\cH^\ast} = 0 \quad ,\quad
ad^\ast (E_{-})\bigl( Ad^\ast (g^{-1}) \cE_{+}\bigr) \bgv_{\cH^\ast} = 0
\lab{separ-vanish}
\ee
or, written in an equivalent form:
\be
\llangle \sh (Y_t (g^{-1}) \bv X_H \rrangle = 0   \quad ,\quad
\llangle ad^\ast (E_{-})\bigl( Ad^\ast (g^{-1}) \cE_{+}\bigr) 
\bv X_H \rrangle = - 
\llangle Ad^\ast (g^{-1}) \cE_{+} \bv ad (E_{-}) X_H \rrangle = 0
\lab{separ-vanish-1}
\ee
The consistency of the second condition in Eq.\rf{separ-vanish-1} is
guaranteed because of the properties \rf{Ker-Im} of the fixed elements
$\cE_{+} \in \cG^\ast$ and $E_{-} \in \cH \subset \cG$.
Note also that due to properties \rf{Ker-Im} the projection along $\cH^\ast$
of the second term on the l.h.s.of Eq.\rf{g-eqmotion-gf} similarly vanishes:
$ad^\ast \bigl( Ad(g) E_{-}\bigr)\cE_{+}\bgv_{\cH^\ast} = 0$.
Thus, taking into account \rf{separ-vanish}, the final 
form of the gauge-fixed equations of motion of the gauged geometric action 
\rf{gauged-gaction} reads:
\be
\pa_t S(g) + ad^\ast \bigl( Ad(g) E_{-}\bigr)\cE_{+} = 0    \quad ,\quad
\pa_t S(g) \bgv_{\cH^\ast} = 0 \quad ,\quad 
\sh \bigl( Y_t (g^{-1})\bigr) \bgv_{\cH^\ast} = 0 
\lab{eqmotion-gf}
\ee
\mskp
{\bf 5. ``Zero-Curvature'' Representation on Coadjoint Orbits}
\mskp
Let us now show that the first Eq.\rf{eqmotion-gf} together with the
constraints (second and thirs Eqs.\rf{eqmotion-gf}) possess a generalized
``zero-curvature'' representation. To this end we introduce, in the spirit
of Chapter 4 of ref.\ct{FT}, an additional loop-grading (``current-algebra'')
structure on the underlying (infinite-dimensional) algebra $\cG$ (recall
Section 3 for the various interesting examples of algebras $\cG$) such that
the original $\cG$ is the zero-grade subalgebra :
\be
{\widehat \cG} = \oplus_{i \in \IZ} \cG^{(i)}  \quad ,\quad
\Sbr{\cG^{(i)}}{\cG^{(j)}} \subset \cG^{(i+j)} \quad,\quad \cG^{(0)}= \cG
\lab{G-grading}
\ee
The loop-grading structure is induced on the corresponding dual space:
\be
{\widehat \cG}^\ast = \oplus_{i \in \IZ} \cG^\ast_{(i)} \quad ,\quad
\Bigl\langle \cG^\ast_{(i)} \bgv \cG^{(j)}\Bigr\rangle = 0  
\;\;\; {\rm for}\; i+j \neq 0
\lab{G-dual-grading}
\ee
The definitions of the basic objects -- the central extension operator
$\sh (\cdot)$, the group cocycle $S(g)$ and the fundamental Maurer-Cartan
form $Y(g)$ from sect. 2.1, naturally are carried over to the whole graded
algebra ${\widehat \cG}$ and the corresponding group ${\widehat G}$. Note that
$\sh (\cdot)$ preserves the grading structure.

Let us consider a group element ${\widehat g} \in {\widehat G}$ of the following
form:
\be
{\widehat g} = g\, \O \quad ,\quad
g=e^{X^{(0)}} \in G \quad ,\quad  \O = e^{\sum_{i\geq 1} \om^{(i)}}
\lab{monodromy-gen}
\ee
where $\om^{(j)} \in \cG^{(j)}$, and let us introduce
the following overdetermined system of equations for ${\widehat g}$ given
entirely in terms of the basic objects connected with a ${\widehat G}$-coadjoint 
orbit:
\be
S({\widehat g}) - S(g) + \cE^{(1)}_{+} = 0   \quad ,\quad
Y_t \bigl({\widehat g}^{-1}\bigr) + 
Ad \bigl(g\,{\widehat g}^{-1}\bigr) E^{(-1)}_{-} - E^{(-1)}_{-} = 0
\lab{monodromy-eqs}
\ee
Here $E^{(-1)}_{-} \in \cG^{(-1)}$ and $\cE^{(1)}_{+} \in \cG^\ast_{(1)}$
are fixed elements possessing the same properties as \rf{Ker-Im}
but extended to the whole graded algebra:
\be
{\widehat \cH} \subset {\widehat \cG}   \quad ,\quad 
{\widehat \cH} = 
{\rm Ker} \Bigl(\m \bigl(\cE^{(1)}_{+}\bigr)\Bigr)  \quad ,\quad 
E^{(-1)}_{-} \in {\widehat \cH}  \quad ,\quad 
\sh \bigl( E^{(-1)}_{-}\bigr) = 0 
\lab{Ker-Im-graded}
\ee
The element ${\widehat g}$ \rf{monodromy-gen} is a generalization of the notion 
of ``monodromy matrix'', and the system \rf{monodromy-eqs} can be viewed as
generalization of the ``zero-curvature'' equations on coadjoint orbits of
arbitrary (infinite-dimensional) groups with central extensions.

To solve the system \rf{monodromy-eqs} we insert the grade-expansion 
\rf{monodromy-gen} and use the general cocycle properties \rf{cocycle} and
\rf{ycocycle}. The lowest grade${}\! =\!1$ projection in the first 
Eq.\rf{monodromy-eqs} and the lowest (non-trivial) grade${}\! =\! 0$ projection 
in the second Eq.\rf{monodromy-eqs} yield the following overdetermined 
{\em linear} system of equations for the lowest positive-grade Lie-algebraic
component $\om^{(1)}$ in ${\widehat g}$ \rf{monodromy-gen} :
\be
\sh \bigl(\om^{(1)}\bigr) + Ad^\ast (g^{-1}) \cE^{(1)}_{+} = 0  \quad ,\quad
Y_t (g^{-1}) - ad \bigl(\om^{(1)}\bigr) E^{(-1)}_{-} = 0
\lab{linear-monodromy-eqs}
\ee
Now, acting with the operator $\sh (\cdot )$ on the second 
Eq.\rf{linear-monodromy-eqs} and using the first Eq.\rf{linear-monodromy-eqs}
together with \rf{jacob} and last Eq.\rf{Ker-Im-graded} we obtain:
\be
\sh \bigl( Y_t (g^{-1})\bigr) -
ad^\ast (E^{(-1)}_{-})\bigl( Ad^\ast (g^{-1}) \cE^{(1)}_{+}\bigr) = 0
\lab{g-eqmotion-gf-graded}
\ee
which upon using \rf{comvary-1} can be rewritten in the equivalent form:
\be
\pa_t S(g) + ad^\ast \bigl( Ad(g) E^{(-1)}_{-}\bigr)\cE^{(1)}_{+} = 0 
\lab{eqmotion-gf-graded}
\ee
The latter equation coincides with first Eq.\rf{eqmotion-gf} upon identifying
$\cE^{(1)}_{+}=\cE_{+}$ and $E^{(-1)}_{-}=E_{-}$. The constraints
(second and third Eqs.\rf{eqmotion-gf}) are recovered from
\rf{g-eqmotion-gf-graded}--\rf{eqmotion-gf-graded} by projecting along
${\widehat \cH}$ and using properties \rf{Ker-Im-graded}. 


In the special case of Kac-Moody group $G$ (see example 3.1), 
Eqs.\rf{monodromy-eqs} acquire the form:
\be
\(\pa_x + E^{(1)}_{+} + A^{(0)}\) {\widehat g} = 0   \quad ,\quad
A^{(0)} \equiv - \pa_x g \, g^{-1} 
\lab{ZS-linear}
\ee
\be
\bpa {\widehat g} + {\widehat g} E^{(-1)}_{-} - 
\( g E^{(-1)}_{-} g^{-1}\) {\widehat g} = 0    \quad ,\quad \bpa \equiv \pa_t
\lab{neg-flow-1-a}
\ee
where
$E^{(\pm 1)}_{\pm}\in{\widehat \cH} = {\rm Ker}\bigl( ad(E^{(1)}_{+})\bigr)$.
The last equation can be written, using the grade structure
\rf{monodromy-gen} of ${\widehat g}$, entirely in terms of the latter:
\be
\(\bpa + \({\widehat g} E^{(-1)}_{-} {\widehat g}^{-1}\)_{+}\) {\widehat g} = 0
\lab{neg-flow-1}
\ee
with the subscript $(+)$ indicating projection along the non-negative grade
part.
The zero-curvature (compaibility) condition for the system 
\rf{ZS-linear},\rf{neg-flow-1} :
\be
\Sbr{\bpa + \({\widehat g} E^{(-1)}_{-} {\widehat g}^{-1}\)_{+}}{
\pa_x + E^{(1)}_{+} + A^{(0)}} = 0
\lab{zero-curv}
\ee
upon inserting the grade-expansion of ${\widehat g}$ \rf{monodromy-gen}
reduces to the form:
\be
\Sbr{\bpa - g E^{(-1)}_{-} g^{-1}}{\pa_x + E^{(1)}_{+} + A^{(0)}} = 0
\lab{zero-curv-0}
\ee
which yields:
\be
\bpa (\pa_x g\, g^{-1}) - \Sbr{E^{(1)}_{+}}{g E^{(-1)}_{-} g^{-1}} = 0 
\lab{eqmotion-gf-KM}
\ee
\be
\bpa \bigl( \pa_x g\, g^{-1}\bigr)\bgv_{\cH} = 0 \quad \longrightarrow \quad
\pa_x g\, g^{-1} \bgv_{\cH} =  - A^{(0)}\bgv_{\cH} = \xi_H (x) 
\lab{constr-gen-a}
\ee
\be
\pa \bigl( g^{-1} \bpa g\bigr) \bgv_{\cH} = 0  \quad \longrightarrow \quad
g^{-1} \bpa g \bgv_{\cH} = \(\; \Sbr{E^{(-1)}_{-}}{\om^{(1)}} \bgv_{\cH} \;\) 
= \eta_H (t)
\lab{constr-gen-b}
\ee
where $\xi_H (x)$ and $\eta_H (t)$ are some fixed non-dynamical elements of
$\cH \subset \cG$. The equality in brackets in \rf{constr-gen-b} is
a special case of second Eq.\rf{linear-monodromy-eqs}.

In the context of integrable systems, when $E^{(1)}_{+}$ is a semisimple element
\rf{semisimple}, the constraints \rf{constr-gen-a}--\rf{constr-gen-b} can be 
brought to a simpler form  with vanishing $\xi_H (x)$ and $\eta_H (t)$
(cf. Eqs.\rf{constr-int-a}--\rf{constr-int-b} below).


Eq.\rf{zero-curv-0} constitutes the well-known zero-curvature representation for
the equations of motion \rf{eqmotion-gf-KM} of non-Abelian Toda field theories
\ct{Leznov-Saveliev} and more general $G/H$ ~gauged WZNW models 
\ct{Weigt-etal} with various special choices for the fixed 
algebraic elements $E^{(1)}_{+}$ and $E^{(-1)}_{-}$; for a systematic
treatment of non-Abelian Toda field theories as gauge-fixed versions of
gauged WZNW models, see refs.\ct{Toda-FT}. 

There is also another way to view Eqs.\rf{eqmotion-gf-KM}, namely,
we will see below that \rf{eqmotion-gf-KM} arise as {\em additional
symmetry} flow equations of generalized Drinfeld-Sokolov integrable hierarchies.
In the special case of generalized Drinfeld-Sokolov hierarchies based on
${\widehat \cG}={\widehat {SL}}(M+1)$ with standard
homogeneous grading $Q = \l \partder{}{\l}$ and
$E^{(\pm 1)}_{\pm} = H^{(\pm 1)}_{\l_M}$ (where $H_{\l_M} \equiv \l_M . H$
with $\l_M$ being the last fundamental weight), which are equivalent to the
class of constrained (reduced) KP hierarchies ${\sl cKP}_{1,M}$
(Eq.\rf{Lax-1-M} below) within Sato pseudo-differential operator
formulation, we will be able to use the standard \DB techniques to generate
solutions to gauged $SL(M+1)/U(1)\times SL(M)$ WZNW field equations 
\rf{eqmotion-gf-KM}.

%
%
\lskip
{\bf 6. Sato Formalism for Additional Symmetries of KP-Type Integrable Hierarchies}
\mskp
\underline{\em 6.1 Sato Pseudo-Differential Operator Formulation}
\sskp
In what follows $D$ denotes the derivative operator w.r.t. $x$
such that $\sbr{D}{f} = \pa f = \pa f /\pa x$ and the generalized Leibniz
rule holds: $D^n f = \sumi{j=0} {n \choose j} (\pa^j f) D^{n-j}$ with
$n \in \IZ$. In order to avoid confusion we shall employ the following
notations: for any (pseudo-)\-differential operator $A = \sum_k a_k D^k$ and
a function $f$, the symbol $\, A(f)\,$ will indicate application (action) of
$A$ on $f$, whereas the symbol $Af$ will denote simply operator product of
$A$ with the zero-order (multiplication) operator $f$. Projections $(\pm)$
are defined as: $A_{+} = \sum_{k\geq 0} a_k D^k$ and
$A_{-} = \sum_{k\leq -1} a_k D^k$. Conjugation is given by
~$A^\ast = \sum_k (-D)^k \, a_k$. Finally, $\Res A \equiv a_{-1}$.

The general one-component (scalar) KP hierarchy is given by
a pseudo-differential Lax operator $\cL$ obeying Sato evolution equations
(also known as isospectral flow equations; for a systematic exposition, see
\ct{Dickey-book}) :   
\be
\cL = D + \sum_{k=1}^\infty u_k D^{-k}    \quad , \quad
\partder{}{t_n}\cL = \Sbr{\(\cL^n\)_{+}}{\cL} 
\lab{Lax-gen}
\ee
with Sato dressing operator $W$ :
\be
\cL = W D W^{-1} \quad , \quad  \partder{}{t_n} W = - \(W D^n W^{-1}\)_{-} W
\quad ,\quad 
W = \sum_{k=0}^\infty \frac{p_k (-[\pa])\t (t)}{\t (t)} D^{-k}
\lab{Sato-dress}
\ee
and (adjoint) Baker-Akhiezer (BA) wave functions $\psi_{BA}^{(\ast)} (t,\l)$ :
\be
\cL^{(\ast)} \psi_{BA}^{(\ast)} = \l \psi_{BA}^{(\ast)} \quad , \quad
\partder{}{t_n} \psi_{BA}^{(\ast)} = 
\pm \({\cL^{(\ast)}}^n\)_{+} (\psi_{BA}^{(\ast)})
\lab{BA-eqs}
\ee
\be
\psi_{BA}^{(\ast)} (t,\l) = W^{(\ast\, -1)} \( e^{\pm \xi (t,\l)}\) =
\frac{\t (t \mp [\l^{-1}])}{\t (t)}\, e^{\pm \xi (t,\l)} \quad ,\quad
 \xi (t,\l) \equiv \sum_{\ell =1}^\infty t_\ell \l^{\ell}
\lab{BA-def}
\ee
Here and below we employ the following short-hand notations:
$(t) \equiv (t_1 \equiv x, t_2, \ldots )$ for the set of isospectral
time-evolution parameters;
$[\pa] \equiv \Bigl(\partder{}{t_1}, \h \partder{}{t_2},
{1\o 3}\partder{}{t_3}, \ldots \Bigr)$ 
and 
$[\l^{-1}] \equiv \Bigl(\l^{-1}, \h \l^{-2}, {1\o 3} \l^{-3},\ldots \Bigr)$;
$p_k (.)$ indicate the well-known Schur polynomials.

The tau-function $\t (t)$ is related to the coeffients of the Lax operator
\rf{Lax-gen} through the relation:
\be
\pa_x \partder{}{t_n} \ln \t = \Res \cL^n
\lab{tau-basic}
\ee
where the terms on the r.h.s. of \rf{tau-basic} are the densities of the 
conserved quantities.

There exist few other objects in Sato formalism for integrable hierarchies
which play fundamental role in our construction.
(Adjoint) eigenfunctions $\P (t)$ ($\Psi (t)$, respectively) are those
functions of KP ``times'' $(t)$ satisfying:
\be
\partder{}{t_l} \P = (\cL^l)_{+} (\P) \quad ,\quad
\partder{}{t_l} \Psi = - (\cL^l)^\ast_{+} (\Psi)
\lab{EF-def}
\ee
According to second Eq.\rf{BA-eqs}, (adjoint) BA functions are special
cases of (adjoint) eigenfunctions, which in addition satisfy spectral
equations (first Eq.\rf{BA-eqs}). 

It has been shown in ref.\ct{ridge} that any (adjoint) eigenfunction possesses
a spectral representation of the form\foot{Integrals over spectral parameters
are understood as: 
$\int d\l \equiv \oint_{0} \frac{d\l}{2i\pi} ={\rm Res}_{\l = 0}$.} :
\be
\P (t) = \int d\l\, \vp (\l)\, \psi_{BA}(t,\l) \quad ,\quad
\Psi (t) = \int d\l\, \psi (\l)\, \psi^\ast_{BA}(t,\l)
\lab{spec-repr}
\ee
with appropriate spectral densities $\vp (\l)$ and $\psi (\l)$ which are
formal Laurent series in $\l$. Clearly, any KP hierarchy possesses an
infinite set of independent (adjoint) eigenfunctions in one-to-one
correspondence with the space of all independent formal Laurent series in
$\l$.

The next important object is the so called squared eigenfunction potential 
(SEP) \ct{oevela} -- a function $S\(\P (t),\Psi (t)\)$ associated with 
an arbitrary pair of (adjoint) eigenfunctions $\P (t),\Psi (t)$ 
which possesses the following characteristics:
\be
{\pa \o \pa t_n} S \( \P (t) , \Psi (t)\) = 
{\rm Res} \( D^{-1} \Psi (\cL^n)_{+} \P D^{-1} \)
\lab{potentialflo}
\ee
In particular, for $n=1$ Eq.\rf{potentialflo} implies
~$\pa_x S\(\P (t),\Psi (t)\) = \P (t)\,\Psi (t)$
(recall $\pa_x \equiv \pa/\pa t_1$). Eq.\rf{potentialflo} determines
$S\(\P (t),\Psi (t)\) \equiv \pai\(\P (t)\,\Psi (t)\)$ up to a shift by a 
trivial constant which is uniquely fixed by the fact that any SEP obeys the
following double-spectral representation \ct{ridge} :
\br
\pai\(\P (t)\,\Psi (t)\) = -\int\!\int d\l d\m \, \psi (\l)\,\vp (\m) \,
{1\o \l}\psi^\ast_{BA}(t,\l) \psi_{BA} (t + [\l^{-1}],\m) 
\nonu \\
=  -\int\!\int d\l d\m \, \frac{\psi (\l)\,\vp (\m)}{\l - \m}
e^{\xi (t,\m) - \xi (t,\l)} \frac{\t (t+[\l^{-1}]-[\m^{-1}])}{\t (t)}
\lab{SEP-spec}
\er
with $\vp (\l), \,\psi (\l)$ being the respective spectral densities in
\rf{spec-repr}.
It is in this well-defined sense that inverse space derivatives
$\pai$ will appear throughout our construction below.

For later use we recall some further relations obeyed by SEP functions
involving (adjoint) BA functions \ct{ridge} :
\br
\pai \(\psi_{BA} (t,\l)\Psi (t)\) = {1\o \l} \psi_{BA} (t,\l)
\Psi (t - [\l^{-1}]) \nonu \\
\pai \(\psi^{\ast}_{BA} (t,\l)\P (t)\) = 
-{1\o \l} \psi^{\ast}_{BA} (t,\l) \P (t + [\l^{-1}])
\lab{SEP-def}
\er
where $\P (t)$ and $\Psi (t)$ are arbitrary (adjoint) eigenfunctions
\rf{EF-def}.

In what follows we shall make an essential use of the well-known 
pseudo-differential operator identities (cf., {\sl e.g.}, the appendix in 
first ref.\ct{noak-addsym}) :
\br
(\cB \, M)_{-} = \cB (f) D^{-1} g \quad ,\quad
(M \, \cB)_{-} = f D^{-1} \cB^\ast (g)
\nonu \\
M_1 M_2 = M_1 (f_2) D^{-1} g_2 + f_1 D^{-1} M_2^\ast (g_1) 
\lab{pseudo-diff-id}  \\
M \equiv f D^{-1} g \quad ,\quad
M_{1,2} \equiv f_{1,2} D^{-1} g_{1,2} \quad ,\quad
M_1 (f_2) = f_1 \pai\(g_1 f_2\) \;\; etc.
\nonu
\er
where $\cB$ is arbitrary purely differential operator.
\mskp
\underline{\em 6.2 Constrained KP Hierarchies. Inverse Powers of Lax Operators}
\sskp
So far we have considered the general case of unconstrained KP hierarchy.
Now we are interested in symmetries for {\em constrained} KP hierarchies
${\sf cKP}_{R,M}$ with Lax operators (cf. 
\ct{noak-addsym,ridge,UIC-97} and references therein) 
\foot{Originally ${\sf cKP}_{R,M}$ hirarchies appeared in different
disguises from various parallel developments:
(i) symmetry reductions of the general unconstrained KP hierarchy
\ct{symm-red-cKP,oevela}; (ii) free-field realizations, in terms of finite number 
of fields, of both compatible first and second KP Hamiltonian structures
\ct{abel-cKP}; (iii) a method of extracting continuum integrable hierarchies
from generalized Toda-like lattice hierarchies underlying (multi-)matrix
models in string theory \ct{Bonora-Xiong}.}:
\be
\cL \equiv \cL_{R,M} = D^{R} + \sum_{i=0}^{R-2} u_i D^i + 
\sum_{j=1}^M \P_j D^{-1} \Psi_j = L_{M+R} L_M^{-1}
\lab{Lax-R-M}
\ee
where $\lcurl \P_i,\,\Psi_i\rcurl_{i=1}^M$ is a set of (adjoint)
eigenfunctions of $\cL$. The class of ${\sf cKP}_{R,M}$ hierachies
\rf{Lax-R-M} contains various well-known integrable hierarchies as special
cases: mKdV hierarchies (for $M=0$); AKNS hierarchy (for $R=1,\, M=1$);
Fordy-Kulish hierarchies \ct{F-K} (for $R=1$, $M$ arbitrary); Yajima-Oikawa
equations (for $R=2,\, M=1$), etc. .

The second representation of $\cL \equiv \cL_{R,M}$\foot{Henceforth we shall
employ the short-hand notation $\cL$ for $\cL_{R,M}$ \rf{Lax-R-M} whenever
this will not lead to a confusion.}
is in terms of a ratio of two monic 
purely differential operators $L_{M+R}$ and $L_M$ of orders $M+R$ 
and $M$, respectively (see \ct{UIC-97} and references therein).
For 
$\cL \equiv \cL_{R,M}$ the Sato evolution (isospectral flow) Eqs.\rf{Lax-gen},
the equations for (adjoint) BA \rf{BA-eqs} and (adjoint) eigenfunctions
\rf{EF-def} acquire the form:
\be
\partder{}{t_n}\cL = \Sbr{\(\cL^{n\o R}\)_{+}}{\cL} \quad ,\quad
\cL^{(\ast)} \psi_{BA}^{(\ast)} = \l^R \psi_{BA}^{(\ast)} \quad , \quad
\partder{}{t_n} \psi_{BA}^{(\ast)} = 
\pm (\cL^{(\ast)})^{n\o R}_{+} (\psi_{BA}^{(\ast)})
\lab{Lax-BA-eqs-R}
\ee
\be
\partder{}{t_n} \P = (\cL^{n\o R})_{+} (\P) \quad ,\quad
\partder{}{t_n} \Psi = - (\cL^{n\o R})^\ast_{+} (\Psi)
\lab{EF-def-R}
\ee

In what follows we will also need the explicit form of inverse powers of the Lax 
operator $\cL = L_{M+R} L^{-1}_M$ \rf{Lax-R-M}.
First, let us recall that the inverses of the underlying purely differential
operators are given by:
\be
L_M^{-1} = \sum_{i=1}^M \vp_i D^{-1} \psi_i  \quad ,\quad
L_{M+R}^{-1} = \sum_{a=1}^{M+R} {\bar {\vp}}_a D^{-1} {\bar {\psi}}_a
\lab{inverses}
\ee
where the functions $\lcurl \vp_i \rcurl_{i=1}^M$ and 
$\lcurl \psi_i \rcurl_{i=1}^M$ span $Ker (L_M)$ and $Ker(L_M^\ast)$, 
respectively, whereas $\lcurl {\bar {\vp}}_a \rcurl_{a=1}^{M+R}$ and
$\lcurl {\bar {\psi}}_a \rcurl_{a=1}^{M+R}$ span $Ker (L_{M+R})$ and 
$Ker(L_{M+R}^\ast)$, respectively. Therefore we have:
\be
\cL = (\cL )_{+} + \sum_{i=1}^M L_{M+R}(\vp_i) D^{-1} \psi_i 
\quad ,\quad i.e. \;\;
\P_i =  L_{M+R}(\vp_i) \;\; ,\;\; \Psi_i = \psi_i
\lab{Lax-plus-1}
\ee
\be
\cL^{-1} = \sum_{a=1}^{M+R} L_M ({\bar {\vp}}_a) D^{-1} {\bar {\psi}}_a
\lab{L-minus-1}
\ee
\be
\cL^{-N} = \sum_{a=1}^{M+R} \sum_{s=0}^{N-1}
\cL^{-(N-1)+s} \bigl( L_M ({\bar {\vp}}_a)\bigr) D^{-1} 
\Bigl(\cL^{-s}\Bigr)^\ast ({\bar {\psi}}_a)
\lab{L-minus-N}
\ee
The last formula \rf{L-minus-N} is completely analogous in structure with the 
formula \ct{EOR-95} for the negative pseudo-differential part of a positive 
power of $\cL$ \rf{Lax-R-M}:
\be
\Bigl(\cL^N\Bigr)_{-} = \sum_{i=1}^M \sum_{s=0}^{N-1}
\cL^{N-1-s}(\P_i) D^{-1} \Bigl(\cL^s\Bigr)^\ast (\Psi_i)
\lab{L-plus-N}
\ee

Let us also note that the following simple consequences from the 
definitions of the corresponding objects will play essential role for the 
consistency of the constructions involving inverse powers of $\cL$ :
\be
\cL \bigl( L_M ({\bar{\vp}}_a)\bigr) = 0  \quad ,\quad
\cL^\ast ({\bar {\psi}}_a) = 0   \quad ,\quad
\cL^{-1} (\P_i) = 0 \quad ,\quad  \(\cL^{-1}\)^\ast (\Psi_i) = 0
\lab{zero-eqs}
\ee

Applying the isospectral flow Eqs.\rf{Lax-gen} to $\cL^{-1}$, {\sl i.e.}, 
$\pa/\pa t_n \cL^{-1} = \Sbr{\(\cL^{n\o R}\)_{+}}{\cL^{-1}}$ and taking into 
account the explicit form of $\cL^{-1}$ \rf{L-minus-1} we find that 
$L_M ({\bar{\vp}}_a)$ and ${\bar {\psi}}_a$ are (adjoint) eigenfunctions of 
$\cL$ (cf. \rf{EF-def-R}) :
\be
\partder{}{t_n} L_M ({\bar{\vp}}_a)  = (\cL^{n\o R})_{+} 
\bigl( L_M ({\bar{\vp}}_a)\bigr) \quad ,\quad
\partder{}{t_n} {\bar {\psi}}_a = - (\cL^{n\o R})^\ast_{+} ({\bar {\psi}}_a)
\lab{EF-def-inverse}
\ee
\mskp
\underline{\em 6.3 Loop-Algebra Symmetries of KP Hierarchies}
\sskp
Let us consider the following system of $M$ infinite sets of (adjoint)
eigenfunctions of $\cL \equiv \cL_{R,M}$ \rf{Lax-R-M} :
\be
\P^{(n)}_i \equiv \cL^{n-1}(\P_i) \quad ,\quad
\Psi^{(n)}_i \equiv \(\cL^\ast\)^{n-1}(\Psi_i) \quad ,\;\;\; n=1,2,\ldots \; ;
\;\; i=1,\ldots ,M
\lab{EF-cKP-sys}
\ee
which are expressed in terms of the $M$ pairs of (adjoint) eigenfunctions
entering the pseudo-differential part of $\cL \equiv \cL_{R,M}$ \rf{Lax-R-M}.
Using \rf{EF-cKP-sys} we can build the following infinite set of additional 
symmetry flows:
\be
\d^{(n)}_A \cL = \Sbr{\cM^{(n)}_A}{\cL}       \quad ,\quad
\cM^{(n)}_A \equiv \sum_{i,j=1}^M A^{(n)}_{ij} 
\sum_{s=1}^n \P^{(n+1-s)}_j D^{-1} \Psi^{(s)}_i
\lab{flow-n-A}
\ee
where $A^{(n)}$ is an arbitrary constant $M \times M$ matrix, {\sl i.e.},
$A^{(n)} \in Mat(M)$. The flows \rf{flow-n-A} define symmetries since they
commute with the isospectral flows $\partder{}{t_l}$ :
\be
\Sbr{\d_\a}{\partder{}{t_l}} = 0  \quad \longleftrightarrow \quad
\partder{}{t_l}\cM_\a = {\Sbr{(\cL^l)_{+}}{\cM^{(n)}_A}}_{-}
\lab{symm-def}
\ee
The last evolution equation in \rf{symm-def} is directly verified upon using
pseudo-differential identities \rf{pseudo-diff-id} and the fact all the functions entering
the definition of $\cM^{(n)}_A$ \rf{flow-n-A} are (adjoint) eigenfunctions
obeying \rf{EF-def}.

Consistency of the flow action \rf{flow-n-A} with the constrained form
\rf{Lax-R-M} of $\cL \equiv \cL_{R,M}$  implies the following flow action on 
the involved (adjoint) eigenfunctions: 
\br
\d^{(n)}_A \P^{(m)}_i = \cM^{(n)}_A (\P^{(m)}_i) - 
\sum_{j=1}^M A^{(n)}_{ij}\P^{(n+m)}_j 
\nonu \\
\d^{(n)}_A \Psi^{(m)}_i = -\(\cM^{(n)}_A\)^\ast (\Psi^{(m)}_i) + 
\sum_{j=1}^M A^{(n)}_{ji}\Psi^{(n+m)}_j 
\lab{flow-n-A-EF}
\er
The specific form of the inhomogeneous terms on the r.h.s. of 
Eqs.\rf{flow-n-A-EF} is the main ingredient of our symmetry flow construction. 
It is precisely these inhomogeneous terms which yield non-trivial loop-algebra
additional symmetries.

Using the pseudo-differential operator identities \rf{pseudo-diff-id}
and taking into account \rf{flow-n-A-EF} we can show that:
\be
\d^{(n)}_A \cM^{(m)}_B - \d^{(m)}_B \cM^{(n)}_A - 
\Sbr{\cM^{(n)}_A}{\cM^{(m)}_B} = \cM^{(n+m)}_{\lb A,\, B\rb}
\lab{M-commut-A-B}
\ee
Eq.\rf{M-commut-A-B} implies that the symmetry flows \rf{flow-n-A}--\rf{flow-n-A-EF}
span the following infinite-dimensional algebra:
\be
\Sbr{\d^{(n)}_A}{\d^{(m)}_B} = \d^{(n+m)}_{\lb A,\, B\rb} \quad ;\quad
A^{(n)},B^{(m)} \in Mat(M) \;\; ,\;\; n,m =1,2, \ldots
\lab{loop-alg-flows}
\ee
isomorphic to $\({\widehat U}(1) \times {\widehat {SL}}(M)\)_{+}$ where the
subscript $(+)$ indicates taking the positive-grade subalgebra of the
corresponding loop-algebra.
We observe, that in the case of ${\sf cKP}_{R,M}$ models we have
~$\cM^{(n)}_{A=\one} = \(\cL_{R,M}^n\)_{-}$ (insert \rf{EF-cKP-sys} into
first relation \rf{flow-n-A} for $A^{(n)}=\one$ and compare with \rf{L-plus-N}).
Therefore, the flows $\d^{(n)}_{A=\one}$ for ${\sf cKP}_{R,M}$ models
coincide upto a sign with the ordinary isospectral flows modulo $R$:
$\d^{(n)}_{A=\one} = -\partder{}{t_{nR}}$. Thereby the flows
$\d^{(n)}_A$ \rf{flow-n-A} will be called "positive" for brevity.


Now we consider another infinite set of (adjoint) eigenfunctions of
$\cL \equiv \cL_{R,M}$ expressed in terms of the (adjoint) eigenfunctions
entering the inverse power of $\cL^{-1} \equiv \cL_{R,M}^{-1}$ \rf{L-minus-1} :
\br
\P^{(-m)}_a \equiv \cL^{-(m-1)}\bigl( L_M ({\bar \vp}_a)\bigr) \quad ,\quad
\Psi^{(-m)}_a \equiv \(\cL^{-(m-1)}\)^\ast ({\bar \psi}_a) 
\lab{EF-sys-inverse} \\
m=1,2,\ldots \; , \;\; a=1,\ldots , M+R  \phantom{aaaaaaaaaa}
\nonu
\er
Using \rf{EF-sys-inverse} we obtain the following set of ``negative'' symmetry 
flows which parallels completely the set of ``positive'' flows \rf{flow-n-A} :
\be
\d^{(-n)}_\cA \cL = \Sbr{\cM^{(-n)}_\cA}{\cL}     \quad ,\quad
\cM^{(-n)}_\cA \equiv \sum_{a,b=1}^{M+R} \cA^{(-n)}_{ab} 
\sum_{s=1}^n \P^{(-n-1+s)}_b D^{-1} \Psi^{(-s)}_a
\lab{flow-n-A-neg}
\ee
where $\cA^{(-n)}_{ab}$ is an arbitrary constant $(M+R) \times (M+R)$ matrix, 
{\sl i.e.}, $\cA^{(-n)} \in Mat(M+R)$. In fact, since according to 
\rf{L-minus-N} we have $\cM^{(-n)}_{\cA = \one}= \cL^{-n}$, the flows 
$\d^{(-n)}_{\cA = \one}$ vanish identically, {\sl i.e.},
$\d^{(-n)}_{\cA = \one} = 0$, therefore, we restrict $\cA^{(-n)} \in SL(M+R)$.

Consistency of the flow action \rf{flow-n-A-neg} with the constrained form
\rf{Lax-R-M} of $\cL \equiv \cL_{R,M}$ 
and with the constrained form \rf{L-minus-1} of the inverse $\cL^{-1}$
implies the following $\d^{(-n)}_\cA$-flow action on the involved (adjoint)
eigenfunctions (using short-hand notations \rf{EF-cKP-sys} and 
\rf{EF-sys-inverse}) :
\be
\d^{(-n)}_\cA \P^{(m)}_i = \cM^{(-n)}_\cA (\P^{(m)}_i)    \quad ,\quad
\d^{(-n)}_\cA \Psi^{(m)}_i = -\(\cM^{(-n)}_\cA\)^\ast (\Psi^{(m)}_i)
\lab{flow-n-A-neg-EF}
\ee
\br
\d^{(-n)}_\cA \P^{(-m)}_a = \cM^{(-n)}_\cA (\P^{(-m)}_a) - 
\sum_{b=1}^{M+R} \cA^{(-n)}_{ab}\P^{(-n-m)}_b 
\nonu \\
\d^{(-n)}_\cA \Psi^{(-m)}_a = -\(\cM^{(-n)}_\cA\)^\ast (\Psi^{(-m)}_a) + 
\sum_{b=1}^{M+R} \cA^{(-n)}_{ba}\Psi^{(-n-m)}_b 
\lab{flow-n-A-neg-EF-inverse}
\er
Similarly, consistency of ``positive'' $\d^{(n)}_A$-flow action \rf{flow-n-A}
with the constrained form \rf{L-minus-1} of the inverse Lax operator implies:
\be
\d^{(n)}_A \P^{(-m)}_a = \cM^{(n)}_A (\P^{(-m)}_a)   \quad ,\quad
\d^{(n)}_A \Psi^{(-m)}_a = -\(\cM^{(n)}_A\)^\ast (\Psi^{(-m)}_a)
\lab{flow-n-A-EF-inverse}
\ee
Using again the pseudo-differential operator identities \rf{pseudo-diff-id}
we find from \rf{flow-n-A-neg-EF}--\rf{flow-n-A-EF-inverse} 
(cf. Eq.\rf{M-commut-A-B}) :
\be
\d^{(n)}_A \cM^{(-m)}_\cB - \d^{(-m)}_\cB \cM^{(n)}_A -
\Sbr{\cM^{(n)}_A}{\cM^{(-m)}_\cB} = 0
\lab{M-commut-A-B-neg-plus}
\ee
\be
\d^{(-n)}_\cA \cM^{(-m)}_\cB - \d^{(-m)}_\cB \cM^{(-n)}_\cA - 
\Sbr{\cM^{(-n)}_\cA}{\cM^{(-m)}_\cB} = \cM^{(-n-m)}_{\lb \cA,\,\cB\rb}
\lab{M-commut-A-B-neg}
\ee
Eqs.\rf{M-commut-A-B-neg-plus}--\rf{M-commut-A-B-neg} imply that the 
``negative'' symmetry flows \rf{flow-n-A-neg}--\rf{flow-n-A-neg-EF-inverse} 
commute with the ``positive'' flows \rf{flow-n-A}--\rf{flow-n-A-EF}:
\be
\Sbr{\d^{(n)}_A}{\d^{(-m)}_\cB} = 0
\lab{plus-neg-commut}
\ee
and that they themselves span the following infinite-dimensional algebra:
\be
\Sbr{\d^{(-n)}_\cA}{\d^{(-m)}_\cB} = \d^{(-n-m)}_{\lb \cA,\,\cB\rb} 
\quad ;\quad
\cA^{(-n)},\cB^{(-m)} \in SL(M+R) \;\; ,\;\; n,m =1,2, \ldots
\lab{loop-alg-neg-flows}
\ee
which is isomorphic to $\({\widehat {SL}}(M+R)\)_{-}$ (the
subscript $(-)$ indicates taking the negative-grade subalgebra of the
corresponding loop-algebra). 

Therefore, we conclude that the full loop algebra of (additional) symmetries of
${\sl cKP}_{R,M}$ hierarchies \rf{Lax-R-M} is the direct sum:
\be
\({\widehat U}(1)\oplus{\widehat{SL}}(M)\)_{+}\oplus\({\widehat {SL}}(M+R)\)_{-}
\lab{loop-alg-full}
\ee
where also the ordinary isospectral-flow symmetries are included.

Furthermore, starting from relation \rf{tau-basic} and using \rf{flow-n-A} and
\rf{flow-n-A-neg} we find for the transformation laws under the $\d^{(n)}_A$- 
and $\d^{(-n)}_\cA$-flow actions of the tau-function:
\be
\d^{(n)}_A \ln \t = -\pai\({\rm Res} \cM^{(n)}_A\) = - \sum_{i,j=1}^M A^{(n)}_{ij} 
\sum_{s=1}^n \pai\(\P^{(n+1-s)}_j \Psi^{(s)}_i\)
\lab{tau-flow-A-n}
\ee
\be
\d^{(-n)}_\cA \ln \t = -\pai\({\rm Res} \cM^{(-n)}_\cA\) = 
- \sum_{a,b=1}^{M+R} \cA^{(-n)}_{ab} 
\sum_{s=1}^n \pai\(\P^{(-n-1+s)}_b \Psi^{(-s)}_a\)
\lab{tau-flow-A-n-neg}
\ee
\mskp
Using the double spectral representation \rf{SEP-spec} for SEP functions we
can rewrite Eqs.\rf{tau-flow-A-n}--\rf{tau-flow-A-n-neg} in the following
equivalent form, namely, the action of additional symmetry flows 
on the tau-function is given by the action of the ``smeared'' Kyoto school 
bilocal vertex operator $\X$ \ct{Miwa-Jimbo} :
\be
\d^{(n)}_A \t (t) = \int\!\!\int\! d\l d\m\, \rho^{(n)}_A (\l,\m) \X \t (t)
\lab{tau-flow-A-n-DMJ}
\ee
\be
\d^{(-n)}_{\cA} \t (t) = 
\int\!\!\int\! d\l d\m\, \rho^{(-n)}_{\cA} (\l,\m) \X \t (t)
\lab{tau-flow-A-n-neg-DMJ}
\ee
Here:
\be
\rho^{(n)}_A (\l,\m) \equiv \frac{\l^n -\m^n}{\l -\m}
\sum_{i,j=1}^M A^{(n)}_{ij} \psi_i (\l) \vp_j (\m)
\lab{rho-def}
\ee
\be
\rho^{(-n)}_{\cA} (\l,\m) \equiv \frac{\l^{-n} -\m^{-n}}{\l^{-1} -\m^{-1}}
\sum_{a,b=1}^{M+R} \cA^{(-n)}_{ab} \psi^{(-1)}_a (\l) \vp^{(-1)}_b (\m) 
\lab{rho-def-neg}
\ee
with $\vp_i (\l),\,\psi_i (\l)$ and $\vp^{(-1)}_a (\l),\,\psi^{(-1)}_a (\l)$
indicating the spectral densities (cf. \rf{spec-repr}) of the (adjoint)
eigenfunctions $\P_i,\,\Psi_i$ and
$\P^{(-1)}_a \equiv L_M \bigl({\bar \vp}_a\bigr),\,
\Psi^{(-1)}_a \equiv {\bar \psi}_a$, respectively. Further, $\X$ denotes
the Kyoto school bilocal vertex operator \ct{Miwa-Jimbo} :
\be
\X = {1 \o {\l -\m}} : e^{{\hat \th}(\l ) - {\hat \th}(\m )}: =
{1 \o \l -\m}\, e^{\xi \( t, \m \)- \xi(t,\l)} \, e^{\sumi{1} {1 \o l}
\( \l^{-l} - \m^{-l} \) \partder{}{t_l} }
\quad {\rm for} \quad \v \m \v \leq \v \l \v
\lab{defx1}
\ee
where:
\be
{\hat \th}(\l ) \equiv - \sum_{l=1}^{\infty} \l^l {t_l} +
\sum_{l=1}^{\infty} {1\o l} \l^{-l} \partder{}{t_l}
\lab{theta-def}
\ee
and the columns $: \ldots :$ in \rf{defx1} indicate Wick normal ordering w.r.t. 
the creation/annihilation ``modes'' $t_l$ and $\partder{}{t_l}$, respectively.

Since the tau-function contains all solutions of the underlying integrable
hierarchy, Eqs.\rf{tau-flow-A-n}--\rf{tau-flow-A-n-neg} or
Eqs.\rf{tau-flow-A-n-DMJ}--\rf{tau-flow-A-n-neg-DMJ} describe the action of
loop-algebra \rf{loop-alg-full} additional symmetries on the space of 
(soliton-like) solutions of ${\sl cKP}_{R,M}$ hierarchies \rf{Lax-R-M}.

\sskp
\mark
The construction above can be straightforwardly extended to the case of the
general unconstrained KP hierarchy defined by \rf{Lax-gen}. All relations
\rf{flow-n-A}--\rf{loop-alg-flows} and
\rf{flow-n-A-neg}--\rf{loop-alg-neg-flows} remain intact where now:
\be
\lcurl \P^{(n)}_i ,\,\Psi^{(n)}_i \rcurl^{n=1,2,\ldots}_{i=1,\ldots ,M}
\quad ,\quad
\lcurl \P^{(-n)}_a ,\,\Psi^{(-n)}_a \rcurl^{n=1,2,\ldots}_{a=1,\ldots ,M+R}
\lab{EF-sys}
\ee
form an infinite system of independent (adjoint) eigenfunctions of the general
Lax operator \rf{Lax-gen} with $M$, $M+R$ being arbitrary positive integers.
\lskip
\underline{\em 6.4 Multi-Component KP Hierarchies from One-Component Ones}
\sskp
Let us now consider the following subset of ``positive'' flows $\d^{(n)}_{E_k}$ 
\rf{flow-n-A} for the general KP hierarchy \rf{Lax-gen} corresponding to:
\be
E_k = diag(0,\ldots 0,1,0, \ldots, 0) \quad , \quad i.e. \quad
\cM^{(n)}_{E_k} = \sum_{s=1}^n \P^{(n+1-s)}_k D^{-1} \Psi^{(s)}_k
\lab{ghost-k}
\ee
Due to Eq.\rf{loop-alg-flows} the flows $\d^{(n)}_{E_k}$ span an 
infinite-dimensional Abelian algebra and, by construction, they commute with 
the original isospectral flows $\partder{}{t_n}$ as well. Using the extended
set of mutually commuting flows:
\be
\partder{}{t_n} \equiv \pa/\pa \stta{1}{n} \quad , \quad
\d^{(n)}_{E_k} \equiv \pa/\pa\!\!\stta{k+1}{n} \quad ,\quad k=1,\ldots ,M
\lab{multi-comp-KP-flows}
\ee
we can construct the following extended KP integrable hierarchy starting from 
\rf{Lax-gen} :
\be
\pa/\pa t_n \cL = \Sbr{\(\cL^{n}\)_{+}}{\cL}  \quad ,\quad
\pa/\pa\!\!\stta{k}{n} \cL = \Sbr{\cM^{(n)}_{E_k}}{\cL}
\lab{M+1-comp-KP}
\ee
with $\cM^{(n)}_{E_k}$ as in \rf{ghost-k}, where the additional sets of 
``isospectral'' flows act on the constituent (adjoint) eigenfunctions as
(cf. \rf{flow-n-A-EF}) :
\br
\pa/\pa\!\!\stta{k}{n} \P^{(m)}_i = \cM^{(n)}_{E_k} (\P^{(m)}_i)  \quad ,\quad 
\pa/\pa\!\!\stta{k}{n} \Psi^{(m)}_i = -\(\cM^{(n)}_{E_k}\)^\ast (\Psi^{(m)}_i)
\quad ,\quad {\rm for} \; i \neq k
\lab{flow-n-i-EF} \\
\pa/\pa\!\!\stta{k}{n} \P^{(m)}_k = \cM^{(n)}_{E_k} (\P^{(m)}_k) -
\P^{(n+m)}_k 
\quad ,\quad
\pa/\pa\!\!\stta{k}{n} \Psi^{(m)}_k = -\(\cM^{(n)}_{E_k}\)^\ast (\Psi^{(m)}_k) + 
\Psi^{(n+m)}_k 
\lab{flow-n-k-EF}
\er
with $i,k=1,\ldots ,M$ and using short-hand notations \rf{EF-cKP-sys}. 
Such extended KP hierarchies have been previously proposed in 
refs.\ct{multi-comp-KP}. As shown in \ct{multi-comp-KP}, we can identify the 
extended set of ``isospectral'' flows \rf{multi-comp-KP-flows}
with the set of isospectral flows
$\Bigl\{\stta{\ell}{n}\Bigr\}^{\ell =1,\ldots ,M+1}_{n=1,2,\ldots}$ of the
(unconstrained) $M+1$-component matrix KP hierarchy. The latter is defined
in terms of the $M+1 \times M+1$ 
matrix Hirota bilinear identities (see refs.\ct{multi-comp-KP}) :
\br
\sum_{k=1}^{M+1} \vareps_{ik}\, \vareps_{jk}\int\! d\l \,
\l^{\d_{ik} + \d_{jk} -2}\, e^{\xi (\st{k}{t}-\st{k}{t^\pr},\l)}
\t_{ik}\bigl(\ldots ,\st{k}{t}-[\l^{-1}],\ldots \bigr)\,
\t_{kj}\bigl(\ldots ,\st{k}{t^\pr}+[\l^{-1}],\ldots \bigr) = 0
\lab{HBI}
\er
which are obeyed by a set of $M(M+1)+1$ tau-functions $\lcurl \t_{ij}\rcurl$
expressed in terms of the single tau-function $\t$ and the ``positive''
symmetry flow generating (adjoint) eigenfunctions \rf{EF-sys} in the original 
one-component (scalar) KP hierarchy \rf{Lax-gen}--\rf{tau-basic} as follows:
\be
\t_{11}=\t_{ii} = \t \quad ,\quad
\t_{1i} = \t\,\P^{(1)}_{i-1} \quad ,\quad \t_{i1} = - \t\,\Psi^{(1)}_{i-1}
\nonu
\ee
\be
\t_{ij} = \vareps_{ij} \t\,\pai\(\P^{(1)}_{j-1}\Psi^{(1)}_{i-1}\)
\quad ,\quad  i \neq j \;\; ,\;\; i,j = 2, \ldots M+1
\lab{tau-M+1-comp}
\ee
Here $\vareps_{ij} =1$ for $i \leq j$ and $\vareps_{ij} =-1$ for $i>j$, and
$\d_{ij}$ are the usual Kronecker symbols.

The above construction of multi-component (matrix) KP hierarchies out of
ordinary one-component ones can be straighforwardly carried over to the case of
constrained KP models \rf{Lax-R-M} :
\be
\pa/\pa t_n \cL = \Sbr{\(\cL^{n\o R}\)_{+}}{\cL}  \quad ,\quad
\pa/\pa\!\!\stta{k}{n} \cL = \Sbr{\cM^{(n)}_{E_k}}{\cL} \quad ,\quad
k=2,\ldots ,M+1
\lab{M-comp-cKP}
\ee
using the identification \rf{EF-cKP-sys} for the symmetry-generating (adjoint)
eigenfunctions. In this case, however, there is a linear dependence among
the flows \rf{multi-comp-KP-flows}
~$\sum_{k=2}^{M+1} \pa/\pa\!\!\stta{k}{n}=\d^{(n)}_{A=\one}=-\partder{}{t_{nR}}$, 
therefore, the associated
${\sl cKP}_{R,M}$-based extended hierarchy \rf{M-comp-cKP} is equivalent to
$M\times M$ matrix constrained KP hierarchy.

Similarly, we can start with the subset of ``negative'' symmetry flows
$\d^{(-n)}_{E_k}$ \rf{flow-n-A-neg} for ${\sl cKP}_{R,M}$ hierarchy:
\be
\d^{(-n)}_{E_k} \equiv \pa/\pa \st{k}{t}\!\!{}_{-n}  \quad ,\quad
\cM^{(-n)}_{E_k} = \sum_{s=1}^n \P^{(-n-1+s)}_k D^{-1} \Psi^{(-s)}_k
\quad ,\;\; k=2,\ldots, M+R \; ,\;\; n=1,2,\ldots
\lab{ghost-k-neg}
\ee
The flow $\d^{(-n)}_{E_k}$ for $k=1$ is excluded since
$\sum_{k=1}^{M+R} \d^{(-n)}_{E_k} = \d^{(-n)}_{\cA = \one}$ which vanishes
identially as explained in the previous subsection.

The flows \rf{ghost-k-neg} also span an infinite-dimensional Abelian algebra
commuting with the isospectral flows. Using \rf{ghost-k-neg} we now construct
another extended KP-type hierarchy analogous to
\rf{M+1-comp-KP}--\rf{flow-n-k-EF} and based on ${\sl cKP}_{R,M}$ 
\rf{Lax-R-M} :
\be
\pa/\pa t_n \cL = \Sbr{\(\cL^{n\o R}\)_{+}}{\cL}  \quad ,\quad
\pa/\pa\!\!\st{k}{t}\!\!{}_{-n} \cL = \Sbr{\cM^{(-n)}_{E_k}}{\cL}
\lab{M+R-comp-KP-neg}
\ee
\be
\pa/\pa\!\!\st{k}{t}\!\!{}_{-n} \P^{(-m)}_a = 
\cM^{(-n)}_{E_k} (\P^{(-m)}_a)  \quad ,\quad 
\pa/\pa\!\!\st{k}{t}\!\!{}_{-n} \Psi^{(-m)}_a =
-\(\cM^{(-n)}_{E_k}\)^\ast (\Psi^{(-m)}_a)
\quad \;\; {\rm for} \; a \neq k
\lab{flow-n-a-EF-neg}
\ee
\be
\pa/\pa\!\!\st{k}{t}\!\!{}_{-n} \P^{(-m)}_k = 
\cM^{(-n)}_{E_k} (\P^{(-m)}_k) - \P^{(-n-m)}_k 
\quad ,\quad
\pa/\pa\!\!\st{k}{t}\!\!{}_{-n} \Psi^{(-m)}_k =
-\(\cM^{(-n)}_{E_k}\)^\ast (\Psi^{(-m)}_k) + 
\Psi^{(-n-m)}_k 
\lab{flow-n-k-EF-neg}
\ee
where now $a=1,\ldots ,M+R$ , $k=2,\ldots ,M+R$ and we have used short-hand 
notations \rf{EF-sys-inverse}.

Then, following the steps of our construction in
ref.\ct{multi-comp-KP} we arrive at $(M+R)$-component constrained KP
hierarchy given in terms of $(M+R)(M+R-1)+1$ tau-functions
$\lcurl {\wti \t}_{ab} \rcurl$ obeying the corresponding $(M+R)\times(M+R)$
matrix Hirota bilinear identities (cf. \rf{HBI}). The latter tau-functions
are expressed in terms of the original single tau-function $\t$ and the
``negative'' symmetry flow generating (adjoint) eigenfunctions \rf{EF-sys-inverse}
in the original ordinary ${\sl cKP}_{R,M}$ hierarchy \rf{Lax-R-M} as follows:
\be
{\wti \t}_{11}={\wti \t}_{aa} = \t \quad ,\quad
{\wti \t}_{1a} = \t\, L_M ({\bar \vp}_{a}) \quad ,\quad 
{\wti \t}_{a1} = - \t\,{\bar \psi}_{a} 
\nonu
\ee
\be
{\wti \t}_{ab} = \vareps_{ab} \t\,\pai\( L_M ({\bar \vp}_{b}) {\bar \psi}_{a}\)
\quad ,\quad  a \neq b \;\; ,\;\; a,b = 2, \ldots M+R
\lab{tau-M+R-comp}
\ee

Let us note that there exist alternative representation of (constrained)
multi-component KP hierarchies based on matrix generalization of Sato
pseudo-differential operator formalism \ct{oevela} (see also
\ct{Feher-Marshall}). The construction in this Section
of ${\sl cKP}_{R,M}$-based extended (multi-component KP) hierarchies 
\rf{M-comp-cKP} and \rf{M+R-comp-KP-neg} has an advantage over the matrix
Sato formulation since it allows us to employ the well-known \DB
techniques from ordinary one-component (scalar) KP hierarchies (full or
constrained) in order to obtain new soliton-like solutions of
multi-component (matrix) KP hierarchies (see last ref.\ct{multi-comp-KP} and,
especially, Section 8 below).
\lskip
\underline{\em 6.5 Higher-Dimensional Nonlinear Evolution Equations as Symmetry
Flows of ${\sl cKP}_{R,M}$ Hierarchies}
\sskp
Let us recall that multi-component (matrix) KP hierarchies \rf{HBI}
contain various physically interesting nonlinear systems such as
Davey-Stewartson and $N$-wave systems, which now can be written entirely in
terms of objects belonging to ordinary one-component (constrained) KP hierarchy. 
Thereby the lowest-grade additional symmetry flow parameters acquire the meaning
of coordinates for additional space dimensions.

For instance, the $N$-wave resonant system ($N=M(M+1)/2$) is given by:
\be
\pa_k f_{ij} = f_{ik} f_{kj} \quad ,\quad i\neq j\neq k \quad ,\quad
i,j,k=1,\ldots, M+1
\lab{N-wave-def}
\ee
\be
\pa_k \equiv \pa/\pa \st{k}{t}\!\!{}_1 \quad ,\quad
f_{1i} \equiv \P^{(1)}_{i-1} \quad ,\quad f_{i1} \equiv - \Psi^{(1)}_{i-1}
\nonu
\ee
\be
f_{ij} \equiv \vareps_{ij} \pai\(\P^{(1)}_{j-1}\Psi^{(1)}_{i-1}\)
\quad ,\quad  i \neq j \;\; ,\;\; i,j = 2, \ldots M+1
\lab{N-wave-id}
\ee

As a further example, let us demonstrate in some detail that the well-known
Davey-Stewartson system \ct{Dav-Stew} arises as particular subset of symmetry flow
equations obeyed by any pair of adjoint eigenfunctions $\bigl(\P_i,\Psi_i\bigr)$
($i$=fixed) or $\bigl(L_M ({\bar \vp}_a), {\bar \psi}_a\bigr)$ ($a$=fixed).
The derivation for $\bigl(\P_i,\Psi_i\bigr)$ ($i$=fixed) has already been
presented in last ref.\ct{multi-comp-KP}. Here for simplicity we take
${\sl cKP}_{1,M}$ hierarchy \rf{Lax-1-M} (the general case for 
${\sl cKP}_{R,M}$ hierarchy \rf{Lax-R-M} is a straightforward generalization 
of the formulas below) and consider a pair
of ``negative'' symmetry flow generating (adjoint) eigenfunctions
$\bigl(\phi \equiv L_M ({\bar \vp}_a), \psi \equiv {\bar \psi}_a\bigr)$
($a$=fixed), which obeys the following subset of flow equations -- w.r.t. 
$\pa/\pa t_2$, $\bpa \equiv \pa/\pa \st{a}{t}\!\!\!\!{}_{-1}$ and
$\pa/\pa \bt_2 \equiv \pa/\pa \st{a}{t}\!\!\!\!{}_{-2}$ 
(cf. Eqs.\rf{flow-n-A-neg-EF-inverse}) :
\be
\partder{}{t_2} \phi = \(\pa^2 + 2\sum_{i=1}^M \P_i \Psi_i\) \phi \quad ,\quad
\partder{}{t_2} \psi = 
- \(\pa^2 + 2\sum_{i=1}^M \P_i \Psi_i\) \psi 
\lab{EF-eqs-2-inverse}
\ee
\be
\bpa \phi = \cM^{(-1)} (\phi) - \cL^{-1}(\phi)  \quad ,\quad
\bpa \psi = - \(\cM^{(-1)}\)^\ast (\psi) 
+ \(\cL^{-1}\)^\ast (\psi)
\lab{ghost-1-minus}
\ee
\be
\pa/\pa \bt_2 \phi = \cM^{(-2)} (\phi) - \cL^{-2}(\phi)  \quad ,\quad
\pa/\pa \bt_2 \psi = - \(\cM^{(-2)}\)^\ast (\psi) 
+ \(\cL^{-2}\)^\ast (\psi)
\lab{ghost-2-minus}
\ee
where:
\be
\cM^{(-1)} \equiv \phi D^{-1} \psi  \quad ,\quad
\cM^{(-2)} \equiv \cL^{-1}(\phi) D^{-1} (\psi) +
\phi D^{-1} \(\cL^{-1}\)^\ast (\psi)
\lab{M-neg-1-2}
\ee
Using \rf{ghost-1-minus} we can rewrite Eqs.\rf{ghost-2-minus} as purely
differential equation w.r.t. $\bpa$ :
\be
\partder{}{\bt_2} \phi = 
\llb - \bpa^2 + 2\bpa\(\pai\(\phi \psi\)\)\rrb \phi
\quad ,\quad
\partder{}{\bt_2} \psi = 
\llb \bpa^2 - 2\bpa\(\pai\(\phi \psi\)\)\rrb \psi
\lab{ghost-2-minus-diff}
\ee
Now, introducing new time variable $T=t_2 - \bt_2$ and the short-hand notation
$Q \equiv \sum_{i=1}^M \P_i \Psi_i - 2(\phi \psi) -
2 \bpa \bigl(\pai (\phi \psi)\bigr)$, and subtracting
Eqs.\rf{ghost-2-minus-diff} from Eqs.\rf{EF-eqs-2-inverse}, we arrive at the
following system of $(2+1)$-dimensional nonlinear evolution equations:
\be
\partder{}{T} \phi = 
\Bigl\lb \h (\pa^2 + \bpa^2) + Q + 2 \phi \psi \Bigr\rb \phi
\lab{DS-dyn-1}
\ee
\be
\partder{}{T} \psi = 
- \Bigl\lb \h (\pa^2 + \bpa^2) + Q + 2 \phi \psi \Bigr\rb \psi
\lab{DS-dyn-2}
\ee
\be
\pa\bpa Q + (\pa +\bpa)^2 \(\phi \psi\) = 0
\lab{DS-nondyn}
\ee
which is precisely the standard Davey-Stewartson system \ct{Dav-Stew} for the
``negative'' (adjoint) eigenfunction pair
$\bigl(\phi \equiv L_M ({\bar \vp}_a),\, \psi \equiv {\bar \psi}_a\bigr)$
($a$=fixed).
\lskip
%
{\bf 7. Algebraic (Drinfeld-Sokolov) Formulation of ${\bf cKP}_{R,M}$
Hierarchies and Its Relation to the Sato Formulation}
\lskip
\underline{\em 7.1 Symmetry Flows in Algebraic (Drinfeld-Sokolov) Setting}
\sskp
Let us very briefly recall the basics of the algebraic (generalized)
Drinfeld-Sokolov construction of integrable hierarchies \ct{Drin-Sok}. The main
ingredients are:
\begin{itemize}
\item
Loop (or Kac-Moody) algebra with integral grading
${\widehat \cG} = \oplus_{n\in \IZ} {\cG}^{(n)}$;
\item
Fixed semisimple element $E \in \cG$ of positive grade, {\sl e.g.} of grade 1 
($E \equiv E^{(1)}$ -- the case to be considered below), meaning that 
${\widehat \cG}$ splits in a direct sum (as vector space) 
${\widehat \cG} = \cK \oplus \cM$ where $\cK \equiv {\rm Ker}\bigl( ad(E)\bigr)$
and $\cM \equiv {\rm Im} \bigl( ad(E)\bigr)$ with
$\Sbr{\cK}{\cK} \subset \cK$, $\Sbr{\cK}{\cM} \subset \cM$ ;
\item
Dynamical field $A \in \cM$ of non-negative grade smaller than the grade of
$E$, {\sl e.g.} of grade 0 ($A \equiv A^{(0)}$ -- the case to be considered 
below).
\end{itemize}
The next basic object is the transfer (monodromy) matrix $T$ taking values in
the corresponding group ${\widehat G}$ and satisfying the linear (matrix-Lax)
equation:
\be
\(\pa + E^{(1)} + A^{(0)}\) T = 0
\lab{transf-matr}
\ee
There are two types of relevant boundary conditions for $T$. The first one,
called ``regular boundary conditions'', sets $T\bgv_{x=x_0} = \one$ at some
fixed point $x=x_0$ and the corresponding solution $\cT \equiv T_{reg}$
of \rf{transf-matr} reads:
\be
\cT = P\exp\Bigl\{ - \int^x_{x_0} dx^\pr \( E^{(1)} + A^{(0)}(x^\pr)\)\bigr\}
\lab{T-reg}
\ee
which implies that $\cT \equiv T_{reg}$ contains only non-negative-grade terms.
The second type of boundary conditions for the transfer-matrix is called 
``asymptotic boundary conditions'' and the
corresponding solution $T \equiv T_{asy}$ of \rf{transf-matr} has, up to an
explicit factor, asymptotic expansion in negative grades only \ct{dress-tech} :
\be
T = \Th e^{-x E^{(1)}} = U\, S\, e^{-x E^{(1)}}
\lab{T-asy}
\ee
The group-valued factors $U$ and $S$ in \rf{T-asy} :
\be
U = \exp\Bigl\{\sum_{j=1}^\infty u^{(-j)}\Bigr\} \quad ,\quad
u^{(-j)} \in \cM^{(-j)}
\lab{U-factor}
\ee
\be
S = e^{\sg} \quad ,\quad \sg = \sum_{j=1}^\infty \sg^{(-j)} \in \cK
\lab{S-factor}
\ee
have a well-defined meaning of ``gauge-rotating'' the matrix Lax operator in
\rf{transf-matr} to the ``bare'' one:
\be
U\, S \(\pa + E^{(1)}\) S^{-1} U^{-1} = \pa + E^{(1)} + A^{(0)}
\lab{dress-alg}
\ee

Symmetry flows in the algebraic (generalized Drinfeld-Sokolov) formalism are 
given as flows acting on $T\equiv T_{asy}$ \rf{T-asy} and $\cT \equiv T_{reg}$
\rf{T-reg}, which are defined in terms of dressing of constant algebraic 
elements $X^{(\pm n)}$ of positive or negative grades \ct{flows-alg,UIC-ARW}:
\be
\d^{(n)}_X T = \( T X^{(n)}T^{-1}\)_{-} T  \quad , \quad
\d^{(n)}_X \cT = - \( T X^{(n)}T^{-1}\)_{+} \cT  \quad \;\; n\geq 1
\lab{flows-alg}
\ee
with $X^{(n)} \in \cK$ (for $X^{(n)} \in \cM$ the flows \rf{flows-alg} are
not well-defined\foot{This ill-definiteness is due to the right-most
exponential factor in $T\equiv T_{asy}$ \rf{T-asy} carrying an infinite
positive-grade tail of terms. Thus if $X^{(n)} \in \cM$, then in
$T X^{(n)}T^{-1} = \Th e^{-x E^{(1)}} X^{(n)} e^{x E^{(1)}} \Th^{-1}$ the
middle factor on the r.h.s. $e^{-x E^{(1)}} X^{(n)} e^{x E^{(1)}}$ is an
infinite series in positive-grade terms, whereas by construction
$\Theta = U\, S$ \rf{T-asy}--\rf{S-factor} is an infinite series in
negative-grade terms. Therefore, $\( T X^{(n)}T^{-1}\)_{-}$ 
for $X^{(n)} \in \cM$ is ill-defined as loop-algebra element, since any 
fixed-grade term in the grade expansion of the latter will be given as an 
infinite sum.}), and:
\be
\d^{(-n)}_X \cT = - \(\cT X^{(-n)}\cT^{-1}\)_{+} \cT  \quad, \quad
\d^{(-n)}_X T = \(\cT X^{(-n)}\cT^{-1}\)_{-} T
\quad \;\; n\geq 1
\lab{flows-alg-neg}
\ee
where $X^{(-n)}$ may be arbitrary negative-grade element.

Applying the flows on the basic linear problem equation \rf{transf-matr} and
using the fact that:
\be
\Sbr{\pa + E^{(1)} + A^{(0)}}{T X T^{-1}} = 0 \quad ,\quad
\Sbr{\pa + E^{(1)} + A^{(0)}}{\cT X \cT^{-1}} = 0
\lab{resolvent}
\ee
for any $X$, one finds for the flow action of the dynamical field:
\be
\d^{(n)}_X A^{(0)} = \Sbr{\pa + E^{(1)} + A^{(0)}}{\( T X^{(n)} T^{-1}\)_{+}}
= - \Sbr{E^{(1)}}{\( T X^{(n)} T^{-1}\)_{(-1)}}
\lab{flow-A-alg}
\ee
\be
\d^{(-n)}_X A^{(0)} = \Sbr{\pa + E^{(1)} + A^{(0)}}{\(\cT X^{(-n)}\cT^{-1}\)_{+}}
= - \Sbr{E^{(1)}}{\(\cT X^{(-n)}\cT^{-1}\)_{(-1)}}
\lab{flow-A-alg-neg}
\ee
where the subscript $(-1)$ indicates taking the grade$\, =\! -1$ part.

The isospectral flows are special subset of positive-grade flows
\rf{flows-alg} corresponding to dressing of the elements $E^{(n)}$ of the
positive-grade part of the center of $\cK$, {\sl i.e.} :
\be
\partder{}{t_n}\equiv \d^{(n)}_{E^{(n)}} \quad ,\quad
\(\partder{}{t_n} - \( T E^{(n)} T^{-1}\)_{-}\) T = 0 \quad ,\;\; {\rm for}\;
n \geq 2
\lab{isospec-alg}
\ee  
Using the explicit form of the flows equations \rf{flows-alg} and 
\rf{flows-alg-neg} one can show (cf. ref.\ct{UIC-ARW}) that:
\begin{itemize}
\item
The positive-grade flows \rf{flows-alg} span an algebra isomorphic to the
``bare'' algebra $\(\cK\)_{+}$ :
\be
\Sbr{\d^{(n)}_X}{\d^{(m)}_Y} T = \( T \Sbr{X^{(n)}}{Y^{(m)}} T^{-1}\)_{-} T 
= \d^{(n+m)}_{\lb X,Y \rb} T
\lab{flows-alg-plus-comm}
\ee
In particular, they commute with the isospectral flows \rf{isospec-alg} which
justifies their name of symmetry flows.
\item
The negative-grade flows \rf{flows-alg-neg} span an algebra isomorphic to
${\widehat \cG}_{-}$ -- the negative-grade part of the underlying loop algebra
${\widehat \cG}$ :
\be
\Sbr{\d^{(-n)}_X}{\d^{(-m)}_Y} \cT = 
- \(\cT \Sbr{X^{(-n)}}{Y^{(-m)}} \cT^{-1}\)_{+} \cT 
= \d^{(-n-m)}_{\lb X,Y \rb} \cT
\lab{flows-alg-neg-comm}
\ee
\item
Negative-grade flows commute with positive-grade flows:
\be
\Sbr{\d^{(n)}_X}{\d^{(-m)}_Y} \cT = 0
\lab{plus-neg-commut-alg}
\ee
In particular, negative flows commute with the isospectral flows
\rf{isospec-alg} and, therefore, the negative-grade flows are similarly
symmetry flows.
\end{itemize}

$T\equiv T_{asy}$ and $\cT \equiv T_{reg}$ are generating functionals of
{\em local} and {\em non-local} conserved charges of the underlying
integrable hierarchy \ct{FT,dress-sym}, respectively, where conservation is 
understood w.r.t. the isospectral flows \rf{isospec-alg}.

In what follows it will be more convenient to use an object ${\wti T}$ related to
$T\equiv T_{asy}$ \rf{T-asy} as:
\be
{\wti T} = 
\Th \exp\Bigl\{ -x E^{(1)} - \sum_{\ell\geq 2} t_\ell E^{(\ell)}\Bigr\} 
\quad \longrightarrow \quad 
\( \partder{}{t_n} + \({\wti T} E^{(n)} {\wti T}^{-1}\)_{+}\) {\wti T} = 0
\lab{T-asy-wti}
\ee
Using ${\wti T}$ one can extend the definition of positive-grade symmetry
flows \rf{flows-alg} to construct also the Virasoro additonal symmetry
flows within the algebraic (generalized Drinfeld-Sokolov) approach. For instance,
in the case of ${\widehat \cG} = {\widehat {SL}}(M+1)$ with standard
homogeneous grading $Q = \l \pa/\pa \l$, which is the algebraic description of
${\sl cKP}_{1,M}$ hierarchies (see next subsection), the Virasoro symmetry
flows are given as (cf. ref.\ct{UIC-ARW}\foot{For (m)KdV hierarchies, which
are the limiting case $M=0$ of the class of ${\sl cKP}_{R,M}$ hierarchies
\rf{Lax-R-M}, Virasoro additional symmetries via dressing within the
algebraic Drinfeld-Sokolov approach have been considered in
refs.\ct{Marian-etal}.}) :
\be
\d^{(n)}_{Vir} {\wti T} = \({\wti T} \ell_n {\wti T}^{-1}\)_{-} {\wti T}
\quad ,\quad \ell_n \equiv -\l^{n+1}\pa/\pa \l \equiv - \l^n Q
\lab{Vir-flows-alg-1}
\ee
whereas in the more general case of ${\widehat \cG} = {\widehat {SL}}(M+R)$
with non-standard grading $Q_R$ (Eq.\rf{grading-R} below) the Virasoro flows
exists for integer values of the modes modulo $R$ \ct{UIC-ARW} :
\be
\d^{(nR)}_{Vir} {\wti T} = 
\({\wti T} {\wti \ell}_{nR} {\wti T}^{-1}\)_{-} {\wti T}
\quad ,\quad {\wti \ell}_{nR} \equiv -\l^n Q_R 
\lab{Vir-flows-alg-R}
\ee
In terms of the original transfer matrix $T \equiv T_{asy}$ \rf{T-asy} 
the flow Eqs.\rf{Vir-flows-alg-R} amount to ``dressing''
of isospectral-time-dependent ``bare'' Virasoro generators \ct{UIC-ARW} :
\be
\d^{(nR)}_{Vir} T = 
\Bigl( T\bigl({\wti \ell}_{nR} - 
\sum_{k=2}^\infty k t_k E^{(k+nR)}\bigr) T^{-1}\Bigr)_{-} T
\lab{time-dep-dress}
\ee
and similarly for \rf{Vir-flows-alg-1} (where $R=1$). 

Let us note that the full Virasoro algebra of additional 
symmetries for constrained ${\sl cKP}_{R,M}$ 
integrable hierarchies \rf{Lax-R-M} has been first constructed in 
refs.\ct{noak-addsym,virflow} within Sato pseudo-differential approach. This 
construction involves a nontrivial modification of ordinary Orlov-Schulmann 
flows  \ct{Orlov-etal,Dickey-book,moerbeke} generating Virasoro and ${\bf W_{1+\infty}}$ 
symmetries of the general unconstrained KP hierarchy. 
Orlov-Schulman flows do not generate symmetries 
for constrained ${\sl cKP}_{R,M}$ hierarchies since they do not preserve 
the constrained form of the pertinent Sato Lax operator \rf{Lax-R-M}. 

Now, let us consider the system of equations consisting of the basic linear
problem \rf{transf-matr} for $\cT = T_{reg}$ \rf{T-reg} and the lowest
negative-grade flow equation (first Eq.\rf{flows-alg-neg}) with
$X^{(-1)}\equiv E^{(-1)}$ belonging to the center of
$\cK = {\rm Ker}\bigl( ad(E^{(1)})\bigr)$ (denoting
$\d^{(-1)}_{E^{(-1)}} \equiv \bpa$) :
\be
\bigl(\pa + E^{(1)} + A^{(0)}\bigr) \cT = 0  \quad ,\quad
\bigl(\bpa + (\cT E^{(-1)}\cT^{-1})_{+}\bigr) \cT = 0 
\lab{lowest-flow-eqs}
\ee
$\cT$ has a grading representation:
\be
\cT = \cT^{(0)} e^\O  \quad ,\quad \O \equiv \sum_{k=1}^\infty \om^{(k)}
\lab{T-reg-grading-repr}
\ee
The zero-grade orders of the both Eqs.\rf{lowest-flow-eqs} yield accordingly:
\be
\bigl(\pa + A^{(0)}\bigr) \cT^{(0)} = 0 \quad {\rm or} \quad
\pa\cT^{(0)} \,\cT^{(0)\, -1} = -A^{(0)}  \quad \longrightarrow \quad
\cT^{(0)} = P\exp \bigl\{ - \int^x dx^\pr A^{(0)} (x^\pr)\bigr\}
\lab{T-reg-0-A}
\ee
\be
\Bigl(\bpa + \cT^{(0)}\lb \om^{(1)},\, E^{(-1)}\rb \cT^{(0)\, -1}\Bigr)
\cT^{(0)} = 0   \quad \longrightarrow \quad 
\cT^{(0)\, -1} \bpa \cT^{(0)} = - \lb \om^{(1)},\, E^{(-1)}\rb
\lab{T-reg-0-bar}
\ee
In particular, we obtain the explicit $\bpa$-flow equation for $A$:
\be
\bpa A^{(0)} = - \Sbr{E^{(1)}}{\cT^{(0)} E^{(-1)}\cT^{(0)\, -1}}
\lab{A-flow-eq-bar}
\ee
It is straightforward to check, using \rf{lowest-flow-eqs}, \rf{T-reg-0-A}
and \rf{A-flow-eq-bar}, that the zero-curvature condition:
\be
\Sbr{\bpa + (\cT E^{(-1)}\cT^{-1})_{+}}{\pa + E^{(1)} + A^{(0)}} = 0
\lab{zero-curv-a}
\ee
is identically satisfied. Note, that due to the first Eq.\rf{lowest-flow-eqs},
Eq.\rf{zero-curv-a} can be equivalently written entirely in terms of $\cT^{(0)}$ as:
\be
\Sbr{\bpa - \cT^{(0)} E^{(-1)}\cT^{(0)\, -1}}{\pa + E^{(1)} + A^{(0)}} = 0
\lab{zero-curv-1}
\ee

Now, let us observe that substituting $A^{(0)} = -\pa\cT^{(0)} \,\cT^{(0)\, -1}$ 
(according to \rf{T-reg-0-A}) in Eq.\rf{A-flow-eq-bar}, the latter acquires
the form:
\be
\bpa \bigl(\pa\cT^{(0)} \,\cT^{(0)\, -1}\bigr) - 
\Sbr{E^{(1)}}{\cT^{(0)} E^{(-1)}\cT^{(0)\, -1}} = 0
\lab{gauged-WZNW-eq}
\ee
together with the constraints:
\be
\pa\cT^{(0)} \,\cT^{(0)\, -1} \bgv_{\cH} \;\;\(\; =
- A^{(0)}\bgv_{\cH} \;\)\;\; = 0 \quad ,\quad {\rm since}\;\;\; A^{(0)}\in\cM
\lab{constr-int-a}
\ee
\be
\cT^{(0)\, -1} \bpa \cT^{(0)} \bgv_{\cH} \;\;\(\; =
\Sbr{E^{(-1)}}{\om^{(1)}} \bgv_{\cH} \;\)\;\; = 0 \quad ,\quad {\rm since}\;\;\;
\Sbr{E^{(-1)}}{\om^{(1)}} \in \cM
\lab{constr-int-b}
\ee
Setting $g \equiv \cT^{(0)}$ in Eq.\rf{gauged-WZNW-eq} the latter is
identified with Eq.\rf{eqmotion-gf-KM} with $E^{(1)}_{+} \equiv E^{(1)}$
and $E^{(1)}_{-} \equiv E^{(-1)}$ (recall that 
$\d^{(-1)}_{E^{(-1)}} \equiv \bpa$).
Moreover, the constraints \rf{constr-int-a}--\rf{constr-int-b}
are particular cases of the general constraints on the WZNW ``currents'' 
\rf{constr-gen-a}--\rf{constr-gen-b}.
Therefore, we conclude that the field equations of motion (in ``light-cone''
coordinates) of the gauged $G/H$ WZNW model can be identified as a particular
case of symmetry flow equations of a generalized Drinfeld-Sokolov integrable
hierarchy based on the loop algebra ${\widehat \cG}$ and where the subalgebra
${\widehat \cH}$ is such that 
${\widehat \cH} = {\rm Ker} \bigl( ad(E^{(1)}_{+})\bigr)$, where $E^{(1)}_{+}$
is semisimple element and both $E^{(\pm 1)}_{\pm}$ belong to the center of 
${\widehat \cH}$.

In ref.\ct{AFGZ-00} it was shown that Eq.\rf{gauged-WZNW-eq} in the special
case of ${\widehat {SL}}(2)$ with $E^{(1)} = {\l \o 2}\s_3$, which
corresponds in Sato formalism to ${\sl cKP}_{1,1}$ hierarchy
(Eq.\rf{Lax-R-M} with $R=1$ and $M=1$), contains the complex Sine-Gordon
equations. Also, in ref.\ct{AFGZ-00} the more general
Drinfeld-Sokolov hierarchies based on $\cG = {\widehat {SL}}(R)$ with:
\be
E^{(1)} = -\Bigl( \sum_{i=1}^{R-1} E^{(1)}_{\a_{i}} +
E^{(0)}_{-(\a_{1}+\cdots +\a_{R-1})} \Bigr) \quad ,\quad
A^{(0)} = - \sum_{i=1}^{R-1} \( \pa\vp_1 + \cdots + \pa\vp_i\) H_{\a_i}
\lab{E-A-def-0}
\ee
(cf. Eq.\rf{E-gen-def} below)
has been considered, which corresponds in Sato formalism to the limiting case
${\sl cKP}_{R,0}$ of ${\sl cKP}_{R,M}$ hierarchies (Eq.\rf{Lax-R-M} with
$M=0$, {\sl i.e.}, the purely differential mKdV Lax operator). It has been
shown in \ct{AFGZ-00} that Eqs.\rf{gauged-WZNW-eq} in the latter special case 
reduce to the the equations of motion of affine Toda field theories. In what 
follows we extend this discussion to the whole class of ${\sl cKP}_{R,M}$ 
hierarchies.
\mskp
\underline{\em 7.2 Algebraic (Drinfeld-Sokolov) Formulation of ${\bf cKP}_{R,M}$
Hierarchies}
\sskp
We start with the subclass of ${\bf cKP}_{1,M}$ hierarchies 
(cf. \rf{Lax-R-M}) with Sato Lax operator:
\be
\cL_{1,M} = D + \sum_{i=1}^M \P_i D^{-1}\Psi_i
\lab{Lax-1-M}
\ee
Let us introduce the column vector:
\be
\ndcol{\psi_1}{\psi_M}{\psi_{M+1}} =
e^{-{1\o{M+1}} \xi(t,\l)} \ndcol{S_M}{S_1}{\psi_{BA}}
\lab{BA-col}
\ee
where the following short-hand notations are used (recall \rf{SEP-def}) :
\br
S_i (t,\l) \equiv \pai \(\psi_{BA} (t,\l)\Psi_i (t)\) = {1\o \l} \psi_{BA} (t,\l)
\Psi_i (t - [\l^{-1}]) \nonu \\
S_i^\ast (t,\l) \equiv \pai \(\psi^{\ast}_{BA} (t,\l)\P_i (t)\) = 
-{1\o \l} \psi^{\ast}_{BA} (t,\l) \P_i (t + [\l^{-1}])
\lab{SEP-def-i}
\er
with $i=1,\ldots ,M$ . It is well-known \ct{AFGZ-97} that the linear spectral Lax
equation (first Eq.\rf{BA-eqs}) for the BA wave function $\psi_{BA}$ of 
\rf{Lax-1-M} can be represented in the following algebraic 
(generalized Drinfeld-Sokolov) form:
\be
\Bigl\lb D + E + A \Bigr\rb \ndcol{\psi_1}{\psi_M}{\psi_{M+1}} = 0
\lab{Lax-SL-M+1}
\ee
with:
\be
E \equiv E^{(1)} = {\l \o {M+1}}\nxnrmat{1}{0}{0}{0}{1}{0}{0}{0}{-M}
\equiv H^{(1)}_{\l_M}
\quad ,\quad
A \equiv A^{(0)} = \nxnrmat{0}{0}{-\Psi_M}{0}{0}{-\Psi_1}{\P_M}{\P_1}{0}
\lab{E-A-def-M+1}
\ee
Here the underlying loop algebra is ${\widehat {SL}}(M+1)$ with standard
homogeneous gradation $Q = \l \pa/\pa \l$ and the corresponding kernel
$\cK \equiv {\rm Ker} \bigl( ad(E)\bigr)$ and image 
$\cM \equiv {\rm Im} \bigl( ad(E)\bigr)$ are given by:
\be
\cK = \lcurl E^{(n)}\equiv H^{(n)}_{\l_M},\,
H^{(n)}_1,\ldots , H^{(n)}_{M-1}, \, 
E^{(n)}_{\pm (\a_{k_1} + \ldots + \a_{k_s})} \rcurl_{n\in \IZ} 
\lab{Ker-SL-M+1}
\ee
\be
\cM = \lcurl E^{(n)}_{\pm \a_M},\, 
E^{(n)}_{\pm (\a_{k_1} + \ldots + \a_{k_s} + \a_M)} \rcurl_{n\in \IZ} 
\lab{Im-SL-M+1}
\ee
$\l_M$ is the last $SL(M+1)$ fundamental weight,
$1 \leq k_1 \leq \ldots \leq k_s \leq M-1$ and $s=1,\ldots ,M-1$.
The center of $\cK$ generating the isospectral flows via \rf{isospec-alg} is
$\cC (\cK) = \lcurl E^{(n)}\equiv H^{(n)}_{\l_M}\rcurl_{n\in \IZ}$.

In order to establish the equivalence between the algebraic and Sato
formulations, we need to establish the opposite transition, {\sl i.e.}, 
the transition from the basic object in the algebraic framework -- the 
transfer matrix \rf{transf-matr}, to the objects characterizing the integrable
hierarchy in Sato formalism. This transition is provided by the following formula 
(cf. definition \rf{BA-col}):
\be
\ndcol{S_M}{S_1}{\psi_{BA}} \equiv 
e^{{1\o{M+1}} \xi(t,\l)} \ndcol{\psi_1}{\psi_M}{\psi_{M+1}} =
e^{{1\o{M+1}} \xi(t,\l)} {\wti T} \ndcol{0}{0}{1} = 
e^{\xi (t,\l)} \Theta\ndcol{0}{0}{1} 
\lab{BA-T-M}
\ee
and similarly for the adjoint objects: 
\be
\ndcol{S^\ast_M}{S^\ast_1}{\psi^\ast_{BA}} = 
e^{-{1\o{M+1}} \xi(t,\l)} {\wti T}^{\ast \, -1} \ndcol{0}{0}{1} = 
e^{-\xi (t,\l)} \Theta^{\ast \, -1}\ndcol{0}{0}{1} 
\lab{adj-BA-T-M}
\ee
where ${\wti T}$ is the ``asymptotic'' transfer matrix \rf{T-asy-wti}.
The derivation of \rf{BA-T-M}--\rf{adj-BA-T-M} uses the special grading
properties of the constituent group factors in ${\wti T}$ \rf{T-asy-wti}
and compares them with the $\l$ dependence of $\psi^{(\ast)}_{BA} (t,\l )$ 
\rf{BA-def} and $S^{(\ast)}_i (t,\l )$ \rf{SEP-def}.

The group factor $S$ in the decomposition of ${\wti T}$ \rf{T-asy-wti}
has the form:
\be
S = e^{\sg (\l)} \quad ,\quad
\sg (\l) = \sum_{k=1}^M \sg_k (\l) H_k + \sum_{\b\, ,\, \b \neq \a_M}
\sg_{\pm\b}(\l) E_{\pm\b}    \quad \longrightarrow \quad
S \ndcol{0}{0}{1} = e^{-\sg_M (\l)}\ndcol{0}{0}{1}
\lab{kernel-exp-1}
\ee
Accordingly, the other group factor $U$ is of the form:
\be
U = \exp \nxnrmat{0}{0}{b_M}{0}{0}{b_1}{a_M}{a_1}{0}  
\quad \longrightarrow \quad
U \ndcol{0}{0}{1} = \cosh(\sqrt{\vec{a}.\vec{b}})\ndcol{0}{0}{1} +
\frac{\sinh(\sqrt{\vec{a}.\vec{b}})}{\sqrt{\vec{a}.\vec{b}}}
\ndcol{b_M}{b_1}{0}
\lab{U-action-M}
\ee

Thus, for the BA wave functions \rf{BA-def} in the algebraic framework we obtain:
\be
\psi_{BA}^{(\ast)} (t,\l) = 
e^{\pm \xi (t,\l)} \frac{\t (t \mp [\l^{-1}])}{\t (t)} =
e^{\pm \xi (t,\l)} e^{\mp \sg_M (\l)}
\Bigl\langle (0,0,\ldots ,1)\bv U^{(\ast\, -1)}\bv \ndcol{0}{0}{1}\Bigr\rangle
\lab{BA-T-alg-M}
\ee
Eqs.\rf{BA-T-alg-M}, taking into account \rf{U-action-M}, imply our main result
about the algebraic formalism's expression for the tau-function of
${\sl cKP}_{1,M}$ hierarchy \rf{Lax-1-M} :
\be
\frac{\t (t-[\l^{-1}])}{\t (t+[\l^{-1}])} = e^{-2\sg_M (\l)} 
\lab{tau-sg-rel-M}
\ee
where $\sg_M (\l)$ is the coefficient in front of $H_M$ in the expansion
\rf{kernel-exp-1} of $\sg (\l) $ in the pertinent $\cK$ \rf{Ker-SL-M+1}.

Taking the scalar product of both colomn vectors \rf{BA-T-M} and \rf{adj-BA-T-M} 
we find the following constraint relation:
\be
\psi_{BA}(t,\l) \psi^\ast_{BA} (t,\l) + 
\sum_{i=1}^M S_i (t,\l) S_i^\ast (t,\l) = 1
\lab{tau-SEP-SL-M+1-rel}
\ee
For the general constraint relation involving BA and SEP functions, valid for any
constrained ${\sl cKP}_{R,M}$ model, see Eq.\rf{tau-SEP-rel} below.

In the particular case of ${\sl cKP}_{1,1}$ hierarchies, {\sl i.e.}, with Sato
Lax operator $\cL = D + \P D^{-1} \Psi$ which corresponds to a
(generalized) Drinfeld-Sokolov hierarchy based on ${\widehat {SL}}(2)$ with
standard homogeneous grading and:
\be
E \equiv E^{(1)} = \h \l\s_3 \quad ,\quad
A \equiv A^{(0)} = \P \s_{-} - \Psi \s_{+}
\lab{SL-2}
\ee
we can, using \rf{BA-T-M}--\rf{adj-BA-T-M} for $M=1$ , explicitly express all 
matrix elements of the ``asymptotic'' transfer matrix \rf{T-asy} in terms of 
Sato-formalism objects:
\be
{\wti T} = \twomat{ \frac{\t (t+[\l^{-1}])}{\t(t)} }{
{1\o \l}\Psi (t-[\l^{-1}]) \frac{\t (t-[\l^{-1}])}{\t(t)} }{
{1\o \l}\P (t+[\l^{-1}]) \frac{\t (t+[\l^{-1}])}{\t(t)} }{
\frac{\t (t-[\l^{-1}])}{\t(t)} }\,
e^{\bigl(-\h \l x - \h \sum_{j=2}^\infty \l^j t_j\bigr)\s_3 }
\lab{T-asy-tau}
\ee

The algebraic formulation of the general ${\sf cKP_{R,M}}$ constrained KP
hierarchies \rf{Lax-R-M} with $R \geq 2$ is given in terms of 
${\widehat {SL}}(M+R)$ with the following non-standard grading:
\be
Q_R = R\l\pa /\pa \l + \sum_{i=1}^{R-1} H^{(0)}_{\l_{M+i}} \equiv
\m \pa /\pa \m + \sum_{i=1}^{R-1} H^{(0)}_{\l_{M+i}} \quad ,\quad  \l = \m^{R}  
\lab{grading-R}  
\ee
and the following choice for the fixed semisimple element:
\be
E \equiv E^{(1)} = -\Bigl( \sum_{i=1}^{R-1} E^{(1)}_{\a_{M+i}} +
E^{(0)}_{-(\a_{M+1}+\cdots +\a_{M+R-1})} \Bigr)
\lab{E-gen-def}
\ee
Also, in \rf{grading-R} we introduced a new spectral parameter $\m$ of ordinary
grade one for later convenience. The linear problem for the monodromy
(transfer) matrix is given by the matrix Lax operator \ct{AFGZ-97} :
\be
L= \left(\begin{array}{cccccccccc}
D & 0  &\cdots &0         & -\Psi_1   & 0      &\cdots& \cdots & \cdots  &0 \\
0 & D&0      &\cdots      & -\Psi_2   & 0      &\cdots& \cdots & \cdots &0 \\
\vdots &    &\ddots &     & \vdots  & 0      &\cdots &\cdots & \cdots
&\vdots \\
0    &    &      &D       & -\Psi_M    & 0      &\cdots &\cdots & \cdots &0 \\
\vp_1 & \vp_2 &\cdots&\vp_M & D- v_1   &  -1   &  0     &\cdots &\cdots & 0 \\
0      &    & \cdots & 0    & 0     & D-v_2 & -1 & 0 &  \cdots &\vdots \\
0      &    & \cdots & 0  & 0        & 0    & D-v_3 &  -1 & \cdots &\vdots \\
\vdots &    & \cdots & 0  & 0  &\cdots &   0    & \ddots &  \ddots &0\\
\vdots &    & \cdots & 0  & 0  &\cdots &   0    & \ddots &  \ddots &-1 \\
0      &   &\cdots   & 0  & -\l & 0 & \cdots &0 & \cdots &  D-v_R
\end{array} \right)
\equiv D + E + A
\lab{alg-Lax-M-R}
\ee
The relation between the coefficients in \rf{Lax-R-M} and those in
\rf{alg-Lax-M-R} is as follows:
\br
(D - v_{R})\ldots (D - v_{1}) = D^R + \sum_{i=1}^{R-2} u_i D^i 
\quad ,\quad  v_R \equiv - \sum_{j=1}^{R-1} v_j
\nonu \\
\P_i = (\pa - v_{R})\ldots (\pa - v_2)\vp_i   \phantom{aaaaaaaaa}
\lab{rel-coeff}
\er

Recall the splitting of 
$\cK = {\rm Ker} \bigl( ad(E)\bigr) = \oplus_{m\in\IZ} \cK^{(m)}$ according to the 
grading \rf{grading-R} \ct{AFGZ-97} :
\be
\cK^{(nR)} = \lcurl \frac{M+R}{M} H^{(n)}_{\l_M}\equiv b_{nR}\, ,\,  
H^{(n)}_1,\ldots , H^{(n)}_{M-1},\, E^{(n)}_{\pm (\a_{k_1} + \ldots + \a_{k_s})} 
\rcurl 
\lab{Ker-SL-M+R-a}
\ee
with the same notations as in Eq.\rf{Ker-SL-M+1}, and:
\be
\cK^{(nR+\ell)} = \lcurl 
\sum_{i=1}^{R-\ell} E^{(n+1)}_{\a_{M+i}+\cdots +\a_{M+\ell -1+i}} +
\sum_{i=1}^{\ell} E^{(n)}_{-(\a_{M+i}+\cdots +\a_{M+R-1-\ell +i})} 
\equiv - b_{nR+\ell} \rcurl
\lab{Ker-SL-M+R-b}
\ee
where $\ell = 1,\ldots ,R-1$. The center of $\cK$ generating the isospectral
flows according to \rf{isospec-alg} is
$\cC (\cK) = \lcurl b_n \rcurl_{n \in \IZ}$ with $b_n$ as defined in
\rf{Ker-SL-M+R-a}--\rf{Ker-SL-M+R-b}.

Now we are interested in the opposite transition: from the algebraic 
(Drinfeld-Sokolov) to Sato formulation of ${\sl cKP}_{R,M}$ hierarchies 
\rf{Lax-R-M}. This transition is established in a
way similar to the simpler case for ${\sl cKP}_{1,M}$ hierarchies 
(cf. Eqs.\rf{BA-T-M}--\rf{adj-BA-T-M}) :
\be
\nmcol{S_1}{S_M}{\psi_{BA}}{\psi_{M+2}}{\psi_{M+R}} =
T \nmcol{0}{0}{1}{\m}{\m^{R-1}} = e^{\xi (t,\m)}\,U\, S\,\nmcol{0}{0}{1}{\m}{\m^{R-1}}
\lab{BA-T-M-R}
\ee
where:
\be
\psi_{M+j} = \m^{R} \(\pa - v_j\)^{-1} \(\pa - v_{j+1}\)^{-1} \ldots 
\(\pa - v_R\)^{-1} \psi_{BA}(t,\m) \quad ,\;\; j=2,\ldots ,R
\lab{psi-M+j}
\ee
and for the corresponding adjoint quantities:
\be
\nmcol{{\wti S}_1}{{\wti S}_M}{\psi^\ast_{M+1}}{
\psi^\ast_{M+2}}{\psi^\ast_{BA}} =
T^{\ast\, -1} \nmcol{0}{0}{\m^{R-1}}{\m^{R-2}}{1} = 
e^{-\xi (t,\m)}\,U^{\ast\, -1} S^{\ast\, -1} \nmcol{0}{0}{\m^{R-1}}{\m^{R-2}}{1}
\lab{adj-BA-T-M-R}
\ee
where:
\be
\psi^\ast_{M+j} = (-1)^{R-j} \(\pa + v_{j+1}\) \ldots \(\pa + v_{R}\)
\psi^\ast_{BA}(t,\m)  \quad ,\;\; j=1,\ldots ,R-1
\lab{adj-psi-M+j}
\ee
\be
{\wti S}_i \equiv (-1)^{R-1} \pai\(\vp_i \psi^\ast_{M+1}\) =
(-1)^{R-1} \pai\,\Bigl(\vp_i \prod_{j=1}^{R}\(\pa + v_j\)\psi^\ast_{BA}\Bigr)
\lab{def-wti-SEP}
\ee
In obtaining the last relation in \rf{BA-T-M-R} use was made of:
\be
b_N \nmcol{0}{0}{1}{\m}{\m^{R-1}} = - \m^N \nmcol{0}{0}{1}{\m}{\m^{R-1}}
\lab{b-N-eqs}
\ee
with $b_N$ as defined in \rf{Ker-SL-M+R-a}--\rf{Ker-SL-M+R-b}.

Taking the scalar product of the column vectors \rf{BA-T-M-R} and 
\rf{adj-BA-T-M-R} we arrive at the following identity generalizing identity
\rf{tau-SEP-SL-M+1-rel}:
\be
\langle \(\psi^\ast\)^T \bv \psi \rangle \equiv
\sum_{i=1}^M {\wti S}_i S_i + \psi^\ast_{M+1} \psi_{BA} +
\sum_{j=2}^{R-1} \psi^\ast_{M+j}\psi_{M+j} + \psi_{BA}^\ast \psi_{M+R} =
R \m^{R-1}
\lab{tau-SEP-rel-alg}
\ee
The same identity can be also directly derived within Sato
pseudo-differential approach by using the relation \ct{oevela} :
\be
\psi^\ast_{BA} \cL_{+}(\psi_{BA}) - \cL^\ast_{+}(\psi^\ast_{BA})\, \psi_{BA}
=\pa \Res \( D^{-1} \psi^\ast_{BA} \cL \psi_{BA} D^{-1}\)
\lab{Oevel-Lemma}
\ee
and using the explicit splitting of $\cL \equiv \cL_{R,M}$ \rf{Lax-R-M} into
differential and purely pseudo-differential parts. Its form in Sato
formalism reads:
\be
\Res \( D^{-1} \psi^\ast_{BA}(t,\m) \cL \psi_{BA}(t,\m) D^{-1}\) + 
\sum_{i=1}^M S^\ast_i (t,\m) S_i (t,\m) = R \m^{R-1}
\lab{tau-SEP-rel}
\ee
Inserting the expressions for $\psi^{(\ast)}_{M+j}$ and ${\wti S}_i$
(Eqs.\rf{psi-M+j} and \rf{adj-psi-M+j}--\rf{def-wti-SEP})
in the l.h.s. of \rf{tau-SEP-rel-alg}, one can show after some algebra that
both forms \rf{tau-SEP-rel-alg} and \rf{tau-SEP-rel} coincide. 
\mskp
\mark The identity \rf{tau-SEP-rel} (or \rf{tau-SEP-rel-alg}) is a constraint 
on the pertinent (adjoint) Baker-Akhiezer functions and can be viewed as 
alternative definition of the constrained  ${\sl cKP}_{R,M}$ hierarchy
\rf{Lax-R-M}. 
\lskip
\underline{\em 7.3 Symmetry Flows of ${\sl cKP}_{R,M}$ Hierarchies in the
Algebraic Setting}
\sskp
Upon inspection of the kernel $\cK$ (Eqs.\rf{Ker-SL-M+R-a}--\rf{Ker-SL-M+R-b})
we deduce that all ${\sl cKP_{R,M}}$ models \rf{Lax-R-M},
{\sl i.e.}, those defined in the algebraic (generalized Drinfeld-Sokolov)
framework through ${\widehat {SL}}(M+R)$ loop algebras with non-standard 
gradings \rf{grading-R}--\rf{alg-Lax-M-R}, share {\em the same}
$\({\widehat {SL}}(M)\)_{+}$ algebra of positive-grade additional symmetries 
irrespective of the value of $R \geq 2$, which is the same as the algebra of
positive-grade additional symmetries of the subclass of models 
${\sl cKP}_{1,M}$ \rf{Lax-1-M}--\rf{Im-SL-M+1}. For the latter models 
these symmetries are generated via dressing of the positive-grade kernel 
elements (notations as in \rf{Ker-SL-M+1}) :
\be
\lcurl H^{(n)}_1,\ldots , H^{(n)}_{M-1},\, 
E^{(n)}_{\pm (\a_{k_1} + \ldots + \a_{k_s})} \rcurl_{n\geq 1}
\lab{SL-M-addsym}
\ee
with the ``asymptotic'' transfer matrix according to \rf{flows-alg}.
Accordingly, for ${\sl cKP_{R,M}}$ hierarchies with $R\geq 2$ 
the $\({\widehat {SL}}(M)\)_{+}$ symmetry flows are generated via dressing of
the positive-modulo-$R$-grade kernel elements in $\cK^{(nR)}$ \rf{Ker-SL-M+R-a}.

Furthermore, ${\sl cKP_{R,M}}$ models within the algebraic framework possess
another algebra of additional symmetries $\({\widehat {SL}}(M+R)\)_{-}$, 
commuting with $\({\widehat {SL}}(M)\)_{+}$ symmetry algebra, which is obtained
via dressing of the whole negative-modulo-$R$-grade part of the underlying loop 
algebra in the generalized Drinfeld-Sokolov scheme with the ``regular'' 
transfer (monodromy) matrix according to \rf{flows-alg-neg}.
Therefore, there is a complete agreement of additional symmetries both
within Sato pseudo-differential operator formulation (cf. \rf{loop-alg-full}) and 
algebraic (generalized Drinfeld-Sokolov) formulation of ${\sl cKP}_{R,M}$ 
hierarchies. 

\lskip
{\bf 8. \DB Transformations and Multiple-Wronskian Solutions of
${\bf cKP}_{R,M}$ Hierarchies}
\sskp
After having shown in subsection 7.1 that the gauge-fixed equations of motion 
of gauged WZNW models are additional-symmetry flow equations for 
generalized Drinfeld-Sokolov hierarchies within the algebraic framework, and 
after establishing the equivalence between algebraic 
(generalized Drinfeld-Sokolov) and Sato 
formulations of the KP-type hierarchies, we can now employ the well-known
\DB techniques in Sato formalism to generate solutions for gauged WZNW field 
equations.

Below we will consider explicitly the case of gauged 
$SL(M+1)/U(1)\times SL(M)\,$ WZNW models
where $E^{(\pm)}_{\pm} = H^{(\pm 1)}_{\l_M}$ in the pertinent field equations
\rf{eqmotion-gf-KM} or, equivalently, Eq.\rf{gauged-WZNW-eq} :
\be
\bpa \bigl(\pa\cT^{(0)} \,\cT^{(0)\, -1}\bigr) - 
\Sbr{H^{(1)}_{\l_M}}{\cT^{(0)} H^{(-1)}_{\l_M}\cT^{(0)\, -1}} = 0
\lab{gauged-WZNW-eq-M}
\ee
with the constraints:
\be
\pa\cT^{(0)} \,\cT^{(0)\, -1} = -A^{(0)} =
- \nxnrmat{0}{0}{-\Psi_M}{0}{0}{-\Psi_1}{\P_M}{\P_1}{0} \quad ,\quad
i.e. \quad \pa\cT^{(0)} \,\cT^{(0)\, -1} \bgv_{U(1)\oplus SL(M)} = 0
\lab{constr-1}
\ee
and:
\be
\cT^{(0)\, -1} \bpa \cT^{(0)}\bgv_{U(1)\oplus SL(M)} = 0
\lab{constr-2}
\ee
due to second Eq.\rf{T-reg-0-bar} (with $E^{(-1)} = H^{(-1)}_{\l_M}$).
Also recall that $\bpa \equiv \d^{(-1)}_{E^{(-1)}} \equiv \d^{(-1)}_{H_{\l_M}}$
is the lowest negative-grade additional symmetry flow in the algebraic 
(generalized Drinfeld-Sokolov)
formulation of the underlying ${\sl cKP}_{1,M}$ integrable hierarchy.

Comparing the corresponding negative-grade symmetry flows in the algebraic
and Sato formulations (cf. \rf{flow-n-A-neg}) and taking into account the 
explicit form of $H^{(n)}_{\l_M}$ (first relation \rf{E-A-def-M+1}), we find:
\be
\d^{(-n)}_{H_{\l_M}} = 
- \pa/\pa \!\! \st{M+1}{t_{-n}} + {1\o {M+1}}\d^{(-n)}_{\cA = \one}
\;\; \stackrel{\sim}{=}\;\; - \pa/\pa \!\! \st{M+1}{t_{-n}}
\lab{flows-n-equiv}
\ee
since, as explained above, the flows $\d^{(-n)}_{\cA = \one}$ vanish
identically. Therefore, to obtain solutions for \rf{gauged-WZNW-eq-M}
it is sufficient to
consider \DB transformations of ${\sl cKP}_{1,M}$-based extended KP-type 
hierarchy of the form \rf{M+R-comp-KP-neg}--\rf{flow-n-k-EF-neg} (for $R=1$)
where the ``light-cone'' time derivative $\bpa$ from \rf{gauged-WZNW-eq-M} is 
identified, upto an overall sign according to \rf{flows-n-equiv}, with the 
lowest additional ``isospectral'' flow 
$\bpa \equiv \d^{(-1)}_{H_{\l_M}} = - \pa/\pa \!\! \st{M+1}{t_{-1}}$.
\mskp
\underline{\em 8.1 \DB Transformations Preserving Additional Symmetries}
\sskp
Let us recall that \DB (DB) transformations within Sato pseudo-differential
approach are defined as ``gauge'' transformations of special kind on the
pertinent Lax operator of the general (unconstrained) KP hierarchy:
\be
\cL \;\;\; \to\;\;\; {\wti \cL} = T_\phi \cL T_\phi^{-1} \quad ,\quad  
T_\phi \equiv \phi D \phi^{-1}
\lab{DB-def}
\ee
which preserve the isospectral (Sato evolution) equations \rf{Lax-gen} :
\be
\partder{}{t_n} {\wti \cL} = \Sbr{\partder{}{t_n}T_\phi\, T_\phi^{-1} 
+ T_\phi \cL^n_{+} T_\phi^{-1}}{{\wti \cL}} 
= \Sbr{{\wti \cL}^n_{+}}{{\wti \cL}} 
\lab{DB-isospec-preserv}
\ee
For the second equality in \rf{DB-isospec-preserv} to be true, the function 
$\phi$ must be an eigenfunction \rf{EF-def}. Similarly, one can define 
adjoint-DB transformation:
\be
\cL \;\;\; \to\;\;\; {\widehat \cL} = T^{\ast\, -1}_\psi \cL T_\psi^{\ast} 
\quad ,\quad   T_\psi \equiv \psi D \psi^{-1}
\lab{adj-DB-def}
\ee
where the function $\psi$ is an adjoint eigenfunction \rf{EF-def}. From
Eq.\rf{tau-basic} one finds the (adjoint) DB transformations of the
tau-function:
\be
\t \;\; \to \;\; {\wti \t} = \t \, \phi \quad ,\quad
\t \;\; \to \;\; {\widehat \t} = - \t \, \psi
\lab{DB-tau}
\ee

In the case of constrained ${\sl cKP}_{R,M}$ hierarchies \rf{Lax-R-M},
the (adjoint) DB transformations \rf{DB-def},\rf{adj-DB-def} must in
addition preserve also the constrained form of $\cL \equiv \cL_{R,M}$.
The transformed Lax operator and its inverse are of the form (using the 
pseudo-differential operator identities \rf{pseudo-diff-id}) :
\be
{\wti \cL} \equiv {\wti \cL}_{R,M} = {\wti \cL}_{+} +
\Bigl( T_\phi \cL (\phi)\Bigr) D^{-1} \phi^{-1} +
\sum_{i=1}^M \Bigl( T_\phi (\P_i)\Bigr) D^{-1} 
\Bigl( T^{\ast\, -1}_\phi (\Psi_i)\Bigr)
\lab{DB-Lax-R-M}
\ee
\be
{\wti \cL}^{-1} = \Bigl( T_\phi \cL^{-1} (\phi)\Bigr) D^{-1} \phi^{-1} +
\sum_{a=1}^{M+R} \Bigl( T_\phi (\P^{(-1)}_a)\Bigr) D^{-1} 
\Bigl( T^{\ast\, -1}_\phi (\Psi^{(-1)}_a)\Bigr)
\lab{DB-Lax-R-M-inverse}
\ee
Similarly for adjoint DB transformations we have:
\be
{\widehat \cL} \equiv {\widehat \cL}_{R,M} = {\widehat \cL}_{+}
- \psi^{-1} D^{-1}\Bigl( T_\psi \cL^\ast (\psi)\Bigr) +
\sum_{i=1}^M \Bigl( T^{\ast\, -1}_\psi (\P_i)\Bigr) D^{-1} 
\Bigl( T_\psi (\Psi_i)\Bigr)
\lab{adj-DB-Lax-R-M}
\ee
\be
{\widehat \cL} = - \psi^{-1} D^{-1}\Bigl( T_\psi \cL^{\ast\, -1} (\psi)\Bigr) +
\sum_{a=1}^{M+R} \Bigl( T^{\ast\, -1}_\psi (\P^{(-1)}_a)\Bigr) D^{-1} 
\Bigl( T_\psi (\Psi^{(-1)}_a)\Bigr)
\lab{adj-DB-Lax-R-M-inverse}
\ee
where the short-hand notations \rf{EF-sys-inverse} have been used.
Since for generic (adjoint) eigenfunctions $\phi$ ($\psi$) the negative
pseudo-differential part of the transformed Lax operator
\rf{DB-Lax-R-M}--\rf{adj-DB-Lax-R-M} has one more term
than the original $\cL \equiv \cL_{R,M}$, there are two types of conditions
implying the vanishing of one superfluous term on the r.h.s. of 
Eqs.\rf{DB-Lax-R-M}--\rf{adj-DB-Lax-R-M}, so that the constrained form
of $\cL \equiv \cL_{R,M}$ is preserved:
\be
{\rm either} \quad \cL (\phi) = 0  \quad ,\quad {\rm or} \quad
\phi = \P_{i_0} \;\; \to \;\; T_\phi (\P_{i_0}) = 0
\lab{DB-cond-EF}
\ee
for some fixed index $i_0$ between $1$ and $M$, and for adjoint-DB transformations:
\be
{\rm either} \quad \cL^\ast (\psi) = 0  \quad ,\quad {\rm or} \quad
\psi = \Psi_{j_0} \;\; \to \;\; T_\psi (\Psi_{j_0}) = 0
\lab{adj-DB-cond-adj-EF}
\ee
for some fixed index $j_0$ between $1$ and $M$.
Due to relations \rf{zero-eqs} the first type of conditions
\rf{DB-cond-EF}--\rf{adj-DB-cond-adj-EF} are satisfied by the (adjoint)
eigenfunctions entering inverse powers of the Lax operator:
\be
\phi = L_M ({\bar \vp}_{a_0}) \quad ,\quad \psi = {\bar \psi}_{b_0}
\lab{DB-choice-2}
\ee
for some fixed indices $a_0,\, b_0$ between $2$ and $M+R$.

Here we are interested in (adjoint) DB transformations which, in addition to
preserving the constrained form of ${\sl cKP}_{R,M}$ Lax operators, preserve
also their additional symmetries. The general case is discussed in the
Appendix. Here we will concentrate on 
(adjoint) DB transformations which preserve the extended set of
``isospectral'' flow equations of ${\sl cKP}_{R,M}$-based extended KP-type
hierarchies \rf{M+R-comp-KP-neg}-\rf{flow-n-k-EF-neg}. For the (adjoint) 
DB-transformed flow-generating operators $\cM^{(-n)}_{E_k}$ we get:
\br
\d^{(-n)}_{E_k} T_\phi \, T^{-1}_\phi + 
T_\phi \cM^{(-n)}_{E_k} T^{-1}_\phi =  \phantom{aaaaaaaaaaaaaaaaaaa}
\nonu \\
\Bigl( T_\phi \bigl( \cM^{(-n)}_{E_k}(\phi) - \d^{(-n)}_{E_k}\phi\bigr)\Bigr)
D^{-1} \phi^{-1} + 
\sum_{s=1}^n T_\phi (\P_k^{(-n-1+s)}) D^{-1} 
T^{\ast\, -1}_\phi (\Psi_k^{(-s)})
\lab{DB-ghost-n}
\er
\br
\Bigl(\d^{(-n)}_{E_k} T^{\ast\, -1}_\psi\Bigr) \, T^{\ast}_\psi + 
T^{\ast\, -1}_\psi \cM^{(-n)}_{E_k} T^{\ast}_\psi =
\phantom{aaaaaaaaaaaaaaaaaaa}
\nonu \\
- \psi^{-1} D^{-1} \biggl( T_\psi \Bigl(\bigl(\cM^{(-n)}_{E_k}\bigr)^\ast (\psi)
+ \d^{(-n)}_{E_k}\psi\Bigr)\biggr) +
\sum_{s=1}^n T^{\ast\, -1}_\psi (\P_k^{(-n-1+s)}) D^{-1} 
T_\psi (\Psi_k^{(-s)})
\lab{adj-DB-ghost-n}
\er
Here again we have one more term in the transformed operators $\cM^{(-n)}_{E_k}$
and again we obtain two types of conditions for vanishing of one superfluous
term:
\be
{\rm either} \quad  \d^{(-n)}_{E_k}\phi = \cM^{(-n)}_{E_k}(\phi)
\quad ,\quad {\rm or} \quad
\phi = \P^{(-1)}_{a_0} \equiv L_M ({\bar \vp}_{a_0}) 
\;\; \to \;\; T_\phi (\P^{(-1)}_{a_0}) = 0
\lab{DB-cond-EF-1}
\ee
for some fixed index $a_0$ between 2 and $M+R$, 
and for adjoint-DB transformations:
\be
{\rm either} \quad  
\d^{(-n)}_{E_k}\psi = - \bigl(\cM^{(-n)}_{E_k}\bigr)^\ast (\psi)
\quad ,\quad {\rm or} \quad
\psi = \Psi^{(-1)}_{b_0} \equiv {\bar \psi}_{b_0} 
\;\; \to \;\; T_\psi (\Psi^{(-1)}_{b_0}) = 0
\lab{adj-DB-cond-adj-EF-1}
\ee
for some fixed index $b_0$ between 2 and $M+R$. The first type of conditions
in \rf{DB-cond-EF-1}--\rf{adj-DB-cond-adj-EF-1} are satisfied, due to 
Eqs.\rf{flow-n-A-neg-EF}, by:
\be
\phi = \P_{i_0} \quad ,\quad \psi = \Psi_{j_0}
\lab{DB-choice-1}
\ee
for some fixed indices $i_0,\, j_0$ between 1 and $M$.

Relations \rf{DB-cond-EF}--\rf{DB-choice-2} and 
\rf{DB-cond-EF-1}--\rf{DB-choice-1} define the set of allowed (adjoint)
DB transformations \rf{DB-def} and \rf{adj-DB-def}, which preserve the form
of the  ${\sl cKP}_{R,M}$-based extended integrable
hierarchies \rf{M+R-comp-KP-neg}-\rf{flow-n-k-EF-neg}. From
\rf{DB-Lax-R-M}--\rf{adj-DB-Lax-R-M-inverse}
we find the (adjoint) DB-transformations of the corresponding
(adjoint) eigenfunctions -- building blocks of $\cL$ \rf{Lax-R-M} and
$\cM^{(-n)}_{E_k}$ \rf{ghost-k-neg} :
\begin{itemize}
\item
For the first choice of DB-generating (adjoint) eigenfunctions \rf{DB-choice-1}
we have:
\be
{\wti \P}^{(n)}_{i_0} = T_\phi (\P^{(n+1)}_{i_0})  \quad, \quad
{\wti \Psi}^{(n)}_{i_0} = \( T_\phi\)^{\ast\, -1} (\Psi^{(n-1)}_{i_0})
\quad {\rm for}\;\; n \geq 2\;\; , \quad
{\wti \Psi}_{i_0} = {1\o \phi} \equiv {1\o {\P_{i_0}}} 
\lab{DB-EF-1-1}
\ee
\be
{\wti \P}^{(n)}_i = T_\phi (\P^{(n)}_i) \quad, \quad
{\wti \Psi}^{(n)}_i = T^{\ast\, -1}_\phi (\Psi^{(n)}_i) \quad 
{\rm for} \;\; i \neq i_0
\lab{DB-EF-1-2}
\ee
\be
{\wti \P}^{(-n)}_a = T_\phi (\P^{(-n)}_a) \quad, \quad
{\wti \Psi}^{(-n)}_a = T^{\ast\, -1}_\phi (\Psi^{(-n)}_a)
\lab{DB-EF-neg-1}
\ee
\be
{\widehat \P}_{j_0} = - {1\o \psi} \equiv - {1\o {\Psi_{j_0}}} \quad ,\quad
{\widehat \P}^{(n)}_{j_0} = - T^{\ast\, -1}_\psi (\P^{(n-1)}_i) \quad 
{\rm for}\;\; n \geq 2\;\; , \quad 
{\widehat \Psi}^{(n)}_{j_0} = - T_\psi (\Psi^{(n+1)}_{j_0}) 
\lab{adj-DB-EF-1-1}
\ee
\be
{\widehat \P}^{(n)}_i = - T^{\ast\, -1}_\psi (\P^{(n)}_i)  \quad ,\quad
{\widehat \Psi}^{(n)}_i = - T_\psi (\Psi^{(n)}_i)  \quad ,\quad i \neq j_0
\lab{adj-DB-EF-1-2}
\ee
\be
{\widehat \P}^{(-n)}_a = - T^{\ast\, -1}_\psi (\P^{(-n)}_a)  \quad ,\quad
{\widehat \Psi}^{(-n)}_a = - T_\psi (\Psi^{(-n)}_a)  
\lab{adj-DB-EF-neg-1}
\ee
\item
For the second choice of DB-generating (adjoint) eigenfunctions 
\rf{DB-choice-2} we obtain:
\be
{\wti \P}^{(n)}_i = T_\phi (\P^{(n)}_i) \quad ,\quad 
{\wti \Psi}^{(n)}_i = T^{\ast\, -1}_\phi (\Psi^{(n)}_i) \quad , \quad 
\phi \equiv L_M ({\bar \vp}_{a_0}) \equiv \P^{(-1)}_{a_0}
\lab{DB-EF-2}
\ee
\be
{\wti \P}^{(-n)}_{a_0} = T_\phi \bigl(\P^{(-n-1)}_{a_0}\bigr)  \;\; ,\;\; 
{\wti \Psi}^{(-n)}_{a_0} = \( T_\phi\)^{\ast\, -1} \bigl(\Psi^{(-n+1)}_{a_0}\bigr)
\;\; {\rm for}\;\; n\geq 2 \;\; ,\;\;
{\wti \Psi}^{(-1)}_{a_0} = {1\o {\phi}} \equiv {1\o {\P^{(-1)}_{a_0}}}
\lab{DB-EF-neg-2-1}
\ee
\be
{\wti \P}^{(-n)}_a = T_\phi \bigl(\P^{(-n)}_a\bigr) \quad ,\quad
{\wti \Psi}^{(-n)}_a = T^{\ast\, -1}_\phi \bigl(\Psi^{(-n)}_a\bigr) \quad 
{\rm for} \;\; a \neq a_0
\lab{DB-EF-neg-2-2}
\ee
\be
{\widehat \P}^{(n)}_i = - T^{\ast\, -1}_\psi (\P^{(n)}_i) \quad ,\quad 
{\widehat \Psi}^{(n)}_i = - T_\psi (\Psi^{(n)}_i) \quad , \quad 
\psi \equiv {\bar \psi}_{b_0} \equiv \Psi^{(-1)}_{b_0}
\lab{adj-DB-EF-2}
\ee
\be
{\widehat \P}^{(-1)}_{b_0} = - {1\o \psi} \equiv
- {1\o {\Psi^{(-1)}_{b_0}}} \;\; ,\;\;
{\widehat \P}^{(-n)}_{b_0} =- T^{\ast\, -1}_\psi (\P^{(-n+1)}_{b_0}) 
\;\; {\rm for} \;\; n\geq 2\;\; , \;\;
{\widehat \Psi}^{(-n)}_{b_0} = 
- T_\psi \bigl(\Psi^{(-n-1)}_{b_0}\bigr)
\lab{adj-DB-EF-neg-2-1}
\ee
\be
{\widehat \P}^{(-n)}_a = - T^{\ast\, -1}_\psi \bigl(\P^{(-n)}_a\bigr) 
\quad, \quad
{\widehat \Psi}^{(-n)}_a = - T_\psi \bigl(\Psi^{(-n)}_a\bigr) 
\quad {\rm for} \;\; a \neq b_0
\lab{adj-DB-EF-neg-2-2}
\ee
\end{itemize}
\mskp
\underline{\em 8.2 Iterations of \DB Transformations}
\sskp
The general \DB orbit consists of successive applications of the allowed
(adjoint) DB transformations \rf{DB-EF-1-1}--\rf{adj-DB-EF-neg-2-2}. 
During iteration of the these DB transformations the following generalization of
Wronskian determinants appears:
\be
{\wti W}_{m;l}\lb \phi_1,\ldots ,\phi_m;\psi_1,\ldots ,\psi_l\rb =
\det \left\Vert 
\begin{array}{ccc}
\phi_1 & \ldots  & \phi_m \\
\pa \phi_1 & \ldots & \pa \phi_m \\
\vdots  &  \ddots  &  \vdots  \\
\pa^{m-l-1} \phi_1 & \ldots & \pa^{m-l-1} \phi_m \\
\pai (\phi_1 \psi_1) & \ldots & \pai (\phi_m \psi_1) \\
\vdots  &  \ddots  &  \vdots \\
\pai (\phi_1 \psi_l) & \ldots & \pai (\phi_m \psi_l)
\end{array}  \right\Vert \quad , \quad m \geq l
\lab{wti-Wronski}
\ee
the ordinary Wronskians being $W_m \lb \phi_1,\ldots ,\phi_m\rb \equiv 
\det \left\Vert \pa^{\a-1} \phi_\b \right\Vert_{\a,\b =1,\ldots ,m}$.

Let us recall that iterations of DB transformations with $T_\phi$ on any
function $f$ are given by:
\be
T^{(k-1;0)}_{\phi_k} \ldots T^{(0;0)}_{\phi_1} (f) =
\frac{W_{k+1}\lb \phi_1,\ldots ,\phi_k,f\rb}{W_k \lb\phi_1,\ldots ,\phi_k\rb}
\lab{DB-iter}
\ee
where:
\be
T^{(l-1;0)}_{\phi_l} \equiv \phi_l^{(l-1;0)} D \bigl(\phi_l^{(l-1;0)}\bigr)^{-1}
\quad ,\quad
\phi_l^{(l-1;0)} = T^{(l-2;0)}_{\phi_{l-1}} \ldots T^{(0;0)}_{\phi_1} (\phi_l)
\lab{DB-iter-1}
\ee
Eq.\rf{DB-iter} follows from the identity (Jacobi expansion theorem for 
Wronskians) :
\be
W_k \lb\phi_1,\ldots ,\phi_{k-1},f\rb \stackrel{\leftrightarrow}{\pa}
W_k \lb\phi_1,\ldots ,\phi_{k-1},g\rb =
W_{k-1} \lb\phi_1,\ldots ,\phi_{k-1}\rb\,
W_{k+1} \lb\phi_1,\ldots ,\phi_{k-1},f,g\rb
\lab{Jacobi-exp}
\ee
where $f,g$ are arbitrary functions.

Iterations of adjoint DB transformations (defined with $T^{\ast\, -1}_\phi$)
on any function $f$ read:
\be
\Bigl( T^{(k-1;0)}_{\phi_k}\Bigr)^{\ast\, -1} \ldots 
\Bigl( T^{(0;0)}_{\phi_1}\Bigr)^{\ast\, -1} (f) =
- \frac{{\wti W}_{k;1}\lb \phi_1,\ldots ,\phi_k;f\rb}{
W_k \lb\phi_1,\ldots ,\phi_k\rb}
\lab{DB-adj-iter}
\ee
Eq.\rf{DB-adj-iter} follows from an identity generalizing \rf{Jacobi-exp}
and involving generalized Wronskian-like determinants \rf{wti-Wronski} :
\be
W_k \lb\phi_1,\ldots ,\phi_k\rb \stackrel{\leftrightarrow}{\pa}
{\wti W}_{k+1;1} \lb \phi_1,\ldots ,\phi_k, f; g\rb =
- W_{k+1} \lb\phi_1,\ldots ,\phi_k,f\rb
{\wti W}_{k;1} \lb \phi_1,\ldots ,\phi_k ; g\rb
\lab{adj-Jacobi-exp}
\ee
Both Wronskian(-like) identities \rf{Jacobi-exp} and \rf{adj-Jacobi-exp} can 
be further generalized to:
\br
{\wti W}_{m;l+1}\lb \phi_1,\ldots ,\phi_m;\psi_1,\ldots ,\psi_l, f\rb
\stackrel{\leftrightarrow}{\pa}
{\wti W}_{m;l+1}\lb \phi_1,\ldots ,\phi_m;\psi_1,\ldots ,\psi_l, g\rb =
\nonu \\
{\wti W}_{m;l}\lb \phi_1,\ldots ,\phi_m;\psi_1,\ldots ,\psi_l\rb\,
{\wti W}_{m;l+2}\lb \phi_1,\ldots ,\phi_m;\psi_1,\ldots ,\psi_l, f,g\rb
\lab{gen-Jacobi-exp}
\er
\br
{\wti W}_{m;n} \lb\phi_1,\ldots ,\phi_m ;\psi_1,\ldots ,\psi_n\rb 
\stackrel{\leftrightarrow}{\pa}
{\wti W}_{m+1;n+1} \lb \phi_1,\ldots ,\phi_m, f; \psi_1,\ldots ,\psi_n, g\rb =
\nonu \\
- {\wti W}_{m+1;n} \lb\phi_1,\ldots ,\phi_m, f; \psi_1,\ldots ,\psi_n \rb
{\wti W}_{m;n+1} \lb \phi_1,\ldots ,\phi_m ; \psi_1,\ldots ,\psi_n, g\rb
\lab{gen-adj-Jacobi-exp}
\er
Using \rf{gen-Jacobi-exp}--\rf{gen-adj-Jacobi-exp} we are able to find the
explicit Wronskian-like expression for arbitrary mixed iterations of DB and 
adjoint-DB transformations of the type
\rf{DB-EF-1-1}--\rf{adj-DB-EF-neg-2-2} :
\br
\Bigl( T^{(m;n-1)}_{\psi_n}\Bigr)^{\ast\, -1} \ldots 
\Bigl( T^{(m;0)}_{\psi_1}\Bigr)^{\ast\, -1}
T^{(m-1;0)}_{\phi_m} \ldots T^{(0;0)}_{\phi_1} (f)
\nonu \\
= \frac{{\wti W}_{m+1;n} \lb\phi_1,\ldots ,\phi_m, f; \psi_1,\ldots ,\psi_n \rb}{
{\wti W}_{m;n} \lb\phi_1,\ldots ,\phi_m ; \psi_1,\ldots ,\psi_n \rb}
\quad ,\quad {\rm for}\;\; m \geq n
\lab{DB-adj-DB-iter-EF-1}
\er
\br
T^{(m;n-1)}_{\psi_n} \ldots T^{(m;0)}_{\psi_1}
\Bigl( T^{(m-1;0)}_{\phi_m}\Bigr)^{\ast\, -1}
\ldots \Bigl( T^{(0;0)}_{\phi_1}\Bigr)^{\ast\, -1} (f)
\nonu \\
= - \frac{{\wti W}_{m;n+1} \lb\phi_1,\ldots ,\phi_m ; \psi_1,\ldots ,\psi_n,f \rb}{
{\wti W}_{m;n} \lb\phi_1,\ldots ,\phi_m ; \psi_1,\ldots ,\psi_n \rb}
\quad ,\quad {\rm for}\;\; m \geq n+1
\lab{DB-adj-DB-iter-adj-EF-1}
\er
where $T^{(l-1;0)}_{\phi_l}$ is the same as in \rf{DB-iter-1}, and where:
\br
\Bigl( T^{(m;l-1)}_{\psi_l}\Bigr)^{\ast\, -1} \equiv 
- \bigl(\psi_l^{(m;l-1)}\bigr)^{-1} D^{-1} \psi_l^{(m;l-1)} \phantom{aaaaaaaaaaaa}
\nonu \\
\psi_l^{(m;l-1)} = T^{(m;l-2)}_{\psi_{l-1}} \ldots T^{(m;0)}_{\psi_1} 
\Bigl( T^{(m-1;0)}_{\phi_m}\Bigr)^{\ast\, -1} \ldots 
\Bigl( T^{(0;0)}_{\phi_1}\Bigr)^{\ast\, -1} (\psi_l)
\lab{adj-DB-iter-1}
\er
Similarly we have:
\br
\Bigl( T^{(m-1;n)}_{\phi_m}\Bigr)^{\ast\, -1} \ldots 
\Bigl( T^{(0;n)}_{\phi_1}\Bigr)^{\ast\, -1}
T^{(0;n-1)}_{\psi_n} \ldots T^{(0;0)}_{\psi_1} (f)
\nonu \\
= \frac{{\wti W}_{n+1;m} \lb\psi_1,\ldots ,\psi_n, f; \phi_1,\ldots ,\phi_m \rb}{
{\wti W}_{n;m} \lb\psi_1,\ldots ,\psi_n ; \phi_1,\ldots ,\phi_m \rb}
\quad ,\quad {\rm for}\;\; n \geq m
\lab{DB-adj-DB-iter-adj-EF-2}
\er
\br
T^{(m-1;n)}_{\phi_m} \ldots T^{(0;n)}_{\phi_1}
\Bigl( T^{(0;n-1)}_{\psi_n}\Bigr)^{\ast\, -1}
\ldots \Bigl( T^{(0;0)}_{\psi_1}\Bigr)^{\ast\, -1} (f)
\nonu \\
= - \frac{{\wti W}_{n;m+1} \lb\psi_1,\ldots ,\psi_n ; \phi_1,\ldots ,\phi_m,f \rb}{
{\wti W}_{n;m} \lb\psi_1,\ldots ,\psi_n ; \phi_1,\ldots ,\phi_m \rb}
\quad ,\quad {\rm for}\;\; m \geq n+1
\lab{DB-adj-DB-iter-EF-2}
\er
where:
\be
T^{(0;l-1)}_{\psi_l} \equiv \psi_l^{(0;l-1)} D \bigl(\psi_l^{(0;l-1)}\bigr)^{-1}
\quad ,\quad
\psi_l^{(0;l-1)} = T^{(0;l-2)}_{\psi_{l-1}} \ldots T^{(0;0)}_{\psi_1} (\psi_l)
\lab{DB-iter-2}
\ee
\br
\Bigl( T^{(l-1;n)}_{\phi_l}\Bigr)^{\ast\, -1} \equiv 
- \bigl(\phi_l^{(l-1;n)}\bigr)^{-1} D^{-1} \phi_l^{(l-1;n)} \phantom{aaaaaaaaaaaa}
\nonu \\
\phi_l^{(l-1;n)} = T^{(l-2;n)}_{\phi_{l-1}} \ldots T^{(;n0)}_{\phi_1} 
\Bigl( T^{(0;n-1)}_{\psi_n}\Bigr)^{\ast\, -1} \ldots 
\Bigl( T^{(0;0)}_{\psi_1}\Bigr)^{\ast\, -1} (\phi_l)
\lab{adj-DB-iter-2}
\er

\mark
Before proceeding lets us recall the following simple property.
A pair of successive DB and adjoint-DB transformations
of the first type \rf{DB-EF-1-1}--\rf{adj-DB-EF-neg-1} w.r.t.
$\phi \equiv \P_{i_0}$ and $\psi \equiv \Psi_{i_0}$, respectively, yield an
identity combined transformation due to \rf{DB-EF-1-1}. Similar property
applies also to the pair of successive DB and adjoint-DB transformations
of the second type \rf{DB-EF-2}--\rf{adj-DB-EF-neg-2-2} w.r.t.
$\phi \equiv \P^{-1}_{a_0}$ and $\psi \equiv \Psi^{-1}_{a_0}$, respectively,
due to \rf{DB-EF-neg-2-1}. 
\sskp
Therefore, the general DB orbit can be labeled by two non-negative 
integer-valued vectors:
\be
\vec{m} \equiv \Bigl( m_1,\ldots ,m_M, {\bar m}_2,\ldots ,{\bar m}_{M+R}\Bigr)
\lab{vec-m}
\ee
\be
\vec{n} \equiv\Bigl( n_1,\ldots ,n_M , {\bar n}_2,\ldots ,{\bar n}_{M+R}\Bigr)
\lab{vec-n}
\ee
where each enrty $m_s$ in
\rf{vec-m} indicates $m_s$ DB steps w.r.t. $\phi = \P_{i_s}$, and each
entry ${\bar m}_s$ indicates ${\bar m}_s$ DB steps w.r.t.
$\phi = \P^{(-1)}_{a_s}$. Similarly, each entry $n_s$ in
\rf{vec-n} indicates $n_s$ adjoint-DB steps w.r.t. $\psi = \Psi_{i_s}$,
and each entry ${\bar n}_s$ indicates ${\bar n}_s$ adjoint-DB steps w.r.t.
$\psi = \Psi^{(-1)}_{a_s}$. According to the above remark one of the two
integers in each pair $(m_s, n_s)$ and $({\bar m}_s, {\bar n}_s)$ must be zero.

Taking into account \rf{DB-EF-1-1}--\rf{adj-DB-EF-neg-2-2}, the (adjoint) DB
iterations along the general DB orbit are given (up to overall signs) by 
\rf{DB-adj-DB-iter-EF-1}--\rf{DB-adj-DB-iter-adj-EF-1} and
\rf{DB-adj-DB-iter-adj-EF-2}--\rf{DB-adj-DB-iter-EF-2} where
(using the short-hand notations \rf{EF-cKP-sys},\rf{EF-sys-inverse}) :
\br
\lcurl \phi_1,\ldots ,\phi_m \rcurl \equiv \lcurl \P \(\vec{m}\)\rcurl =
\phantom{aaaaaaaaaaaaaaaaaaaa}
\nonu \\
\lcurl \P^{(1)}_{1},\ldots ,\P^{(m_1)}_{1}; \ldots ; \P^{(1)}_{M},\ldots
,\P^{(m_M)}_{M}; \P^{(-1)}_{2},\ldots ,\P^{(-{\bar m}_2)}_{2}; \ldots ;
\P^{(-1)}_{M+R},\ldots ,\P^{(-{\bar m}_{M+R})}_{M+R} \rcurl 
\lab{DB-orbit}
\er
\br
\lcurl \psi_1,\ldots ,\psi_n \rcurl \equiv \lcurl \Psi\(\vec{n}\)\rcurl =
\phantom{aaaaaaaaaaaa}
\nonu \\
\lcurl \Psi^{(1)}_{1},\ldots ,\Psi^{(n_1)}_{1}; \ldots ; 
\Psi^{(1)}_{M},\ldots ,\Psi^{(n_M)}_{M}; \Psi^{(-1)}_{2},\ldots ,
\P^{(-{\bar n}_2)}_{2} ; \ldots ; \Psi^{(-1)}_{M+R},\ldots ,
\P^{(-{\bar n}_{M+R})}_{M+R} \rcurl
\lab{adj-DB-orbit}
\er
with:
\be
|\vec{m}| \equiv \sum_{k=1}^M m_k + \sum_{l=2}^{M+R} {\bar m}_l
\quad ,\quad
|\vec{n}| \equiv \sum_{k=1}^M n_k + \sum_{l=2}^{M+R} {\bar n}_l
\lab{mod-m-n}
\ee
In what follows, (adjoint) DB-transformed tau-function and (adjoint) 
eigenfunctions of the ${\sl cKP}_{R,M}$ hierarchy \rf{Lax-R-M} along the DB 
orbit \rf{DB-orbit}--\rf{adj-DB-orbit} will be denoted as 
$\t_{(\vec{m};\vec{n})}$ , $\P_{i,(\vec{m};\vec{n})}$ and 
$\Psi_{i,(\vec{m};\vec{n})}$.

Using the short-hand notations from \rf{DB-orbit}--\rf{adj-DB-orbit} and the
notation for the generalized Wronskian-like determinants \rf{wti-Wronski},
the general DB solutions for the tau-function and the constituent (adjoint)
eigenfunctions of ${\sl cKP}_{R,M}$ hierarchies \rf{Lax-R-M} can be written
in the following compact form:
\be
\frac{\t_{(\vec{m};\vec{n})}}{\t^{(0;0)}} = (-1)^{|\vec{n}|(|\vec{n}|-1)/2}\;
{\wti W}_{\vec{m};\vec{n}} \Bigl\lb 
\lcurl \P \(\vec{m}\)\rcurl; \lcurl \Psi \(\vec{n}\)\rcurl
\Bigr\rb  \quad ,\quad {\rm for}\;\; |\vec{m}| \geq |\vec{n}|
\lab{tau-orbit-1}
\ee
\be
\P_{i,(\vec{m};\vec{n})} =  (-1)^{|\vec{n}|}
\frac{{\wti W}_{\vec{m}^{(i)}_{+};\vec{n}} \Bigl\lb 
\lcurl \P \(\vec{m}^{(i)}_{+}\)\rcurl; \lcurl \Psi \(\vec{n}\)\rcurl\Bigr\rb}{
{\wti W}_{\vec{m};\vec{n}} \Bigl\lb 
\lcurl \P \(\vec{m}\)\rcurl; \lcurl \Psi \(\vec{n}\)\rcurl\Bigr\rb}
\quad ,\quad {\rm for}\;\; |\vec{m}| \geq |\vec{n}| \quad ,\quad n_i =0
\lab{EF-orbit-1-a}
\ee
\be
\P_{i,(\vec{m};\vec{n})} =  (-1)^{|\vec{n}|-1}
\frac{{\wti W}_{\vec{m};\vec{n}^{(i)}_{-}} \Bigl\lb 
\lcurl \P \(\vec{m}\)\rcurl; \lcurl \Psi \(\vec{n}^{(i)}_{-}\)\rcurl\Bigr\rb}{
{\wti W}_{\vec{m};\vec{n}} \Bigl\lb 
\lcurl \P \(\vec{m}\)\rcurl; \lcurl \Psi \(\vec{n}\)\rcurl\Bigr\rb}
\quad ,\quad {\rm for}\;\; |\vec{m}| \geq |\vec{n}| \quad ,\quad n_i \geq 1
\lab{EF-orbit-1-b}
\ee
\be
\Psi_{i,(\vec{m};\vec{n})} = (-1)^{|\vec{n}|-1}
\frac{{\wti W}_{\vec{m};\vec{n}^{(i)}_{+}} \Bigl\lb 
\lcurl \P \(\vec{m}\)\rcurl; \lcurl \Psi \(\vec{n}^{(i)}_{+}\)\rcurl\Bigr\rb}{
{\wti W}_{\vec{m};\vec{n}} \Bigl\lb 
\lcurl \P \(\vec{m}\)\rcurl; \lcurl \Psi \(\vec{n}\)\rcurl\Bigr\rb}
\quad ,\quad {\rm for}\;\; |\vec{m}| \geq |\vec{n}|+1 \quad ,\quad m_i =0
\lab{adj-EF-orbit-1-a}
\ee
\be
\Psi_{i,(\vec{m};\vec{n})} = (-1)^{|\vec{n}|}
\frac{{\wti W}_{\vec{m}^{(i)}_{-};\vec{n}} \Bigl\lb 
\lcurl \P \(\vec{m}^{(i)}_{-}\)\rcurl; \lcurl \Psi \(\vec{n}\)\rcurl\Bigr\rb}{
{\wti W}_{\vec{m};\vec{n}} \Bigl\lb 
\lcurl \P \(\vec{m}\)\rcurl; \lcurl \Psi \(\vec{n}\)\rcurl\Bigr\rb}
\quad ,\quad {\rm for}\;\; |\vec{m}| \geq |\vec{n}| \quad ,\quad m_i \geq 1
\lab{adj-EF-orbit-1-b}
\ee
In Eqs.\rf{EF-orbit-1-a}--\rf{adj-EF-orbit-1-b} the following notations are
used: the vectors $\vec{m}^{(i)}_{\pm}$ are obtained from $\vec{m}$ 
\rf{vec-m} by the shift $m_i \to m_i \pm 1$ and, similarly, the vectors 
$\vec{n}^{(i)}_{\pm}$ are obtained from $\vec{n}$ \rf{vec-n} by the shift 
$n_i \to n_i \pm 1$.

Similarly we obtain: 
\be
\frac{\t_{(\vec{m};\vec{n})}}{\t^{(0;0)}} = (-1)^{|\vec{n}|(|\vec{n}|-1)/2}\;
{\wti W}_{\vec{n};\vec{m}} \Bigl\lb 
\lcurl \Psi \(\vec{n}\)\rcurl; \lcurl \P \(\vec{m}\)\rcurl\Bigr\rb  
\quad ,\quad {\rm for}\;\; |\vec{n}| \geq |\vec{m}|
\lab{tau-orbit-2}
\ee
\be
\P_{i,(\vec{m};\vec{n})} = (-1)^{|\vec{n}|}
\frac{{\wti W}_{\vec{n};\vec{m}_{+}} \Bigl\lb 
 \lcurl \Psi \(\vec{n}\)\rcurl;\lcurl \P\(\vec{m}_{+}\)\rcurl\Bigr\rb}{
{\wti W}_{\vec{n};\vec{m}} \Bigl\lb 
\lcurl \Psi \(\vec{n}\)\rcurl; \lcurl \P \(\vec{m}\)\rcurl\Bigr\rb}
\quad ,\quad {\rm for}\;\; |\vec{n}| \geq |\vec{m}|+1  \quad ,\quad n_i = 0
\lab{EF-orbit-2-a}
\ee
\be
\P_{i,(\vec{m};\vec{n})} = (-1)^{|\vec{n}|-1}
\frac{{\wti W}_{\vec{n}^{(i)}_{-};\vec{m}} \Bigl\lb 
 \lcurl \Psi \(\vec{n}^{(i)}_{-}\)\rcurl;\lcurl \P\(\vec{m}\)\rcurl\Bigr\rb}{
{\wti W}_{\vec{n};\vec{m}} \Bigl\lb 
\lcurl \Psi \(\vec{n}\)\rcurl; \lcurl \P \(\vec{m}\)\rcurl\Bigr\rb}
\quad ,\quad {\rm for}\;\; |\vec{n}| \geq |\vec{m}|+1  \quad ,\quad n_i \geq 1
\lab{EF-orbit-2-b}
\ee
\be
\Psi_{i,(\vec{m};\vec{n})} = (-1)^{|\vec{n}|-1}
\frac{{\wti W}_{\vec{n}^{(i)}_{+};\vec{m}} \Bigl\lb 
\lcurl \Psi \(\vec{n}^{(i)}_{+}\)\rcurl; \lcurl \P \(\vec{m}\)\rcurl\Bigr\rb}{
{\wti W}_{\vec{n};\vec{m}} \Bigl\lb 
\lcurl \Psi \(\vec{n}\)\rcurl; \lcurl \P \(\vec{m}\)\rcurl\Bigr\rb}
\quad ,\quad {\rm for}\;\; |\vec{n}| \geq |\vec{m}|  \quad ,\quad m_i = 0
\lab{adj-EF-orbit-2-a}
\ee
\be
\Psi_{i,(\vec{m};\vec{n})} = (-1)^{|\vec{n}|}
\frac{{\wti W}_{\vec{n};\vec{m}^{(i)}_{-}} \Bigl\lb 
\lcurl \Psi \(\vec{n}\)\rcurl; \lcurl \P \(\vec{m}^{(i)}_{-}\)\rcurl\Bigr\rb}{
{\wti W}_{\vec{n};\vec{m}} \Bigl\lb 
\lcurl \Psi \(\vec{n}\)\rcurl; \lcurl \P \(\vec{m}\)\rcurl\Bigr\rb}
\quad ,\quad {\rm for}\;\; |\vec{n}| \geq |\vec{m}|  \quad ,\quad m_i \geq 1
\lab{adj-EF-orbit-2-b}
\ee

\mskp
\underline{\em 8.3 Multiple-Wronskian Solutions}
\sskp
Let us take a closer look at the DB solution for the tau-function
\rf{tau-orbit-1} of ${\sl cKP}_{R,M}$ integrable hierarchies \rf{Lax-R-M}
in the case of DB orbits with ${\bar m}_2 =\ldots = {\bar m}_{M+R}=0$ and
$n_1 =\ldots = n_M = 0$ in \rf{DB-orbit}--\rf{adj-DB-orbit} 
(below ${\bar M} \equiv M+R$) :
\be
{\wti W}_{\vec{m},\vec{n}} \Bigl\lb \P^{(1)}_1,\ldots ,\P^{(m_1)}_1, \ldots ,
\P^{(1)}_M,\ldots ,\P^{(m_M)}_M ; \Psi^{(-1)}_2,\ldots ,\Psi^{(-{\bar n}_2)}_2,
\ldots ,\Psi^{(-1)}_{{\bar M}} ,\ldots ,
\Psi^{(-{\bar n}_{{\bar M}})}_{{\bar M}} \Bigr\rb =
\lab{tau-orbit-0-det}
\ee
\br
\phantom{aaaaaaaaaaaaaaaaaaaaaaaaaaaaaaaaaaaaaaaaaa}
\nonu \\
\det\left\Vert
\begin{array}{cccccccc}
\P^{(1)}_1 & \cdots & \P^{(m_1)}_1 & \cdots & \P^{(1)}_M & \cdots &
\P^{(m_M)}_M   \\   
\vdots & \ddots & \vdots & \ddots & \vdots & \ddots & \vdots \\
\pa^{m-n-1} \P^{(1)}_1 & \cdots & \pa^{m-n-1} \P^{(m_1)}_1 & \cdots & 
\pa^{m-n-1} \P^{(1)}_M & \cdots & \pa^{m-n-1} \P^{(m_M)}_M   \\   
\pai \bigl(\P^{(1)}_1 \Psi^{(-1)}_2\bigr) & \cdots &
\pai \bigl(\P^{(m_1)}_1 \Psi^{(-1)}_2\bigr) & \cdots &
\pai \bigl(\P^{(1)}_M \Psi^{(-1)}_2\bigr) & \cdots &
\pai \bigl(\P^{(m_M)}_M \Psi^{(-1)}_2\bigr) \\  
\vdots & \ddots & \vdots & \ddots & \vdots & \ddots & \vdots \\
\pai \bigl(\P^{(1)}_1 \Psi^{(-{\bar n}_2)}_2\bigr) & \cdots &
\pai \bigl(\P^{(m_1)}_1 \Psi^{(-{\bar n}_2)}_2\bigr) & \cdots &
\pai \bigl(\P^{(1)}_M \Psi^{(-{\bar n}_2)}_2\bigr) & \cdots &
\pai \bigl(\P^{(m_M)}_M \Psi^{(-{\bar n}_2)}_2\bigr) \\  
\vdots & \ddots & \vdots & \ddots & \vdots & \ddots & \vdots \\
\vdots & \ddots & \vdots & \ddots & \vdots & \ddots & \vdots \\
\pai \bigl(\P^{(1)}_1 \Psi^{(-1)}_{{\bar M}}\bigr) & \cdots &
\pai \bigl(\P^{(m_1)}_1 \Psi^{(-1)}_{{\bar M}}\bigr) & \cdots &
\pai \bigl(\P^{(1)}_M \Psi^{(-1)}_{{\bar M}}\bigr) & \cdots &
\pai \bigl(\P^{(m_M)}_M \Psi^{(-1)}_{{\bar M}}\bigr) \\  
\vdots & \ddots & \vdots & \ddots & \vdots & \ddots & \vdots \\
\pai \bigl(\P^{(1)}_1 \Psi^{(-{\bar n}_{{\bar M}})}_{{\bar M}}\bigr) & \cdots &
\pai \bigl(\P^{(m_1)}_1 \Psi^{(-{\bar n}_{{\bar M}})}_{{\bar M}}\bigr) & \cdots &
\pai \bigl(\P^{(1)}_M \Psi^{(-{\bar n}_{{\bar M}})}_{{\bar M}}\bigr) & \cdots &
\pai \bigl(\P^{(m_M)}_M \Psi^{(-{\bar n}_{{\bar M}})}_{{\bar M}}\bigr)  
\end{array}
\right\Vert
\nonu
\er

In refs.\ct{multi-comp-KP} the following
properties were shown to hold for SEP (square eigenfunction potential, cf. 
\rf{potentialflo}--\rf{SEP-def}) functions of the type entering in
\rf{tau-orbit-0-det}, namely, such that
the second adjoint eigenfunction belongs to the set of additional 
``isospectral'' flow generating (adjoint) eigenfunctions 
\rf{ghost-k-neg},\rf{flow-n-k-EF-neg} whereas the first eigenfunction
transforms homogeneously w.r.t. the corresponding additional 
``isospectral'' flows as in \rf{flow-n-a-EF-neg}.
SEP functions of the form (using short-hand notations
\rf{EF-cKP-sys},\rf{EF-sys-inverse}) :
\be
\st{k}{F}\!\!\!{}_{i,N} \equiv \pai \bigl( \P^{(N)}_i \Psi^{(-1)}_k\bigr)
\qquad N \geq 1 \;\; ,\;\; i=1,\ldots, M \;\; ,\;\; k=2,\ldots , M+R
\lab{EF-k}
\ee
which enter \rf{tau-orbit-0-det}, obey eigenfunction-type purely differential
equations (analogous to \rf{EF-def}) w.r.t. the additional ``isospectral'' 
parameters $\Bigl\{ \st{k}{t}\!\!\!{}_{-n}\Bigr\}_{n=1}^\infty$ :
\be
\pa/\pa \st{k}{t}\!\!\!{}_{-n} \st{k}{F}\!\!\!{}_{i,N} =
\Bigl\lb \st{k}{\pa}\!{}^n + \ldots \Bigr\rb \st{k}{F}\!\!\!{}_{i,N}
\quad ,\quad
\st{k}{\pa} \equiv \pa/\pa \st{k}{t}\!\!\!{}_{-1}
\lab{EF-k-eqs}
\ee
where the dots indicate lower-order derivative terms w.r.t. $\st{k}{\pa}$
with coefficients depending on the additional 
``isospectral'' flow generating (adjoint) eigenfunctions 
\rf{ghost-k-neg},\rf{flow-n-k-EF-neg}. Furthermore, for the rest of the SEP's
in \rf{tau-orbit-0-det} we have:
\be
\pai \bigl( \P^{(N)}_i \Psi^{(-{\bar n})}_k\bigr) = 
\st{k}{\pa}\!{}^{{\bar n} -1} \st{k}{F}\!\!\!{}_{i,N}  + \ldots
\quad ,\quad {\bar n} \geq 2
\lab{SEP-higher}
\ee
where the dots indicate terms which yield vanishing contribution in the
Wronskian-like determinant \rf{tau-orbit-0-det}. Thereby 
\rf{tau-orbit-0-det} acquires the following {\em multiple-Wronskian} form:
\br
\cW_{\vec{m},\vec{n}}
\Bigl\lb \P^{(1)}_1,\ldots ,\P^{(m_M)}_M, \st{2}{F}\!\!\!{}_{1,1},\ldots ,
\st{2}{F}\!\!\!{}_{M,m_M},\ldots ,\st{{\bar M}}{F}\!\!\!{}_{1,1},\ldots ,
\st{{\bar M}}{F}\!\!\!{}_{M,m_M} \Bigr\rb =
\nonu \\
\phantom{aaaaaaaaaaaaaaaaa}  
\nonu \\
\det\left\Vert
\begin{array}{cccccccc}
\P^{(1)}_1 & \cdots & \P^{(m_1)}_1 & \cdots & \P^{(1)}_M & \cdots &
\P^{(m_M)}_M   \\   
\vdots & \ddots & \vdots & \ddots & \vdots & \ddots & \vdots \\
\pa^{m-n-1} \P^{(1)}_1 & \cdots & \pa^{m-n-1} \P^{(m_1)}_1 & \cdots & 
\pa^{m-n-1} \P^{(1)}_M & \cdots & \pa^{m-n-1} \P^{(m_M)}_M   \\   
\st{2}{F}\!\!\!{}_{1,1} & \cdots & \st{2}{F}\!\!\!{}_{1,m_1}& \cdots &
\st{2}{F}\!\!\!{}_{M,1}& \cdots & \st{2}{F}\!\!\!{}_{M,m_M}\\  
\vdots & \ddots & \vdots & \ddots & \vdots & \ddots & \vdots \\
\st{2}{\pa}\!\!\!{}^{{\bar n}_2 -1}\!\! \st{2}{F}\!\!\!{}_{1,1} & \cdots &
\st{2}{\pa}\!\!\!{}^{{\bar n}_2 -1}\!\! \st{2}{F}\!\!\!{}_{1,m_1}& \cdots &
\st{2}{\pa}\!\!\!{}^{{\bar n}_2 -1}\!\! \st{2}{F}\!\!\!{}_{M,1} & \cdots &
\st{2}{\pa}\!\!\!{}^{{\bar n}_2 -1}\!\! \st{2}{F}\!\!\!{}_{M,m_M} \\  
\vdots & \ddots & \vdots & \ddots & \vdots & \ddots & \vdots \\
\vdots & \ddots & \vdots & \ddots & \vdots & \ddots & \vdots \\
\st{{\bar M}}{F}\!\!\!{}_{1,1}  & \cdots &
\st{{\bar M}}{F}\!\!\!{}_{1,m_1} & \cdots &
\st{{\bar M}}{F}\!\!\!{}_{M,1} & \cdots &
\st{{\bar M}}{F}\!\!\!{}_{M,m_M} \\  
\vdots & \ddots & \vdots & \ddots & \vdots & \ddots & \vdots \\
\st{{\bar M}}{\pa}\!\!\!{}^{{\bar n}_{\bar M} -1}\!\!
\st{{\bar M}}{F}\!\!\!{}_{1,1} & \cdots &
\st{{\bar M}}{\pa}\!\!\!{}^{{\bar n}_{\bar M} -1}\!\!
\st{{\bar M}}{F}\!\!\!{}_{1,m_1} & \cdots &
\st{{\bar M}}{\pa}\!\!\!{}^{{\bar n}_{\bar M} -1}\!\!
\st{{\bar M}}{F}\!\!\!{}_{M,1} & \cdots &
\st{{\bar M}}{\pa}\!\!\!{}^{{\bar n}_{\bar M} -1}\!\!
\st{{\bar M}}{F}\!\!\!{}_{M,m_M}   
\end{array}
\right\Vert
\lab{multi-Wronski}
\er
where:
\be
{\bar M} \equiv M+R \quad, \quad
\vec{m} = (m_1,\ldots ,m_M) \quad, \quad 
\vec{n} = ({\bar n}_2,\ldots ,{\bar n}_{{\bar M}}) \quad, \quad 
m = \sum_{i=1}^M m_i   \quad, \quad  n = \sum_{a=2}^{\bar M} {\bar n}_a
\lab{multi-Wronski-notat}
\ee
and notations from \rf{EF-k-eqs} are used. Accordingly, the expressions 
\rf{EF-orbit-1-a} and \rf{adj-EF-orbit-1-b} for the DB-transformed (adjoint) 
eigenfunctions can be written, using notations in 
\rf{multi-Wronski}--\rf{multi-Wronski-notat}, as a ratio of multiple
Wronskians \rf{multi-Wronski} :
\be
\P_{i,(\vec{m};\vec{n})} =  (-1)^{n}
\frac{\cW_{\vec{m}^{(i)}_{+};\vec{n}} \Bigl\lb \ldots ,\P^{(1)}_i, \ldots,
\P^{(m_i)}_i,\P^{(m_i +1)}_i, \ldots \Bigr\rb}{\cW_{\vec{m};\vec{n}} \Bigl\lb 
\ldots ,\P^{(1)}_i, \ldots,\P^{(m_i)}_i,\ldots \Bigr\rb}
\quad ,\quad {\rm for}\;\; m \geq n
\lab{multi-Wronski-EF-1-a}
\ee
\be
\Psi_{i,(\vec{m};\vec{n})} = (-1)^n
\frac{\cW_{\vec{m}^{(i)}_{-};\vec{n}} \Bigl\lb 
\ldots ,\P^{(1)}_i, \ldots,\P^{(m_i -1)}_i,\ldots 
\Bigr\rb}{
\cW_{\vec{m};\vec{n}} \Bigl\lb 
\ldots ,\P^{(1)}_i, \ldots,\P^{(m_i -1)}_i, \P^{(m_i)}_i,\ldots \Bigr\rb}
\quad ,\quad {\rm for}\;\; m \geq n \quad ,\quad m_i \geq 1
\lab{multi-Wronski-adj-EF-1-b}
\ee

In particular, we are interested in DB orbits passing through the ``free'' 
initial point, {\sl i.e.}, where the initial hierarchy is given by the ``free''
Lax operator $\cL = D^R$. In this case $\P^{(N)}_i$ and $\Psi^{(N)}_i$ are 
``free'' (adjoint) eigenfunctions given by:
\be
\P^{(N)}_i (t) = \int d\l\, \l^{NR}\,\vp_i (\l) e^{\xi (t,\l)}
\quad ,\quad
\Psi^{(N)}_i (t) = \int d\l\, \l^{NR}\,\psi_i (\l) e^{-\xi (t,\l)}
\lab{free-EF}
\ee
with the same notations as in \rf{BA-def} and \rf{spec-repr}, where
$\vp_i (\l)$ and $\psi_i (\l)$ are arbitrary spectral densities.
Accordingly, the SEP functions, which appear as matrix elements
of the generalized Wronskian-like determinants \rf{wti-Wronski} entering the
DB solutions \rf{tau-orbit-1}--\rf{adj-EF-orbit-2-b}, acquire the following
``free'' form ({\sl i.e}, set the tau-function $\t (t)=1$ in Eq.\rf{SEP-spec}) :
\be
\pai\(\P^{(N_1)}_i (t)\,\Psi^{(N_2)}_j (t)\) = 
\int\!\int d\l d\m \,\l^{N_1R}\m^{N_2R}\;\frac{\vp_i (\l)\,\psi_j (\m)}{\l - \m}
\; e^{\xi (t,\l) - \xi (t,\m)}
\lab{free-SEP}
\ee
Similarly, in the ``free'' case
the functions $\st{k}{F}\!\!\!{}_{i,N}$ satisfy the linearized
Eqs.\rf{EF-k-eqs} and, therefore, are given by expressions similar to 
\rf{free-EF} :
\be
\st{k}{F}\!\!\!{}_{i,N} (\st{k}{t}) = \int d\m\, \st{k}{f}\!\!\!{}_{i,N}(\m)
e^{\xi (\st{k}{t},\m)}  \quad ,\quad
\xi (\st{k}{t}\!\! ,\m) = \sum_{n=1}^\infty \m^n \st{k}{t}\!\!\!{}_n
\lab{free-EF-k}
\ee
with arbitrary spectral densities $\st{k}{f}\!\!\!{}_{i,N}(\m)$.

Substituting \rf{free-EF}--\rf{free-EF-k} into 
\rf{multi-Wronski-EF-1-a}--\rf{multi-Wronski-adj-EF-1-b}, 
or in the more general DB-orbit expressions
\rf{tau-orbit-1}--\rf{adj-EF-orbit-2-b},
we obtain explicit DB solutions in generalized Wronskian-like form
\rf{wti-Wronski}, in particular, in multiple-Wronskian \rf{multi-Wronski} form
of the whole class of ${\sl cKP}_{R,M}$ reduced KP hierarchies \rf{Lax-R-M}.

In the special case $R=1$, plugging into $A^{(0)}$ \rf{constr-1}
the expressions for $\P_1,\ldots ,\P_M$ and $\Psi_1,\ldots ,\Psi_M$
given by \rf{multi-Wronski-EF-1-a}--\rf{multi-Wronski-adj-EF-1-b}
or more generally by \rf{tau-orbit-1}--\rf{adj-EF-orbit-2-b} with the 
underlying (adjoint) eigenfunctions and SEP functions
of the simple ``free'' forms \rf{free-EF}--\rf{free-EF-k},
we obtain explicit DB solutions for $SL(M+1)/ U(1)\times SL(M)\,$
gauged WZNW field equations of motion \rf{gauged-WZNW-eq-M}. These solutions
depend, apart from the original'' light-cone'' coordinates
$\( x \equiv x_{-}, x_{+}\)$, identified within the context of the underlying
extended ${\sl cKP}_{1,M}$ hierarchy with the lowest ``isospectral'' flow
parameters $x \equiv x_{-} \equiv t_1$ and $x_{+} \equiv - \st{M+1}{t_{-1}}$,
also on the rest of the infinite set of ``isospectral'' times:
\be
\Bigl( t_1 \equiv x \equiv x_{-},t_2,t_3, \ldots ; \st{2}{t}\!\!\!{}_1 ,
\st{2}{t}\!\!\!{}_2, \st{2}{t}\!\!\!{}_3 , \ldots ; \ldots ; 
\st{M+1}{t_{-1}} \equiv -x_{+},\st{M+1}{t_{-2}},\st{M+1}{t_{-3}},\ldots \Bigr)
\lab{isospec-times}
\ee
This additional dependence can be viewed as Abelian symmetry deformations
for the corresponding solutions of gauged WZNW equations of motion. Moreover,
from the results in the Appendix we deduce that
the whole non-Abelian additional symmetry algebra of the pertinent
${\sl cKP}_{1,M}$ hierarchy (\rf{loop-alg-full} with $R=1$) acts as symmetry
deformation algebra on the space of DB solutions of gauged
$SL(M+1)/U(1)\times SL(M)\,$ WZNW equations of motion.

Concluding this section, let us consider as the simplest non-trivial example
the ${\sl cKP}_{1,1}$-based extended KP-type hierarchy with Lax operator
$\cL = D + \P D^{-1} \Psi$ which is equivalently described within the
generalized Drinfeld-Sokolov framework by \rf{SL-2}--\rf{T-asy-tau}. 
In this case the multiple Wronskian tau-function \rf{multi-Wronski} simplifies 
to the following double-Wronskian form:
\be
\cW_{m,n} = \det\left\Vert
\begin{array}{ccc}
\P^{(1)} & \cdots & \P^{(m)} \\
\vdots & \ddots & \vdots  \\
\pa^{m-n-1}\P^{(1)} & \cdots & \pa^{m-n-1}\P^{(m)} \\
F_1 & \cdots & F_m \\
\vdots & \ddots & \vdots  \\
\bpa^{n-1} F_1 & \cdots & \bpa^{n-1} F_m  \end{array} \right\Vert
\lab{double-Wronski}
\ee
where:
\be
\P^{(l)} \equiv \P^{(l)}_1  \quad ,\quad F_l \equiv \st{2}{F}\!\!\!{}_l
\quad ,\quad m \equiv m_1 \;\; ,\;\; n \equiv {\bar n}_2 \quad ,\quad
\bt_l \equiv \st{2}{t}\!\!\!{}_l \;\; ,\;\; \bpa \equiv \st{2}{\pa}
\lab{double-Wronski-notat}
\ee
\be
\P^{(l)} (t) = \int d\l \,\l^l \vp (\l) e^{\xi (t,\l)} \quad ,\quad
F_l (\bt ) = \int d\m f_l (\m) e^{\xi (t,\m)}
\lab{double-Wronski-EF}
\ee
For a special delta-function choice of the spectral densities of
$\P^{(l)} (t)$ and $F_l (\bt )$ \rf{double-Wronski-EF}, the double Wronskian 
tau-function \rf{double-Wronski} coincides with the double Wronskian 
tau-function of first ref.\ct{H-H} which describes the well-known 
(multi-)dromion solutions \ct{Boiti-etal} of Davey-Stewartson system 
\rf{DS-dyn-1}--\rf{DS-nondyn}.

\lskip
{\bf 9. Conclusions and Outlook}
\mskp
In the first part of the present paper we discussed the subject of gauging
of geometric actions on coadjoint orbits of arbitrary (infinite-dimensional)
groups with central extensions (with examples given in Section 3). The main
tool for our construction of gauged geometric actions on general coadjoint
orbits was the group composition law \rf{pw} which generalizes the well-known
Polyakov-Wiegmann composition formula for ordinary WZNW geometric actions.
Furthermore, we have shown that the equations of motions of the pertinent
gauged geometric actions possess a ``zero-curvature'' representation
(Section 5) on the underlying group coadjoint orbit. 

It is a very interesting problem for further study to work out explicit
examples of physically relevant gauged geometric actions on coadjoint orbits of
infinite-dimensional groups with central extensions ({\sl e.g.} for those in
the examples in Section 3) which go beyond the well-known case of Kac-Moody
groups yielding the ordinary gauged WZNW actions. For instance, one can
study gauging of ${\bf W_{\infty}}$-geometric action \rf{dop-action}
with a gauging subgroup whose algebra is the Cartan subalgebra of
${\bf W_{\infty}}$-algebra consisting of differential operators of the type
$\lcurl x^n D^n \rcurl_{n=0}^\infty$ (cf. ref.\ct{dop-tmf}).

In the second part of the paper we showed that gauged $G/H\,$ WZNW field
equations can be identified with the lowest negative-grade additional
symmetry flow equations of generalized Drinfeld-Sokolov integrable
hierarchies based on ${\widehat \cG}$ -- the loop algebra corresponding to
$\cG$, the Lie algebra of $G$. Next we discussed in more detail the case of
generalized Drinfeld-Sokolov hierarchies based on ${\widehat {SL}}(M+R)$ and
explicitly established their equivalence with the class ${\sl cKP}_{R,M}$ of 
constrained (reduced) KP hierarchies. We described in detail the whole loop 
algebra of additional non-isospectral symetries of ${\sl cKP}_{R,M}$. Adding 
the flows from the Cartan subalgebra of the underlying additional-symmetry loop
algebra, we constructed extended ${\sl cKP}_{R,M}$-based integrable hierarchies
(subsection 6.4) with multiple sets of ``isospectral'' flows equivalent to
certain reductions of multicomponent (matrix) KP hierarchies. Apart from the
gauged WZNW field equations mentioned above, other higher-dimensional
nonlinear systems such as Davey-Stewartson and $N$-wave resonant systems
were shown to arise as symmetry flow equations of ${\sl cKP}_{R,M}$
integrable hierarchies.

Finally, we provided in Section 8 a detailed derivation of the general \DB
solutions for ${\sl cKP}_{R,M}$ hierarchies preserving the whole loop
algebra of their additional non-isospectral symmetries. These DB solutions
involve generalized Wronskian-like determinants, in particular, multiple
Wronskians and for $R=1$ they contain among themselves solutions to the 
gauged $SL(M+1)/U(1)\times SL(M)\,$ WZNW field equations.

Another important task for further investigation is to study the
physical properties of the very broad class of the multiple-Wronskian
solutions of ${\sl cKP}_{R,M}$ integrable hierarchies given in Section 8,
which contain as simplest examples the well-known (multi-)dromion solutions. 

Also, it is a very interesting problem to exhibit the explicit relation between 
the present multiple-Wronskian solutions of gauged WZNW models and the 
recently found \ct{Galen-braz} solitonic solutions of singular non-Abelian 
affine Toda field theories.
\lskip
{\bf Appendix}
\mskp
The (adjoint) DB-transformed additional-symmetry \rf{loop-alg-full} generating 
operators \rf{flow-n-A} are of the form:
\br
\d^{(n)}_{A} T_\phi \, T^{-1}_\phi + 
T_\phi \cM^{(n)}_{A} T^{-1}_\phi =  \phantom{aaaaaaaaaaaaaaaaaaa}
\nonu \\
\Bigl( T_\phi \bigl( \cM^{(n)}_{A}(\phi) - \d^{(n)}_{A}\phi\bigr)\Bigr)
D^{-1} \phi^{-1} + 
\sum_{i,j=1}^M A^{(n)}_{ij} \sum_{s=1}^n T_\phi (\P_j^{(n+1-s)}) D^{-1} 
T^{\ast\, -1}_\phi (\Psi_i^{(s)})
\lab{DB-A-n}
\er
\br
\Bigl(\d^{(n)}_{A} T^{\ast\, -1}_\psi\Bigr) \, T^{\ast}_\psi + 
T^{\ast\, -1}_\psi \cM^{(n)}_{A} T^{\ast}_\psi =
\phantom{aaaaaaaaaaaaaaaaaaa}
\nonu \\
- \psi^{-1} D^{-1} \biggl( T_\psi \Bigl(\bigl(\cM^{(n)}_{A}\bigr)^\ast (\psi)
+ \d^{(n)}_{A}\psi\Bigr)\biggr) + \sum_{i,j=1}^M A^{(n)}_{ij} 
\sum_{s=1}^n T^{\ast\, -1}_\psi (\P_j^{(n+1-s)}) D^{-1} 
T_\psi (\Psi_j^{(s)})
\lab{adj-DB-A-n}
\er
and accordingly for \rf{flow-n-A-neg} :
\br
\d^{(-n)}_{\cA} T_\phi \, T^{-1}_\phi + 
T_\phi \cM^{(-n)}_{\cA} T^{-1}_\phi =  \phantom{aaaaaaaaaaaaaaaaaaa}
\nonu \\
\Bigl( T_\phi \bigl( \cM^{(-n)}_{\cA}(\phi) - \d^{(-n)}_{\cA}\phi\bigr)\Bigr)
D^{-1} \phi^{-1} + \sum_{a,b=1}^{M+R} \cA^{(-n)}_{ab} 
\sum_{s=1}^n T_\phi (\P_b^{(-n-1+s)}) D^{-1} 
T^{\ast\, -1}_\phi (\Psi_a^{(-s)})
\lab{DB-A-n-neg}
\er
\br
\Bigl(\d^{(-n)}_{\cA} T^{\ast\, -1}_\psi\Bigr) \, T^{\ast}_\psi + 
T^{\ast\, -1}_\psi \cM^{(-n)}_{\cA} T^{\ast}_\psi =
\phantom{aaaaaaaaaaaaaaaaaaa}
\nonu \\
- \psi^{-1} D^{-1} \biggl( T_\psi \Bigl(\bigl(\cM^{(-n)}_{\cA}\bigr)^\ast (\psi)
+ \d^{(-n)}_{\cA}\psi\Bigr)\biggr) + \sum_{a,b=1}^{M+R} \cA^{(-n)}_{ab}
\sum_{s=1}^n T^{\ast\, -1}_\psi (\P_b^{(-n-1+s)}) D^{-1} T_\psi (\Psi_a^{(-s)})
\lab{adj-DB-A-n-neg}
\er
In order for (adjoint) DB transformations to preserve the additonal
symmetries, {\sl i.e.}, to preserve the form of $\cM^{(n)}_A$ and 
$\cM^{(-n)}_\cA$ one term on the r.h.s. of Eqs.\rf{DB-A-n}--\rf{adj-DB-A-n-neg}
must vanish. The first possibility, as in 
\rf{DB-cond-EF-1}-\rf{adj-DB-cond-adj-EF-1}, is the generating (adjoint)
eigenfunctions to transform homogeneously under the additional symmetries:
\begin{itemize}
\item
For positive-grade symmetry flows:
\be
\d^{(n)}_{A}\phi = \cM^{(n)}_{A}(\phi) \quad ,\quad 
\d^{(n)}_{A}\psi = -\(\cM^{(n)}_{A}\)^\ast (\psi)
\lab{flow-homog-A}
\ee
which is fulfilled for:
\be
\phi = L_M \bigl({\bar \vp}_{a_0}\bigr) \equiv \P^{(-1)}_{a_0} \quad ,\quad
\psi = {\bar \psi}_{a_0} \equiv \Psi^{(-1)}_{a_0} \quad ,\quad {\rm for} \;\;
a_0 = {\rm fixed}
\lab{choice-A-1-1}
\ee
due to \rf{flow-n-A-EF-inverse}.
\item
For the negative-grade symmetry flows:
\be
\d^{(-n)}_{\cA}\phi = \cM^{(-n)}_{\cA}(\phi) \quad ,\quad 
\d^{(-n)}_{\cA}\psi = -\(\cM^{(-n)}_{\cA}\)^\ast (\psi)
\lab{flow-homog-A-neg}
\ee
which is fulfilled for:
\be
\phi = \P_{i_0} \quad ,\quad \psi = \Psi_{i_0} \quad , \quad {\rm for} \;\;
i_0 = {\rm fixed}
\lab{choice-A-1-2}
\ee
due to \rf{flow-n-A-neg-EF}.
\end{itemize}
\noindent
The second possibility is one of the terms in the sums on the r.h.s. of
\rf{DB-A-n}--\rf{adj-DB-A-n-neg} to vanish, which implies to make the
opposite choice w.r.t. \rf{choice-A-1-1}--\rf{choice-A-1-2}: 
\begin{itemize}
\item    
Choose \rf{choice-A-1-2} for positive-grade symmetries in 
\rf{DB-A-n}--\rf{adj-DB-A-n} 
\item
Choose \rf{choice-A-1-1} for 
negative-grade symmteries in \rf{DB-A-n-neg}--\rf{adj-DB-A-n-neg}.
\end{itemize}
\sskp
It is easy to find out that these latter choices for (adjoint) DB
transformations preserve a subalgebra of  the additional loop algebra symmetries
\rf{loop-alg-full}. Namely, for positive-grade symmetries
(where initially $A^{(n)} \in U(1) \oplus SL(M)$), we have to restrict the set
of matrices $A^{(n)}$ as follows ($i_0$=fixed) :
\be
A^{(n)}_{ji_0} = A^{(n)}_{i_0 j} = \d_{i_0 j} \quad \longrightarrow \quad
A^{(n)} \in U(1) \oplus SL(M-1)
\lab{restrict-A-plus}
\ee
for consistency with (adjoint) DB transformations w.r.t. \rf{choice-A-1-2}.
For negative-grade symmetries (where initially $\cA^{(-n)} \in SL(M+R)$)
we have to restrict the set of matrices $\cA^{(-n)}$ as ($a_0$=fixed):
\be
\cA^{(-n)}_{ba_0} = \cA^{(-n)}_{a_0 b} = \d_{a_0 b} 
\quad \longrightarrow \quad  \cA^{(-n)} \in SL(M+R-1)
\lab{restrict-A-neg}
\ee
for consistency with (adjoint) DB transformations w.r.t. \rf{choice-A-1-1}.

In conclusion, we obtain:
\begin{itemize}
\item
(Adjoint) DB transformations w.r.t. \rf{choice-A-1-1} preserve
$\({\widehat U}(1)\oplus {\widehat {SL}}(M-1)\)_{+} \oplus
\({\widehat {SL}}(M+R)\)_{-}$ subalgebra of additional symmetries.
\item
(Adjoint) DB transformations w.r.t. \rf{choice-A-1-2} preserve
$\({\widehat U}(1)\oplus {\widehat {SL}}(M)\)_{+} \oplus
\({\widehat {SL}}(M+R-1)\)_{-}$ subalgebra of additional symmetries.
\end{itemize}
\lskip
{\underline{\bf Acknowledgements.}} 
We gratefully acknowledge support from U.S. National Science Foundation 
grant {\sl INT-9724747}. We thank Prof. H. Aratyn and the Physics Department
of University of Illinois at Chicago for hospitality. We are grateful to H.
Aratyn for collaboration in the initial stage of the project.
This work is also partially supported by Bulgarian NSF grant {\sl F-904/99}.

\end{document}